\documentclass[11pt]{article} 
\usepackage[utf8]{inputenc}
\usepackage[version=4]{mhchem}
\usepackage{siunitx}
\usepackage{longtable,tabularx}
\usepackage{amsmath,amssymb,color,hyperref,graphicx,pdfsync,overpic,color,epstopdf,rotating,dashrule,float,bm,multicol,setspace}
\usepackage{adjustbox}
\usepackage{mathrsfs}
\usepackage{mathtools}
\usepackage{subfig}
\usepackage{textcomp}
\usepackage{listings}
\usepackage{fullpage}
\usepackage{array}
\usepackage{comment}
\usepackage[numbers]{natbib}
\newcolumntype{V}{>{\centering\arraybackslash} m{.4\linewidth} }

\DeclareMathOperator*{\argmin}{argmin}

\newcommand\Real{\mbox{Re}} 

\newsavebox{\astrutbox}
\sbox{\astrutbox}{\rule[-5pt]{0pt}{20pt}}

\newcommand{\mat}[1]{{\bm{#1}}}

\newcommand{\mA}{\mat{A}}
\newcommand{\mB}{\mat{B}}
\newcommand{\mC}{\mat{C}}
\newcommand{\mD}{\mat{D}}

\newcommand{\mU}{\mat{U}}
\newcommand{\mV}{\mat{V}}

\newcommand{\mH}{\mat{H}}
\newcommand{\mSigma}{\mat{\Sigma}}

\newcommand{\bx}{{ \bm{x}}}

\newcommand{\xx}{x} 

\newcommand{\dod}[2]{\frac{\partial #1}{\partial #2}}
\newcommand{\ddodd}[2]{\frac{{\rm d} #1}{{\rm d} #2}}

\graphicspath{{Figures/}}
\title{Improved approximations to the Wagner function using sparse identification of nonlinear dynamics}
\author{Scott T.~M.~Dawson \& Steven L.~Brunton}
\begin{document}

\maketitle

\begin{abstract}
The Wagner function in classical unsteady aerodynamic theory represents the response in lift on an airfoil that is subject to a sudden change in conditions. 
While it plays a fundamental role in the development and application of unsteady aerodynamic methods, explicit expressions for this function are difficult to obtain. The Wagner function requires computation of an inverse Laplace transform, or similar inversion, of a non-rational function in the Laplace domain, which is closely related to the Theodorsen function.  
This has led to numerous proposed approximations to the Wagner function, which facilitate convenient and rapid computations. 
While these approximations can be sufficient for many purposes, their behavior is often noticeably different from the true Wagner function, especially for long-time asymptotic behavior. 
In particular, while many approximations have small maximum absolute error across all times, the relative error of the asymptotic behavior can be substantial. 
As well as documenting this error, we propose an alternative approximation methodology that is accurate for all times, for a variety of accuracy measures. 
This methodology casts the Wagner function as the solution of a nonlinear scalar ordinary differential equation, which is identified using a variant of the  sparse identification of nonlinear dynamics (SINDy) algorithm.  
We show that this approach can give accurate approximations using either first- or second-order differential equations. 
We additionally show that this method can be applied to model the analogous lift response for a more realistic aerodynamic system, featuring a finite thickness airfoil and a nonplanar wake. 
\end{abstract}

\section{Introduction}
\label{sec:intro}
Classical unsteady aerodynamic theory, as developed and refined by Wagner~\cite{wagner:25}, Theodorsen~\cite{Theodorsen:35}, von K\'{a}rm\'{a}n and Sears~\cite{karman1938airfoil}, K\"{u}ssner~\cite{kussner1936} Garrick~\cite{Garrick:1938}, and their contemporaries, gives a means to predict the lift response for an airfoil subject to unsteady conditions. 
Such unsteady conditions can arise from rapid changes in the angle of attack or freestream velocity, or encounters with gusts. 
This body of work assumes inviscid flow, where the flow remains attached, with vorticity shed from the airfoil from the trailing edge. 
This vorticity originates due to the imposition of the Kutta condition, which enforces that the rear stagnation point be located at the trailing edge, which is assumed to be sharp. 
The shed vorticity is subsequently assumed to convect with the freestream velocity, and does not dissipate or translate in a direction transverse to the flow.  
While these assumptions required for application of classical unsteady aerodynamic theory are quite strong, it continues to be successfully applied across a range of applications, such as in the modeling and regulation of lift during gust encounters \cite{leung2018modeling,jones2020gust,sedky2020unsteady,andreu2021unsteady}, and for estimating the circulation of vortices produced by vertical gust generators \cite{hufstedler2019vortical}. 
At the core of classical unsteady aerodynamic theory are a small number of related functions that describe the response of the flow, and particularly the lift and moment on the airfoil, to unsteady conditions.  
This work will focus on the Wagner function $\phi(t)$, which describes the lift response to a sudden change in flow conditions. For example, a step change in the angle of attack, $\alpha$ or freestream velocity, $U$ at time $t=t_0$, will result in a  subsequent change in lift for $t > t_0$ proportional to $\phi(t-t_0)$.

When the system is assumed to be linear, the effect of arbitrary unsteadiness may be predicted through the use of a Duhamel convolution integral of the Wagner function with the unsteady input. 
Note, however, that while classical unsteady aerodynamic theory is linear, the Wagner function itself cannot be formulated as the step response of a finite-dimensional linear system. 
Analogously in the frequency domain, the Theodorsen function cannot be expressed as a rational transfer function. 
Despite this complexity, as was noted by Garrick~\cite{Garrick:1938}, the Wagner and Theodorsen functions (with the inclusion of an additional derivative or integral term) form a Laplace transform pair, in the same manner as a transfer function and step response are related in the case of a linear system. 
The Wagner function is particularly difficult to compute in an efficient manner, with definitions requiring the inverse Laplace or Fourier transform of the Theodorsen function, which itself is most typically represented using Hankel functions (Bessel functions of the third kind).  
  
The lack of a convenient and readily-usable expression for the Wagner function has resulted in the formulation of a variety of simplifying approximations that can be used in place of the exact form (e.g.,~\cite{Jones:38,Garrick:1938,jones1945aerodynamic,venkatesan1986new,peterson1988improved,eversman1991modified,vepa1977finite,brunton:2012a,dowell1980simple}).  
These approximations are often claimed to be sufficiently accurate for most purposes. 
For example, Leishman~\cite{leishman:06} remarks in a footnote that 
  
    \begin{quote}
  ``For some applications the exponential approximation to the indicial response may not be considered adequate. This is usually because the rate of approach to the asymptotic value is not as correct for the exponential approximation compared to the exact behavior. This effect, however, is more of academic interest rather than of any practical importance''.
    \end{quote}  
  
\begin{figure}
    \centering
 \subfloat[]{\includegraphics[width= 0.45\textwidth]{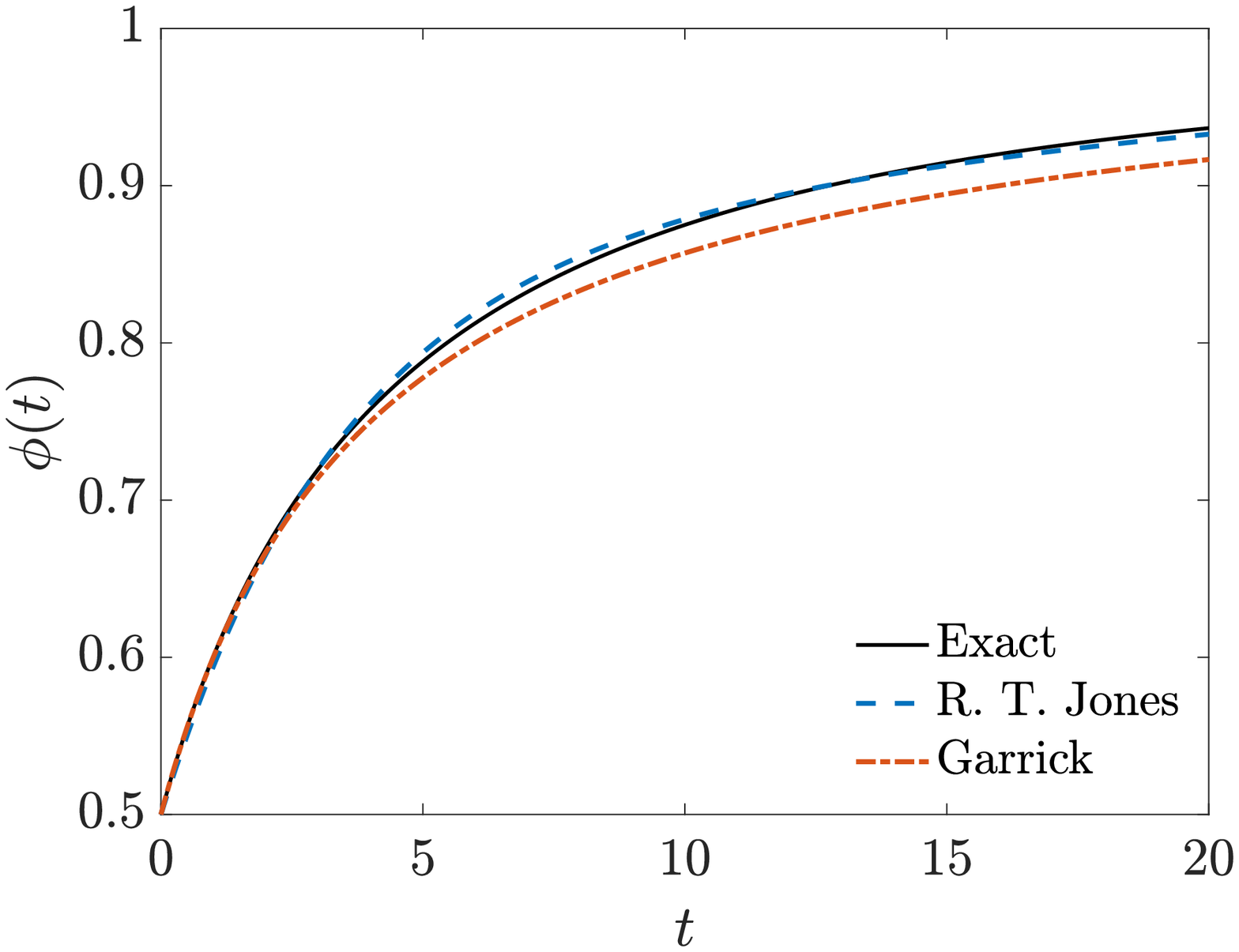}} 
 \subfloat[]{\includegraphics[width= 0.45\textwidth]{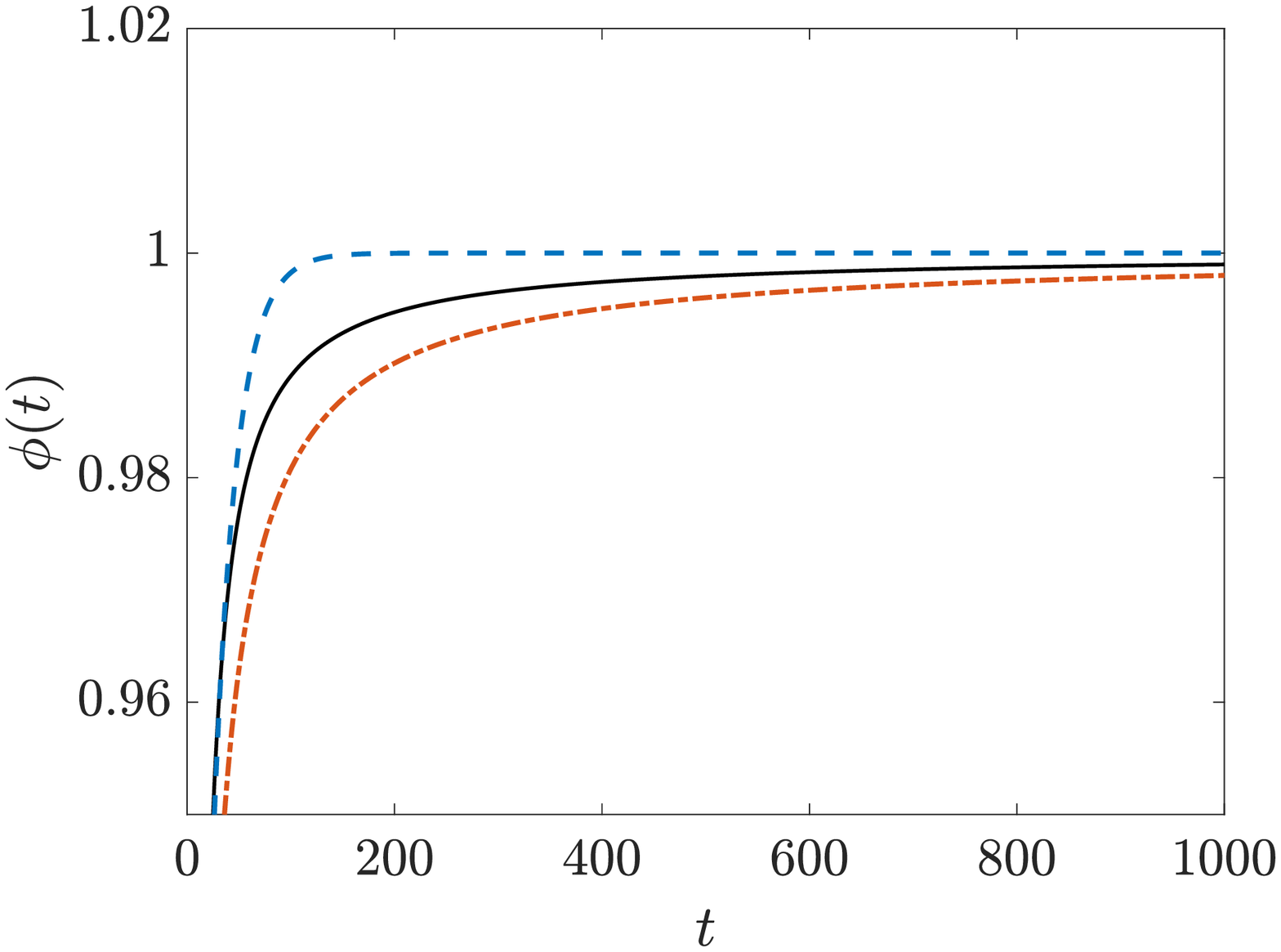}} \\
 \subfloat[]{\includegraphics[width= 0.45\textwidth]{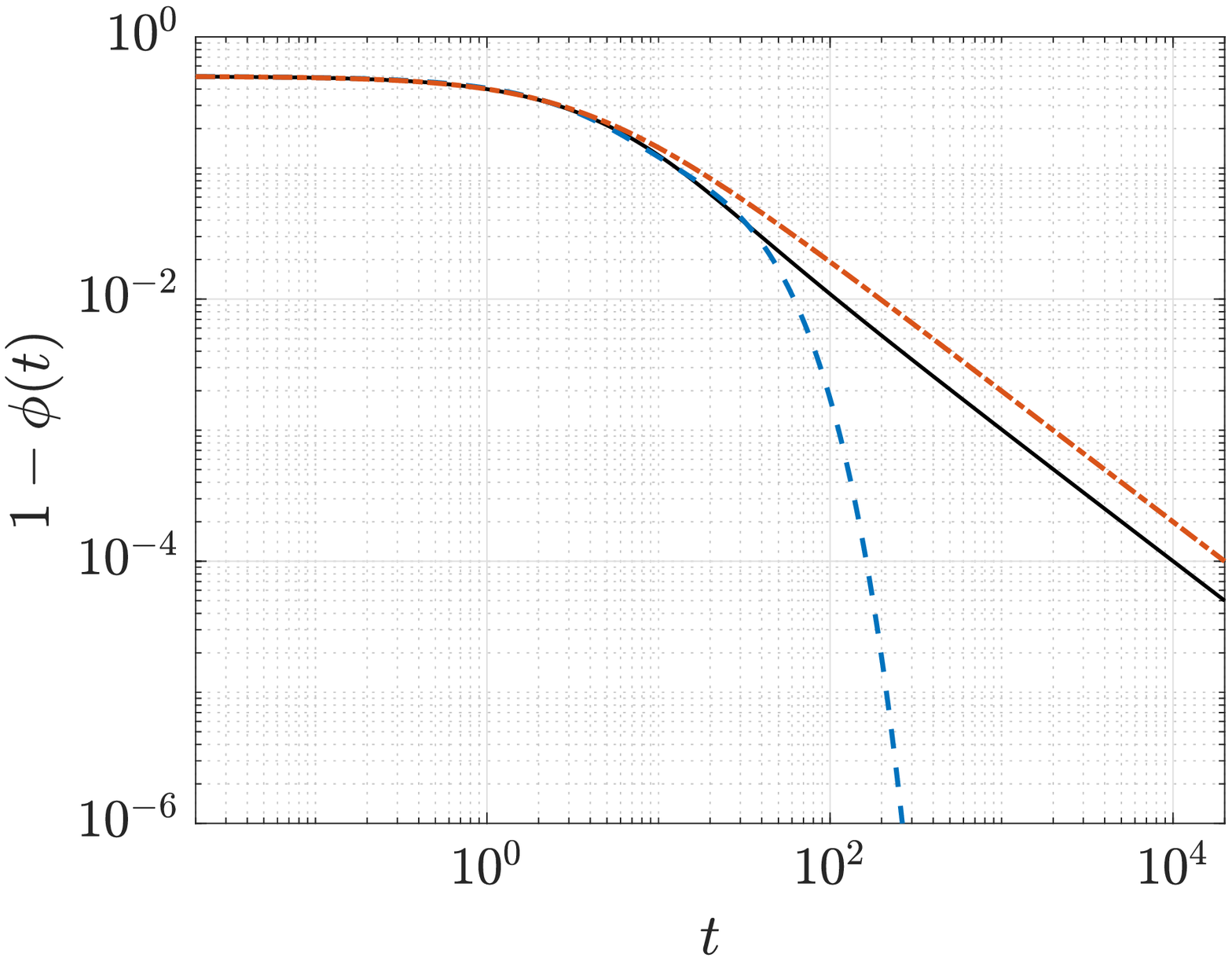}}
\caption{Comparison between the exact Wagner function $\phi(t)$ and approximations due to R.~T.~Jones and Garrick over the domain (a) $0\leq t\leq 20$, and (b)  $0\leq t\leq 1000$, and (c) for the function $1-\phi(t)$ on a log scale.}
\label{fig:Wag1}
\end{figure}

While these approximations are indeed suitable for many typical applications, they are often not as accurate as might be assumed, at least under certain accuracy metrics. 
To demonstrate this explicitly, consider two of the earliest (and perhaps most commonly used) approximations: those due to R.~T.~Jones~\cite{Jones:38}, and I.~E.~Garrick~\cite{Garrick:1938}. 
Jones' approximations can be expressed in the form
\begin{equation}
\label{eq:jones}
\hat\phi(t) = 1 - 0.165\exp(-0.0455t) -0.335\exp(-0.3t),
\end{equation}
while Garrick's approximation\footnote{Which he remarked was simply ``a fortunate choice by the author'' ~\cite{Garrick:1938}} is
\begin{equation}
\label{eq:garrick}
\hat\phi(t) =  1 - \frac{2}{4+t}.
\end{equation}
Henceforth, where needed, we will use $\hat\phi$ to denote an approximation to the Wagner function, $\phi$. 
Here, time is nondimensionalized by the airfoil semichord and the freestream velocity. 
Figure~\ref{fig:Wag1} shows how these two approximations compare to the exact Wagner function, and its additive reciprocal $1-\phi(t)$, on a log scale, at various times.  Both approximations match the correct initial value $\phi(0) = 0.5$ and the correct limit $\lim_{t\to \infty}\phi(t) =1$; however, there are clear discrepancies at intermediate times.  
In particular, Fig.~\ref{fig:Wag1}(c) shows that the asymptotic behavior of the Wagner function does not correspond to exponential decay, as suggested by Jones' approximation. This observation is of particular relevance since the majority of approximations to the Wagner function take the form of a step response of a finite-dimensional linear system, which will never capture the non-exponential decay of the Wagner function. 
Garrick's approximation exhibits the correct asymptotic scaling for large $t$, though is less accurate for earlier times. 
 
\begin{figure}
 \centering 
 \subfloat[]{\includegraphics[width= 0.45\textwidth]{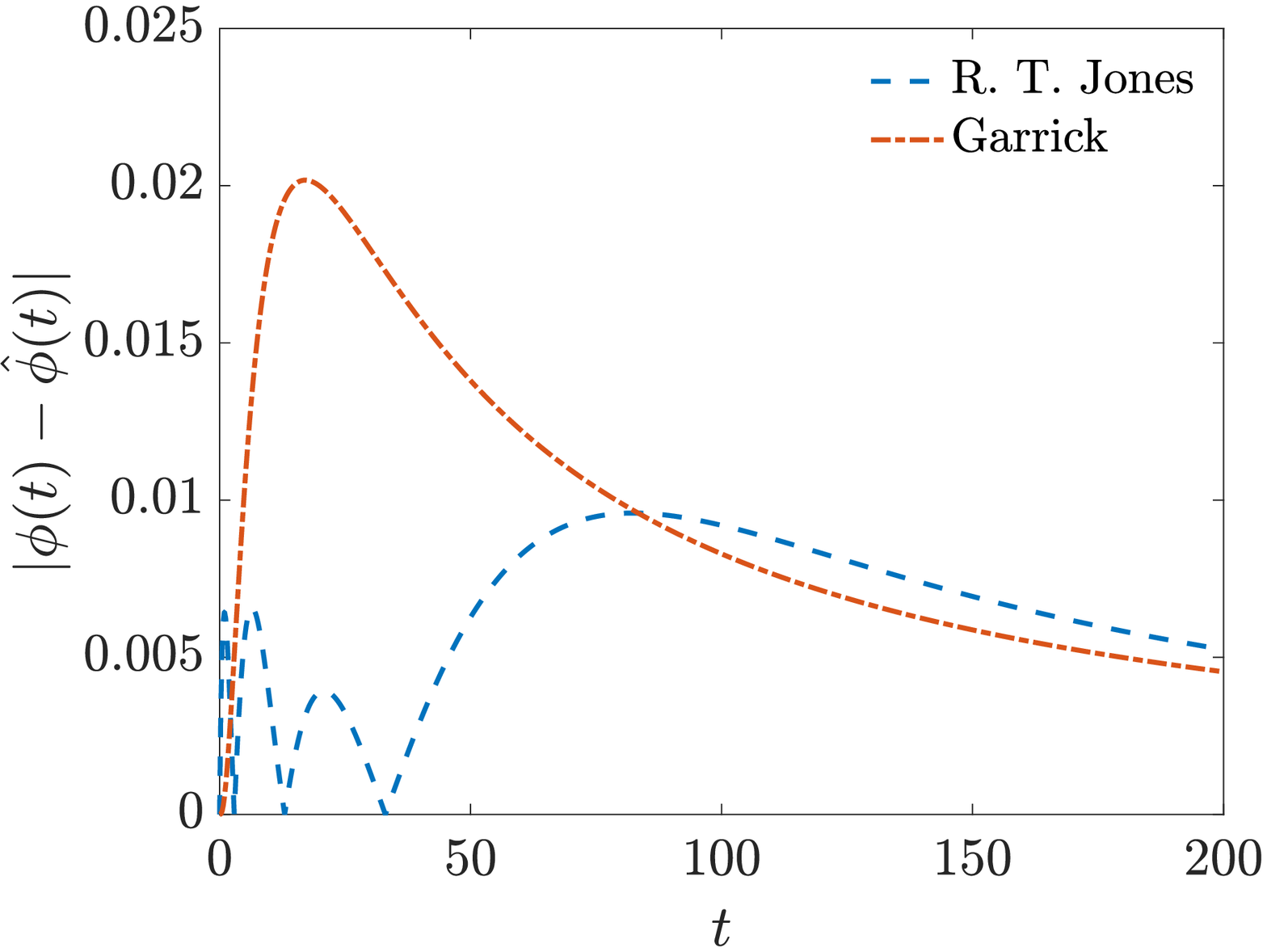}} \ \ 
 \subfloat[]{\includegraphics[width= 0.45\textwidth]{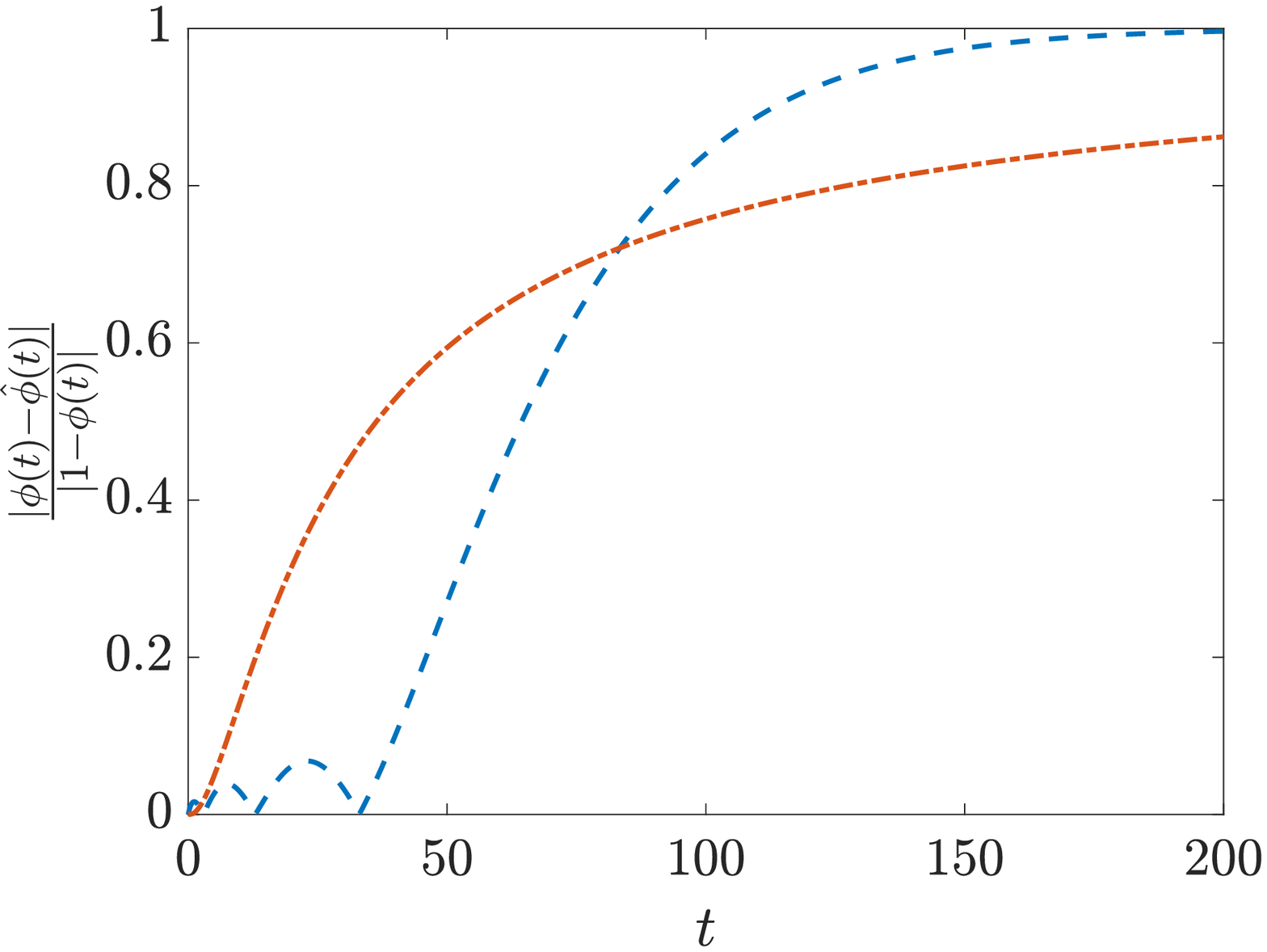}}
\caption{(a) Absolute and (b) relative (to asymptote) error in the approximations of the Wagner function due to  R.~T.~Jones and Garrick.}
\label{fig:WagError1}
\end{figure}

To be more precise, we show the error of these approximations as a function of time in Fig.~\ref{fig:WagError1}.  
Considering the error defined as $|\phi(t) - \hat\phi(t)|$, shown in Fig.~\ref{fig:WagError1}(a), we find that this error is approximated to within less than $1\%$ for Jones' approximation, and $2\%$ for Garrick's approximation, which are figures typically quoted (e.g.~\cite{leishman:06}).  
However, if we instead consider the relative error from the final state by normalizing the error by $1-\phi(t)$, we find that this relative error is significant, particularly at large times (Fig.~\ref{fig:WagError1}(b)). Since $\phi(t)$ approaches a value of unity at large time, this  relative error metric $\frac{|\phi(t)-\hat\phi(t)|}{|1-\phi(t)|}$ gives an indication of whether an approximation to the Wagner function obeys the correct asymptotic behavior. In other words, at a given time, $t$, the approximation $\hat\phi(t)$ will only give an accurate relative measure of the lift deficiency $1-\phi(t)$ if $\frac{|\phi(t)-\hat\phi(t)|}{|1-\phi(t)|} \ll 1$. 
Note also that considering an integrated error, such as $\|\phi(t) - \hat\phi(t)\|_2$, is not particularly enlightening, since $1-\phi(t)$ is not square-integrable over the domain $t \in [0,\infty)$.

Since the Wagner function describes the response to a step change in circulation around the airfoil, it can be used to determine how an aerodynamic system responds to a step change in conditions, and the manner in which this response returns to a steady state. 
For example, it could inform how large a computational domain needs to be, and how long a numerical simulation should be run, in order to capture the full transient behavior induced by a step change in conditions, such as a change in freestream velocity, or in angle of attack. 
From Fig.~\ref{fig:Wag1}(c), we see that to get within $10^{-4}$ of the equilibrium condition, one must run a simulation for approximately $10^4$ time units.  
However, if using the R.~T.~Jones approximation, one would incorrectly infer that the system would take less that 200 time units to reach equilibrium.  
Similarly, such an approximation can also lead to a substantial underestimation of the size of the domain needed to fully capture the transient response, since the influence of a shed vortex convecting downstream decays algebraically, rather than exponentially. 

As mentioned above, the approximation due to R.~T.~Jones in Eq.~\eqref{eq:jones} may be interpreted as the impulse response of a linear system, one state space realization of which is given by
   \begin{equation}
   \label{eq:SSjones}
\begin{aligned}
\frac{{\rm d}}{{\rm d}t}\begin{bmatrix}  
x_1\\ x_2 \\  x_3 \end{bmatrix}
&= 
   \begin{bmatrix}      
   0 & 1 & 0 \\
   0 & 0 & 1  \\
   0 &  -0.01365  &  -0.3455
   \end{bmatrix}
   \begin{bmatrix}  
x_1\\ x_2 \\ x_3 \end{bmatrix}
+
\begin{bmatrix}  
1\\ 0 \\  0 \end{bmatrix}u,\\
\hat\phi(t) &= 
\begin{bmatrix}  
0.01365 & 0.2808 & 0.5000 \end{bmatrix}
   \begin{bmatrix}  
x_1\\ x_2 \\ x_3 \end{bmatrix}.
\end{aligned}.
\end{equation} 
 Here $\bx = (x_1,x_2,x_3)^T$ is the internal system state, and the input $u$ represents the rate-of-change of aerodynamic flow conditions (e.g.~angle of attack or freestream velocity), so that the impulse response of the system represents the response to an instantaneous step change in this condition. 
 As it turns out, the Garrick approximation can also be cast as the response of a differential equation, with the equation  \begin{equation}
\frac{{\rm d}\hat\phi}{{\rm d}t} = - \frac{1}{2}(\hat\phi-1)^2, \quad \hat\phi(0) = 0.5,
\end{equation}
having Eq.~\eqref{eq:garrick} as a solution. 
This suggests that more accurate approximations might similarly be formulated as solutions to nonlinear differential equations. 
   
In this work, we show that the Wagner function can indeed be accurately modeled by a nonlinear scalar differential equation, which we identify using a variant of the sparse identification of nonlinear dynamics (SINDy) methodology~\cite{brunton2016sindy}. 
SINDy-based and related methods have previously been applied and extended to obtain low-dimensional models for various applications, such as for the identification of partial differential equations~\cite{rudy2017data,schaeffer2017learning}, for model selection~\cite{mangan2017model}, and for systems with known energy constraints and conservative properties~\cite{loiseau2018constrained}.
  
Our goal in this work is not just to produce an improved approximation to the Wagner function itself, but also to formulate a methodology that can readily extend to more realistic aerodynamic systems, and capture the effects of factors such as viscosity, large angles of attack, and vortex roll-up.   
In particular, at early times the Wagner function departs significantly from behavior observed in practice, which to accurately capture requires a more detailed analysis of early-time vortex dynamics~\cite{pullin1978large,chow1982initial,graham1983lift}. 
The Wagner function can also be extended to more general flow conditions, such as the presence of additional leading- and trailing-edge vortices~\cite{li2015unsteady}. More generally, classical unsteady aerodynamic theory can additionally be extended to account, for example, for high amplitude motions with the generation of leading edge circulation~\cite{ramesh2013unsteady}, and for viscous effects using triple-deck boundary layer theory~\cite{taha2019viscous}.
  
This paper proceeds as follows. Sect.~\ref{sec:Wagner} derives the Wagner function using unsteady potential flow theory, and also provides details for its accurate computation. In Sect.~\ref{sec:approx}, we discuss and analyze various methods that have been used previously to approximate the Wagner function (and the Theodorsen function, which is its frequency-domain counterpart), in addition to the two motivating examples discussed in this section.  In Sect.~\ref{sec:sindy}, we introduce our proposed modeling procedure, and show that it given a closer approximation to those surveyed in Sect.~\ref{sec:approx}, using a range of metrics, for both the Wagner function itself (in Sect.~\ref{sec:WagSindy}), and for a more general case accounting for a finite thickness airfoil and a non-planar wake (Sect.~\ref{sec:BEM}).

\section{The Wagner function}
\label{sec:Wagner}
This section introduces and discusses the Wagner function. First, we show in Sect.~\ref{sec:derivation}  how the Wagner function may be derived from unsteady potential flow theory. Sect.~\ref{sec:compute} then provides details for the accurate numerical computation of the exact Wagner function, which will be used as a benchmark for the subsequent approximations. 

\subsection{Derivation of the Wagner function}
\label{sec:derivation}
Here we show how the Wagner function may be obtained from unsteady potential flow theory, with appropriate assumptions.  
Since our primary focus is the Wagner function, rather than finding the response to sinusoidal or arbitrary airfoil motions, our derivation differs in places from those given in classical papers (e.g.~\cite{karman1938airfoil}) and textbooks (e.g.~\cite{bisplinghoff2013aeroelasticity}), which often work in the frequency domain.  
Additionally, since our primary goal is the derivation of the Wagner function, which prescribes the ``circulatory'' response of the system to a step change in conditions, we will not need to consider explicitly the ``non-circulatory" portion of the flow that imposes the correct boundary conditions; typically achieved using a distribution of sources and sinks of appropriate strength across the surface. As an aside, note that while the decomposition of unsteady aerodynamic forces into ``circulatory" and ``noncirculatory" components is convenient for their derivation, in some cases this distinction can contain ambiguity \cite{taha2020high}.

We assume two-dimensional potential flow at small angles of attack. 
For simplicity, we assume that lengths have been nondimensionalized such that the airfoil semichord is of unit length, and the freestream velocity is unity. Time is similarly nondimensionalized such that one time unit corresponds to the time taken for the freestream to convect a distance of one half chord. The choice to nondimensionalize using the semichord rather than the chord length is in accordance to the original work in this area.  This semichord scaling also slightly simplifies the Joukowski transformation that is employed to map  the airfoil to the unit circle.

\begin{figure}
 \centering {
  \subfloat[]{\includegraphics[trim=2cm 2cm 2cm 1cm, clip,width= 0.45\textwidth]{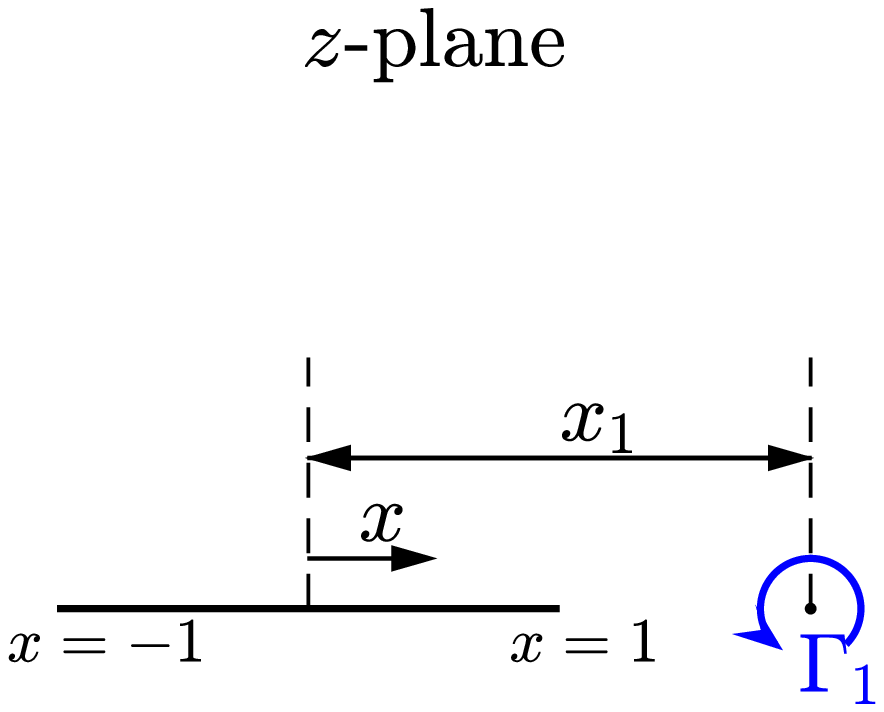} } 
  \subfloat[]{\includegraphics[trim=2cm 2cm 2cm 1cm, clip,width= 0.45\textwidth]{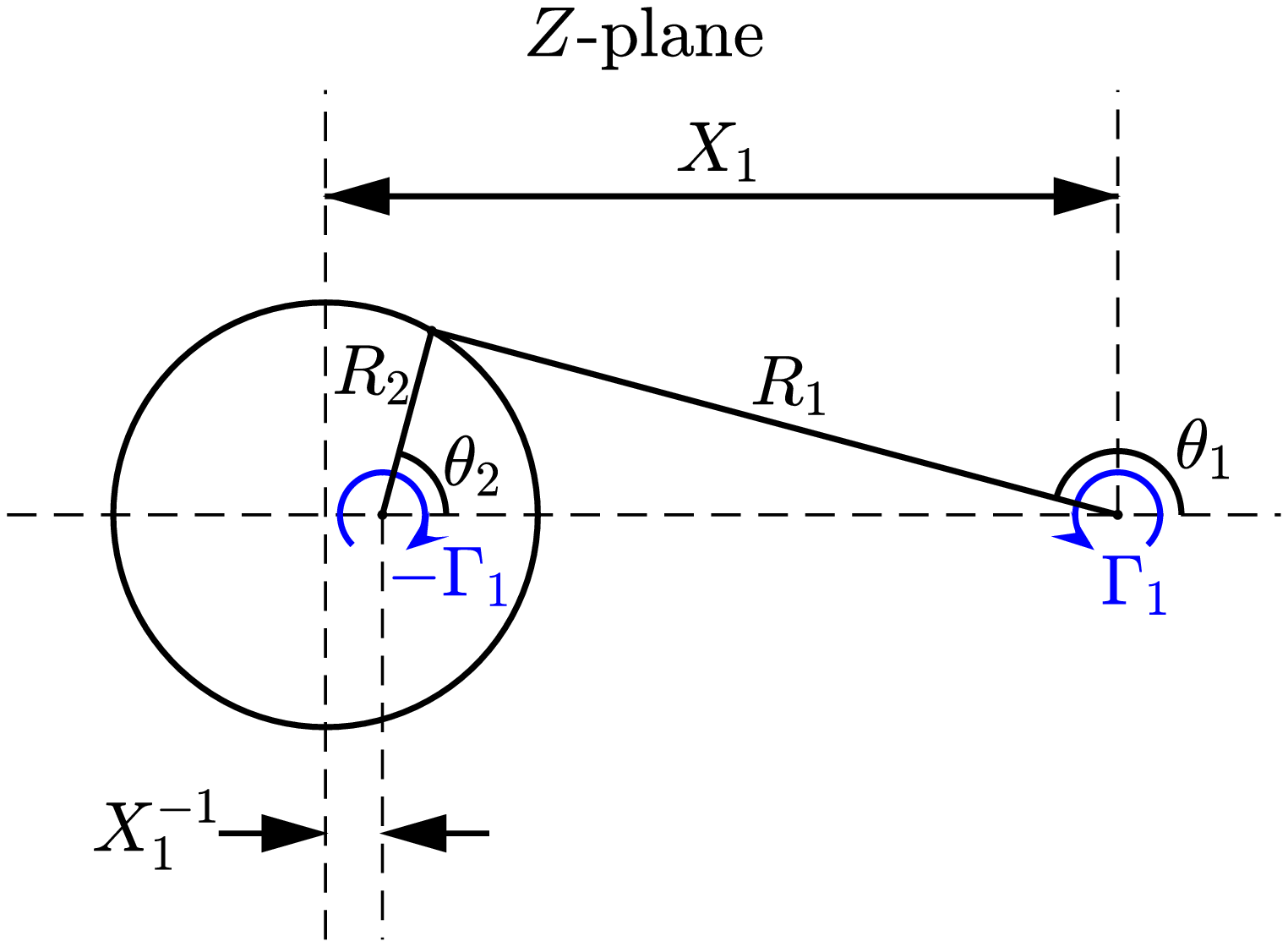}}
}
\caption{ (a) Schematic diagram of a thin airfoil in the $z$-plane with a vortex of strength $\Gamma_1$ located a distance $\xx_1$ from the midchord,  and  (b) the equivalent setup transformed into the $Z$-plane, where the airfoil is mapped to the unit circle.}
\label{fig:diagram}
\end{figure}

For steady potential flow over a thin airfoil at a nonzero angle of attack $\alpha_{0}$, implementation of the Kutta condition requires a bound circulation $\Gamma_{0}$ around the airfoil, resulting in a lift force $L_{0}$, related by 
\begin{equation}
- \Gamma_0 =   \frac{L_0}{\rho U}  = 2\pi U \alpha_0,
\end{equation}
where $\rho$ and $U$ are the fluid density and freestream velocity respectively. For a freestream velocity from left to right, the negative sign on the circulation indicates that a positive angle of attack, generating positive lift, corresponds to negative (clockwise) bound circulation. Flow around a flat plate airfoil may be studied using the Joukowski transformation between the physical $z$-plane and transformed $Z$-plane by
\begin{equation}
z = T(Z) = \frac{1}{2}\left( Z + Z^{-1}\right).
\end{equation}
This transformation maps an infinitely-thin airfoil with leading and trailing edges at $z = \pm 1$ to the unit circle, as shown in Fig.~\ref{fig:diagram}.  The complex potential for steady potential flow about a unit circle centered at the origin, with circulation $-\Gamma_0$, is given by
  \begin{equation}
  \label{eq:F0}
 F_0(Z) = U\left(Z + Z^{-1}\right) - \frac{i \Gamma_0  }{2\pi}\log{Z}.
\end{equation}
This gives the potential flow about an airfoil in equilibrium conditions.  Note that this corresponds to a complex velocity in the $Z$-plane of
\begin{equation}
W_{0}(Z) = \ddodd{F_{0}}{Z} = U\left(1 - Z^{-2}\right) - \frac{i \Gamma_0 }{2\pi Z}.
\end{equation}

The Wagner function describes how the lift approaches such a steady state given a change in conditions, such as a change in angle or attack or freestream velocity. For simplicity, here we will consider flow that is impulsively started, over an airfoil at an angle attack $\alpha_0$.  We seek, under the relevant assumptions, the function $\phi(t)$ such that the transient lift response is given by
\begin{equation}
L(t) = \phi(t) L_0.
\end{equation}
If the flow starts from rest then there is initially no circulation present.  Conservation of circulation means that the total circulation must remain zero for all time.  That is, if $\Gamma_a$ and $\Gamma_w$ denote the total circulation about the airfoil and wake respectively, we have
\begin{equation}
\Gamma_a(t) + \Gamma_w(t) = 0.
\end{equation}
After the flow starts, circulation must develop over the airfoil in order for it to attain its nonzero lift.  We assume that the flow over the airfoil is completely attached, and that vorticity is only shed at the trailing edge, as will be required to satisfy the Kutta condition. 
Under the assumption that the shed vorticity does not translate vertically, we have
\begin{equation}
\Gamma_w = \int_1^\infty \gamma(\xx)d\xx,
\end{equation}
where $\gamma(\xx)$ denotes the distributed sheet of vorticity that has been shed into the wake.  Note in particular that an impulsively-started flow will have vorticity shedding continuously as the flow develops.
Thus, there is not just one instantaneously-shed ``starting vortex", but rather a vortex sheet arising from the impulsively started flow. 

Consider an element of shed circulation $ {\rm d} \gamma_1$ located at $\xx_1$ in the wake.  We can model the effect of this vortex in the $Z$-plane. 
 In particular, define $X_0$ (lying on the $X$-axis of the $Z$-plane) such that $\xx_1 = T(X_1)$. 
If we have a vortex with circulation $\Gamma_1$ at $X_1$ in the $Z$-plane, then it can be shown that for the unit circle to be a streamline, we must have an opposite-strength vortex at $X_1^{-1}$.  The complex potential for this vortex pair is given by
 \begin{equation}
F_V(Z) =  \frac{\Gamma_1i }{2\pi}\left(\log(Z-X_1)-\log(Z-X_1^{-1})\right).
\end{equation}
Note that we can add a freesteam velocity to this vortex pair, giving
 \begin{equation} 
F_1(Z) = U\left( Z + Z^{-1} \right)+  \frac{\Gamma_1i }{2\pi}\left(\log(Z-X_1)-\log(Z-X_1^{-1})\right),
\end{equation}
with the corresponding complex velocity in the $Z$ plane
\begin{equation}
W_{1}(Z) = \ddodd{F_{0}}{Z} = U\left( 1 - Z^{-2}\right)+ \frac{\Gamma_1 i }{2\pi Z}\left(\frac{1}{Z-X_1}-\frac{1}{Z-X_1^{-1}}\right).
\end{equation}
As an aside, note that the complex velocity in the $z$ and $Z$ planes are related via
 \begin{equation}
w = \ddodd{F_{0}}{z} = W  \ddodd{Z}{z},
\end{equation} 
where $w$ is the complex velocity in the physical $z$-plane. 
We can confirm that as $\xx_1$ approaches infinity (and thus so must  $X_1$ under the assumption that $X_1 > 1$), $F_1$ approaches $F_0$ given in Eq.~\eqref{eq:F0}, provided $\Gamma_1 = \Gamma_0$. 

We now seek an expression for the lift on the airfoil with this flow field. 
 Letting $Z - X_1 = R_1e^{i\theta_1}$ and $Z- X_1^{-1} = R_2e^{i\theta_2}$ as indicated in Fig.~\ref{fig:diagram}(b), the velocity potential  due to the vortex pair is given by
 \begin{equation}
 \label{eq:phiV}
\Phi_V(Z) = \Real\left(F_V(Z)\right) = \frac{\Gamma_1 }{2\pi}(\theta_2-\theta_1). 
\end{equation}
We choose the domain $\theta_2 \in [0,2\pi)$, which means that the velocity potential has a branch cut along the $X$ axis between $X_1$ and  $X_1^{-1}$. 
 Letting $Z = Re^{i\theta}$, from trigonometry  it can be shown that
\begin{equation}
\label{eq:tan}
\tan(\theta_2-\theta_1) = \frac{(X_1^{-1}-X_1)\sin\theta}{2 - (X_1+X_1^{-1})\cos\theta}.
\end{equation}
From Eqns.~\eqref{eq:phiV} and \eqref{eq:tan}, we can compute
\begin{equation}
\Phi_V(X_1,\theta) =  \frac{\Gamma_1 }{2\pi} \arctan\left( \left(X_1^{-1}-X_1\right)\sin\theta, \ 2 - \left(X_1+X_1^{-1}\right)\cos\theta\right),
\end{equation}
where $\arctan(Y,X)$ denotes the four-quadrant arctan function of $Z = X + i Y$. 
Mapping back into the $z$-plane and noting that $T(e^{i\theta}) = \cos\theta = \Real(z) = x$,
 we have
 \begin{align}
 \label{eq:phi}
\Phi_{V,u}(\xx_1,x) &=  \frac{\Gamma_1 }{2\pi} \arctan\left(-{\sqrt{\xx_1^2-1}\sqrt{1- x^2}}, \ {1-\xx_1 x}\right), \\
\Phi_{V,l}(\xx_1,x) &=  \frac{\Gamma_1 }{2\pi} \arctan\left({\sqrt{\xx_1^2-1}\sqrt{1- x^2}}, \ {1-\xx_1 x}\right), 
\end{align}
where $ \Phi_{V,u}$ and $\Phi_{V,l}$ denote the velocity potential on the upper and lower surfaces of the airfoil, respectively, which satisfy $\Phi_{V,l}(\xx_1,x) = - \Phi_{V,u}(\xx_1,x)$. 
Using the unsteady Bernoulli equation, the pressure difference between the upper and lower surfaces can be computed to be
\begin{align}
p_u(\xx_1,x)-p_l(\xx_1,x)  &= -\rho\left[ \left(\dod{\Phi_{V,u}}{t}   + U   \dod{\Phi_{V,u}}{x} \right) - \left(\dod{\Phi_{V,l}}{t}   + U   \dod{\Phi_{V,l}}{x} \right)\right] \\ 
&=  -2\rho\left(\dod{\Phi_{V,u}}{t}   + U   \dod{\Phi_{V,u}}{x} \right) 
\label{eq:bern1}
\end{align}
Assuming that the wake vortex convects downstream at the freestream velocity, we have
\begin{equation}
\dod{\Phi_{V,u}}{t}   = U \dod{\Phi_{V,u}}{\xx_1} .
\end{equation}
Eq.~\eqref{eq:bern1} then becomes
\begin{align}
p_u(\xx_1,x)-p_l(\xx_1,x) & = -2\rho U \left(\dod{\Phi_{V,u}}{\xx_1}   +   \dod{\Phi_{V,u}}{x} \right).
\end{align}
 Computing these derivatives from Eq.~\eqref{eq:phi},
\begin{align}
p_u(\xx_1,x)-p_l(\xx_1,x) &=  - \frac{\rho U \Gamma_1 }{\pi} \frac{\xx_1+x}{\sqrt{\xx_1^{2}-1}\sqrt{1-x^2}}.
\end{align}
The lift force can then be computed by integrating the pressure difference across the surface of the airfoil,
\begin{align}
\label{eq:lift}
L(\xx_1)  &= -\frac{\rho U \Gamma_1 }{\pi}  \int_{-1}^{1}  \frac{\xx_1+x}{\sqrt{\xx_1^{2}-1}\sqrt{1-x^2}} {\rm d}x \\
 & = - \frac{\rho U \Gamma_1 \xx_1 }{\sqrt{\xx_1^{2}-1}}.
\end{align}
The Wagner function at a given time will include the effect of all vorticity that has been shed into the wake since startup.  We account for this by considering a distribution of vorticity $\gamma(\xx,t)$ that replaces the discrete vortex $\Gamma_1$ in Eq.~\eqref{eq:lift} using $\Gamma_1 =  \gamma {\rm d}\xx$, recalling that we define  counterclockwise vorticity to be positive. For a given wake length $\xx_{\rm{max}}$, this gives a lift of
\begin{equation} 
L(\xx_{\rm{max}})  = - \rho U \int_{1}^{\xx_{\rm{max}}} \gamma(\xx,t) \frac{\xx}{\sqrt{\xx^{2}-1}} {\rm d}\xx.
\end{equation}
Since all shed vorticity convects with the freestream velocity, we can determine $\gamma(\xx,t)$ at a certain location by the time when the vorticity was shed, $t' = 1+t -\xx$, which we denote by $\mu(t')$. 
Noting that ${\rm d}t' = -{\rm d}\xx$, the lift can be expressed as
\begin{equation}
 L(t)  = \rho U \int_{0}^{t} \mu(t') \frac{1+t-t'}{\sqrt{(t-t')(t-t'+2)}}{\rm d}t'. 
\end{equation}
The Wagner function $\phi(t)$ is then given by the ratio of $L(t)$ to the steady-state lift,
\begin{equation}
\label{eq:Wagint}
\phi(t) = \frac{L(t)}{L_0}  = \int_{0}^{t} \tilde\mu(t') \frac{1+t-t'}{\sqrt{(t-t')(t-t'+2)}} {\rm d}t',
\end{equation}
where $\tilde\mu(t') = \frac{\mu(t')}{L_0}$ denotes the normalized wake vorticity distribution.
To determine $L(t)$ explicitly, it remains to find $\tilde\mu(t')$, which describes how much vorticity is shed into the wake as a function of time.

 We assume that the vorticity shed is that which enforces the Kutta condition at the trailing edge.  
Note that for steady flow with bound circulation $-\Gamma_0$, the velocity at the trailing edge of the cylinder in the $Z$-plane is given by
\begin{equation}
\label{eq:W0}
W_{0}(Z = 1)  = - \frac{i \Gamma_0 }{2\pi}.
\end{equation}
To enforce the Kutta condition, it is sufficient require that the velocity at the trailing edge matches this steady-state value. For a wake vortex with circulation $\Gamma_{1}$ and image vortex with circulation $-\Gamma_{1}$ (as in Fig.~\ref{fig:diagram}),  the trailing-edge velocity in the $Z$-plane
\begin{align}
W_{1}(Z = 1) &=  \frac{i\Gamma_1 }{2\pi }\left(\frac{1}{1-X_1}-\frac{1}{1-X_1^{-1}}\right) \\
& =  - \frac{i\Gamma_1 }{2\pi}\left(\frac{X_{1}+1}{X_1-1}\right) \\
& = - \frac{i\Gamma_1 }{2\pi} \left(\sqrt{\frac{\xx_1+1}{\xx_1-1}} \right).
\end{align}
Integrating over the vorticity shed into the wake and equating with Eq.~\eqref{eq:W0}, we have
\begin{equation}
\label{eq:kuttaGam}
\Gamma_0 = \int_1^{\xx_{\rm{max}}}\gamma(\xx,t)\sqrt{\frac{\xx+1}{\xx-1}}  {\rm d}\xx,
\end{equation}
and thus 
\begin{equation}
\label{eq:kuttaL}
L_0 = -\rho U \Gamma_0 = - \rho U \int_1^{\xx_{\rm{max}}}\gamma(\xx,t)\sqrt{\frac{\xx+1}{\xx-1}}  {\rm d}\xx.
\end{equation}
Once again using the change of variables $t' = 1+t -\xx$,
Eq.~\eqref{eq:kuttaL} can be rewritten as 
\begin{equation}
 \label{eq:kuttat} 
L_0 = \rho U\int_{0}^{t} \mu(t')\frac{\sqrt{(t-t'+2)}}{\sqrt{(t-t')}}{\rm d}t',
\end{equation}
or equivalently, in terms of the normalized vorticity distribution,
\begin{equation}
 \label{eq:kuttat2} 
1 = \int_{0}^{t} \tilde\mu(t')\frac{\sqrt{(t-t'+2)}}{\sqrt{(t-t')}}{\rm d}t'.
\end{equation}
Eq.~\eqref{eq:kuttat2} can be used with \eqref{eq:Wagint} to find an expression for the Wagner function. Note that both of these equations contain convolution integrals, which may be solved via Laplace transform. 
 Letting $Q(s) = \mathcal{L}(\tilde\mu)$, where $\mathcal{L}$ denotes the Laplace transform, we obtain 
  \begin{align}
 \label{eq:Lap1}
\mathcal{L}(\phi) &= Q(s)e^s K_1(s) \\
 \label{eq:Lap2}
  Q(s) e^s(K_0(s) &+ K_1(s)) - s^{-1} = 0, 
\end{align} 
 where $K_\nu$ denotes a modified Bessel function of the second kind, of order $\nu$.  
We can combine Eqns.~\eqref{eq:Lap1} and \eqref{eq:Lap2} to 
eliminate $Q(s)$ and obtain an expression for the Laplace-transformed Wagner function,
\begin{equation}
\label{eq:LapWag}
\mathcal{L}(\phi)  = \frac{K_1(s)}{s(K_0(s) + K_1(s))}.
\end{equation}
Note that modified Bessel functions $K_\nu$ are related to Hankel functions (i.e., Bessel functions of the third kind) by
\begin{equation}
\label{eq:Kdom}
K_{\nu}(s) =\left\{\begin{array}{ll}{\frac{\pi}{2} i^{\nu+1} H_{\nu}^{(1)}(i s)} & {-\pi<\arg(s) \leq \frac{\pi}{2}} \\ {\frac{\pi}{2}(-i)^{\nu+1} H_{\nu}^{(2)}(-i s)} & {-\frac{\pi}{2}<\arg(s) \leq \pi}\end{array}\right.
\end{equation}
This means that, for appropriate values of $s$, Eq.~\eqref{eq:LapWag} can be expressed as 
\begin{equation}
\label{eq:HWag}
\mathcal{L}(\phi)  = \frac{H^{(2)}_1(-is)}{s\left(H^{(2)}_1(-is)+ i H^{(2)}_0(-is)\right)}.
\end{equation}
Eq.~\eqref{eq:HWag} is the most commonly used formulation, though from Eq.~\eqref{eq:Kdom} it is only valid for some $s$ (as has been previously noted, e.g.~\cite{epps2018vortex}). Note in particular that the Theodorsen function $C(k)$ (where $k$ is a frequency variable) is related to the Wagner function via a Laplace transform. In particular, if we express $C(k)$ in terms of the Laplace domain variable $s = ik$ and extend its definition such that it is a transfer function defined on the entire complex plane, we have $C(k) = C(-is)$ and 
\begin{equation}
\label{eq:WagTheo}
\frac{C(-is)}{s} = \int_{0}^{\infty} \phi(t)\exp(-st){\rm d}t,
\end{equation}
with this Theodorsen transfer function thus given explicitly by 
\begin{equation}
\label{eq:Theo}
C(-is)  = \frac{H^{(2)}_1(-is)}{H^{(2)}_1(-is)+ i H^{(2)}_0(-is)}.
\end{equation}
Many derivations of the Wagner function first arrive at the Theodorsen function, and then compute the Wagner function by inverting the integral in Eq.~\eqref{eq:WagTheo}.  Regardless of the approach, the result ends up being a function that must be computed from an inverse Laplace or Fourier transform.  The next section will detail how this computation may be performed accurately in practice. 

\subsection{Numerical computation of the Wagner function}
\label{sec:compute}

Here we describe two methods by which the Wagner function can be computed numerically, by performing either a numerical inverse Fourier transform, or a numerical inverse Laplace transform.  To the best of our knowledge, use of the latter method has not been explicitly discussed in the context of computing the Wagner function previously.

We first describe the inverse Fourier transform approach, which is also discussed, for example, in~\cite{bisplinghoff2013aeroelasticity,peters2008}.  Letting $s = i k$, with $k > 0$ we can express the Theodorsen function in terms of its real and complex components
 \begin{equation}
C(k) = \frac{H^{(2)}_1(k)}{H^{(2)}_1(k)+ i H^{(2)}_0(k)}= F(k) + i G(k),
\end{equation}
where
\begin{align}
G_1(k) &= \frac{J_1(k)\left(J_1(k)+Y_0(k)\right)+ Y_1(k)\left(Y_1(k)-J_0(k)\right)}{\left(J_1(k)+Y_0(k)\right)^2 + \left(Y_1(k)-J_0(k)\right)^2}, \\
G_2(k) &= \frac{Y_1(k)Y_0(k)+ J_1(k)J_0(k)}{\left(J_1(k)+Y_0(k)\right)^2 + \left(Y_1(k)-J_0(k)\right)^2},
\end{align}
and $J_\nu$ and $Y_\nu$ are Bessel functions of the first and second kind, respectively.  We then have the following equivalent expressions for the Wagner function:
\begin{align}
\label{eq:wagF1}
\phi(t) &= 1 + \frac{2}{\pi}\int_{k = 0}^\infty \frac{1}{k}\left(G_1(k)-1\right)\sin(kt){\rm d}k \\
\label{eq:wagF2}& = \frac{1}{2} +  \frac{2}{\pi}\int_{k = 0}^\infty\frac{1}{k}\left(G_1(k)-\frac{1}{2}\right)\sin(kt){\rm d}k \\
\label{eq:wagF3}& = 1 + \frac{2}{\pi}\int_{k = 0}^\infty\frac{1}{k}G_2(k)\cos(kt){\rm d} k.
\end{align}
Peters~\cite{peters2008} notes that Eqns.~\eqref{eq:wagF2} and \eqref{eq:wagF3} are numerically well-behaved for small and large times respectively, while Eq.~\eqref{eq:wagF1} has the best overall numerical behavior across all times. Aside from these known issues, we find that these integrals can be accurately computed in Mathematica using the numerical integration function ``Nintegrate" with default settings, which applies a globally adaptive strategy to determine integration quadrature points. 

For the inverse Laplace transform approach, we approximate the Wagner function using the numerical methods described in~\cite{abate2006unified}, implemented using the Matlab package ``Numerical Inverse Laplace Transform'' \cite{mcclureToolbox}. 
Within the unified framework developed in~\cite{abate2006unified}, the inverse Laplace transform of the Wagner function can be formulated by computing a finite sum similar to the general form
\begin{equation}
\label{eq:numLap}
\phi(t) \approx \frac{1}{t}\sum_{j=0}^n \text{Re}\left[w_j \mathcal{L}\left(\phi\left(\frac{\delta_j}{t} \right) \right)\right], 
\end{equation}
where $\delta_j$ are nodes in the Laplace domain corresponding to weights $w_j$, with the Laplace transform of the Wagner function as given in Eq.~\eqref{eq:HWag}. 
A variety of methods can be shown to fit in this framework. For example,  the Euler algorithm~\cite{abate2000introduction}, which is based on a Fourier expansion with Euler summation for accelerated convergence, can be expressed as the sum
\begin{equation}
\label{eq:numLapE}
\phi(t) \approx \frac{10^{M/3}}{t}\sum_{j=0}^{2M}(-1)^j \xi_j \text{Re}\left[\mathcal{L}\left(\phi\left(\frac{\delta_j}{t} \right) \right)\right],
 \end{equation}
where 
$\delta_j = \frac{M\ln(10)}{3} + i \pi j$, and 
$\xi_0 = 1/2$, $\xi_{2M} = 2^{-M}$,  $\xi_k = 1$ for $1\leq k \leq M$, and  $\xi_{2M-j} = \xi_{2M-j+1} +2^{-M}\binom Mj$ for $1\leq j \leq M-1$. 

Alternatively, the Talbot method~\cite{talbot1979accurate}, which is based on deformations of Bromwich contours to perform the integration, can be computed similarly from the finite sum
\begin{equation}
\label{eq:numLapT}
\phi(t) \approx \frac{2}{5t}\sum_{j=0}^{M-1}\text{Re}\left[w_j \mathcal{L}\left(\phi\left(\frac{\delta_j}{t} \right) \right)\right], 
\end{equation}
where $\delta_0 = \frac{2M}{5}$, $w_0 = \frac{1}{2}\exp(\delta_0)$,  and 
\begin{align*}
\delta_j &= \frac{2j\pi}{5}\left(\cot(j\pi/M)+i\right)\\
 w_j &= \left[ 1 + \frac{ij \pi}{M}\left(1+\cot^2(j\pi/M)\right)- i\cot(j\pi/M)\right]\exp(\delta_j)
 \end{align*}
  for $1\leq j \leq M-1$. These sums converge rapidly, with accurate results typically obtained with $M = 16$.
 Data and code for these computations of the Wagner function will be included in supplementary materials. 



Note that the original values tabulated in Wagner's original work~\cite{wagner:25}  (and also presented in subsequent works, e.g.~\cite{karman1938airfoil,Garrick:1938}) are only accurate for short times ($ t \lessapprox 5$).  The inaccuracy of some of these values was noted as early as the work of von K\'{a}rm\'{a}n and Sears~\cite{karman1938airfoil}, who remark that ``for large values of [time] Wagner's calculations are not accurate, since his curve tends to a false asymptote''.


The Wagner function is also closely related to the K\"ussner function, which represents the response of an airfoil to a sharp-edged gust, subject to the same assumptions used in the derivations in this section. The same methodology employed here can be used to derive and compute the K\"ussner function, though we omit the details. 

\section{Approximations to the Wagner function}
\label{sec:approx}

This section will provide an overview of past work seeking to formulate approximations to the Wagner function, beyond the two motivating examples introduced in Sect.~\ref{sec:intro}. We consider approximations with linear systems in Sect.~\ref{sec:LinApprox}, and alternative approximation methods in Sect.~\ref{sec:otherApprox}. 
Note that a number of the approximations that will be discussed were originally formulated for the Theodorsen function, and they were not conceived to model the asymptotic (long-time) behavior of the Wagner function. Nevertheless, we hope that it is useful to provide a survey of many of these approximations that have been proposed since the original description of the Wagner function.

\subsection{Approximation with linear systems}
\label{sec:LinApprox}
This section seeks to provide a systematic description of linear-systems-based approximations to Wagner's function.
The most commonly used approximations are functions that can be characterized as the impulse response of a (finite-dimensional) linear system, which, assuming distinct eigenvalues, can be expressed in the general form
\begin{equation}
\label{eq:WagnerLin}
\hat\phi(t) =  c_{0} + \sum_{j = 1}^{r} c_{j}\exp{\lambda_{j}t},
\end{equation}
where $\lambda_{j}$ are the poles of the underlying dynamical system.  As the Wagner function is non-oscillatory and remains finite, we have real $\lambda_{j} < 0.$ The fact that $\phi(0) = 0.5$ and $\lim_{t\to\infty}\phi(t) = 1$ means that we also expect to have $c_{0} = 1$ and $\sum_{j = 0}^{r}c_{j} = -0.5$.

Note that in the frequency domain, the associated transfer function for this operator (with an additional factor of a derivative) is an approximation to the Theodorsen function.  Indeed, several approximations of the Wagner function were additionally formulated as approximations to the Theodorsen function in the frequency domain.
 The Theodorsen and Wagner functions are related by the Laplace transform pair
\begin{equation}
\frac{C(s)}{s} = \int_{0}^{\infty} \phi(t)\exp(-st){\rm d}t,
\end{equation}
so the Wagner function approximation $\hat\phi(t)$ corresponding to a Theodorsen function approximation $\hat C(s) $ is given by the inverse Laplace transform
\begin{equation}
\hat\phi(t) = \mathcal{L}^{-1}\left(\frac{\hat C(s)}{s}\right).
\end{equation}

Therefore, if we have a linear transfer function approximation to the Theodorsen function of the general form
\begin{equation}
\label{eq:theolin}
\hat C(s) = \frac{\sum\limits_{j=0}^{r-1}b_{j}s^{j}}{\sum\limits_{j=0}^{r-1}a_{j}s^{j}},
\end{equation}
then the equivalent approximation to the Wagner function is given by
\begin{equation}
\hat\phi(t) = \mathcal{L}^{-1}\left( \frac{\sum_{j=0}^{r-1}b_{j}s^{j}}{\sum_{j=0}^{r} a_{j}s^{j+1}}\right) 
=\mathcal{L}^{-1}\left( \frac{\sum_{j=0}^{r-1}b_{j}s^{j}}{\sum_{j=1}^{r} a_{j-1}s^{j}}\right).
\end{equation}
Note that with reference to Eq.~\eqref{eq:WagnerLin}, we can also express this in partial fraction form as 
\begin{equation}
\hat\phi(t) = \mathcal{L}^{-1}\left( \frac{c_0}{s}  +  \sum_{j = 1}^{r-1}\frac{c_j}{s-\lambda_j}\right),
\end{equation}
thus providing a direct relationship between the constants that define the linear approximations to the Theodorsen and Wagner functions. 

From linear systems theory, assuming that the $a_j$ and $b_j$ coefficients are scaled such that $a_{r-1} = 1$, we can also express approximations to Wagner's function in canonical controllable state-space form by
\begin{equation}
\label{eq:Wagss}
\begin{aligned}
\frac{{\rm d}}{{\rm d}t}\begin{bmatrix}  
x_1\\ x_2 \\ \vdots \\ x_{r-1} \\ x_r \end{bmatrix}
&= 
   \begin{bmatrix}      
   0 & 1 & 0& \cdots & 0 \\
   0 & 0 & 1 & \cdots & 0 \\
   \vdots & \vdots & \vdots & \ddots & \vdots \\
   0 &   0 &   0 &   \cdots &   1\\
   0 & -a_0 & -a_1 & \cdots & -a_{r-2}
   \end{bmatrix}
   \begin{bmatrix}  
x_1\\ x_2 \\ \vdots \\ x_{r-1} \\ x_r \end{bmatrix}
+ 
\begin{bmatrix}  
0\\ 0 \\ \vdots \\ 0 \\ 1 \end{bmatrix}u,\\
\hat\phi(t) &= 
\begin{bmatrix}  
b_0& b_1 & \cdots & b_{r-1} \end{bmatrix}
   \begin{bmatrix}  
x_1\\ x_2 \\ \vdots \\ x_r \end{bmatrix}.
\end{aligned}
\end{equation} 
 As in Eq.~\refeq{eq:SSjones}, here $\bx = (x_1,x_2,\cdots,x_r)^T$ is the internal system state, and the input variable $u$ represents the time-rate-of-change in aerodynamic flow conditions. 
If the Theodorsen function approximation is a proper transfer function, then the additional factor of $1/s$ ensures that the corresponding Wagner function approximation formulated in this manner is a strictly proper dynamical system; i.e., it has no feedthrough term between the input and output. Note that the state-space formulation would not be strictly proper if the input $u$ was not a rate-of-change quantity (e.g.~if it represented the angle of attack, rather than its time derivative). 
 In Eq.~\eqref{eq:Wagss}, each component of $\bx$ is related via $x_j = \dot x_{j-1} = x_1^{(j-1)}$, where $x_1^{(j-1)}$ is the $(j-1)$-th derivative of $x_1$. This means that Eq.~\eqref{eq:Wagss} can also be expressed using a $j$-th-order scalar differential equation as follows
 \begin{equation}
 \begin{aligned}
a_0 \dot x_1 &+ a_1 \ddot x_1 + a_2 x_1^{(3)} + \cdots + a_{r-2} x_1^{(r-1)} + x_1^{(r)} 
= \sum_{j=0}^{r-1}a_j x_1^{(j+1)} = u, \\
\hat\phi(t) &= b_0 x_1 + b_1 \dot x_1 + b_3 \ddot x_1 + b_{r-1} x_1^{(r-1)} = \sum_{j=0}^{r-1}b_j x_1^{(j)}.
\end{aligned}
\end{equation}

We list the parameters $c_j$ and $\lambda_j$ for various linear system approximations to the Wagner function in Table~\ref{tab:lincoeffs}. Several of these works develop numerous approximations of different orders; in such cases we list only one of the approximations (the fourth order approximation when it exists).  Note again that several of these approximations were originally formulated as approximations to the Theodorsen function.

 \begin{table}[hbt!]
 \caption{\label{tab:lincoeffs} Coefficients and eigenvalues (following Eq.~\eqref{eq:WagnerLin}) for various linear system approximations to the Wagner function.}
\centerline{
\tiny
\begin{tabular}
{ l c  c  c  c  c   c  c  c  c  c   }
\hline
Model & $c_0$ & $c_1$ & $c_2$ & $c_3$  & $c_4$ & $\lambda_0$  & $\lambda_1$  & $\lambda_2$ & $\lambda_3$ & $\lambda_4$ \\
\hline
R.~T.~Jones~\cite{Jones:38} & $1$ & $-0.165$ & $-0.335$ &- &-  & $0$  & $-0.0455$  & $-0.3$ & - &-\\
W.~P. ~Jones~\cite{jones1945aerodynamic} & $1$ & $-0.165$ & $-0.335$ & -&-  & $0$  & $-0.04$  & $-0.32$ & -&- \\
Venkatesan \& Friedmann~\cite{venkatesan1986new}& $1$ & $-0.203$ & $-0.236$ & $-0.06$ &-  & $0$  & $-0.072$  & $-0.261$ & $-0.8$&- \\ 
Peterson \& Crawley~\cite{peterson1988improved}  & $1$ & $-0.1058$ & $-0.2877$ & $0.0009$ &$-0.1002 $ & $0$  & $-0.0367$  & $-0.1853$ & $-0.568$1&$-0.5914$ \\ 
Eversman \& Tewari~\cite{eversman1991modified} & $0.9996$ & $-0.10624$ & $-0.30304$ & $1.8665$ &$-1.9386 $ & $0$  & $-0.0371$  & $-0.19142$ & $-1.1106$ & $-1.0768$ \\ 
Vepa~\cite{vepa1977finite} & $1$ & $0.011351$ & $ 0.045273$ & $-0.21479$ &$-0.22859 $ & $0$  & $-0.00044955$  & $-0.025409$ & $-0.10548$ &$-0.39661$ \\ 
Brunton~\cite{brunton:2012a} & $0.99699$ & $0.035611$ & $0.15655$ & $-0.24364$ & $-0.06119 $ & $0$  & $-0.014428$  & $-0.078617$ & $-0.2522$ & $-0.81275$ \\ 
Dowell~\cite{dowell1980simple} & $1$ & $-0.1055$ & $-0.2879$ & $-0.1003$ &-  & $0$  & $-0.0371$  & $-0.1857$ &$ -0.5886$ &- \\ 
\hline
\end{tabular}
}
\end{table}

   \begin{figure}
 \centering {
 \subfloat[]{\includegraphics[width= 0.45\textwidth]{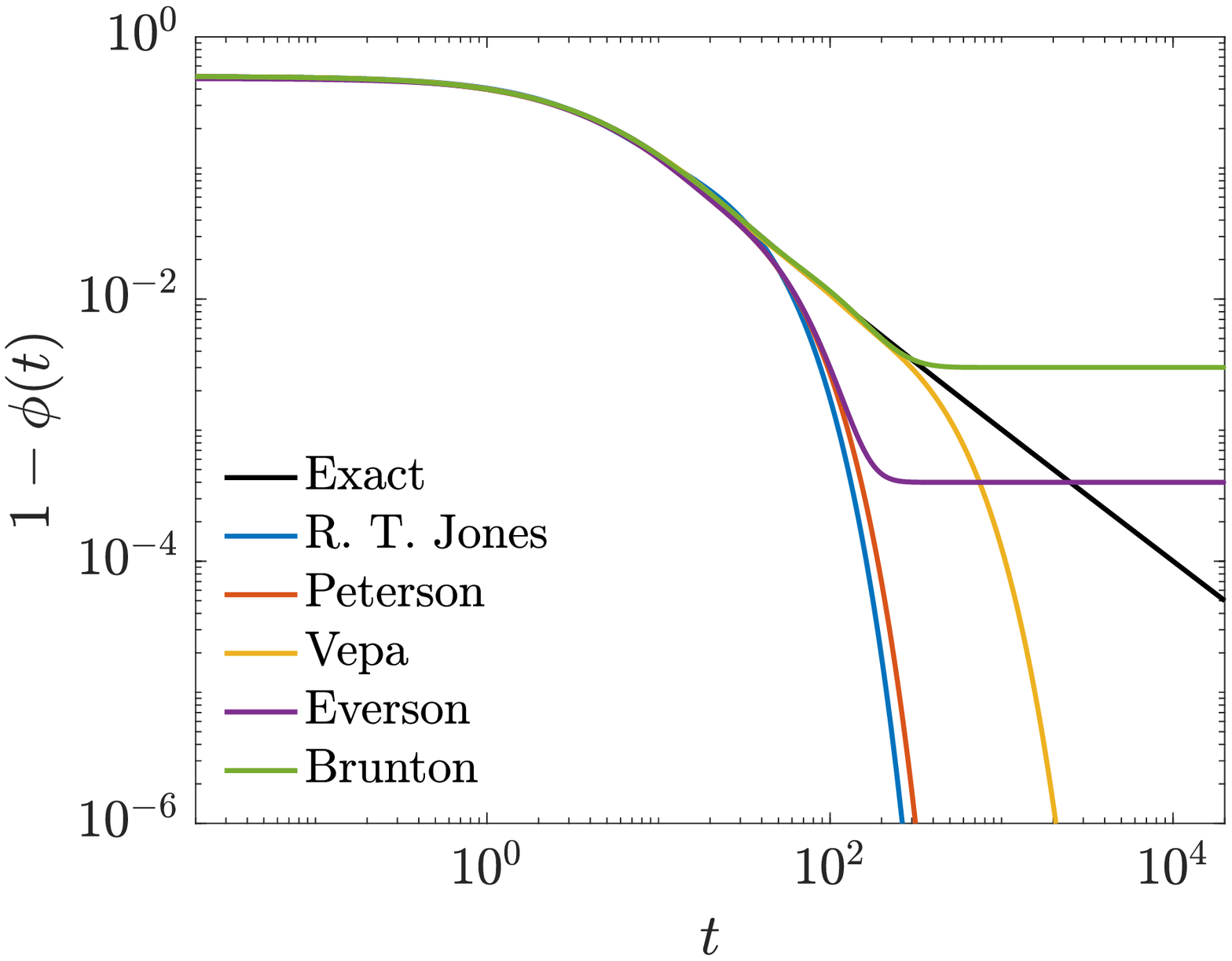}} \ \
 \subfloat[]{\includegraphics[width= 0.45\textwidth]{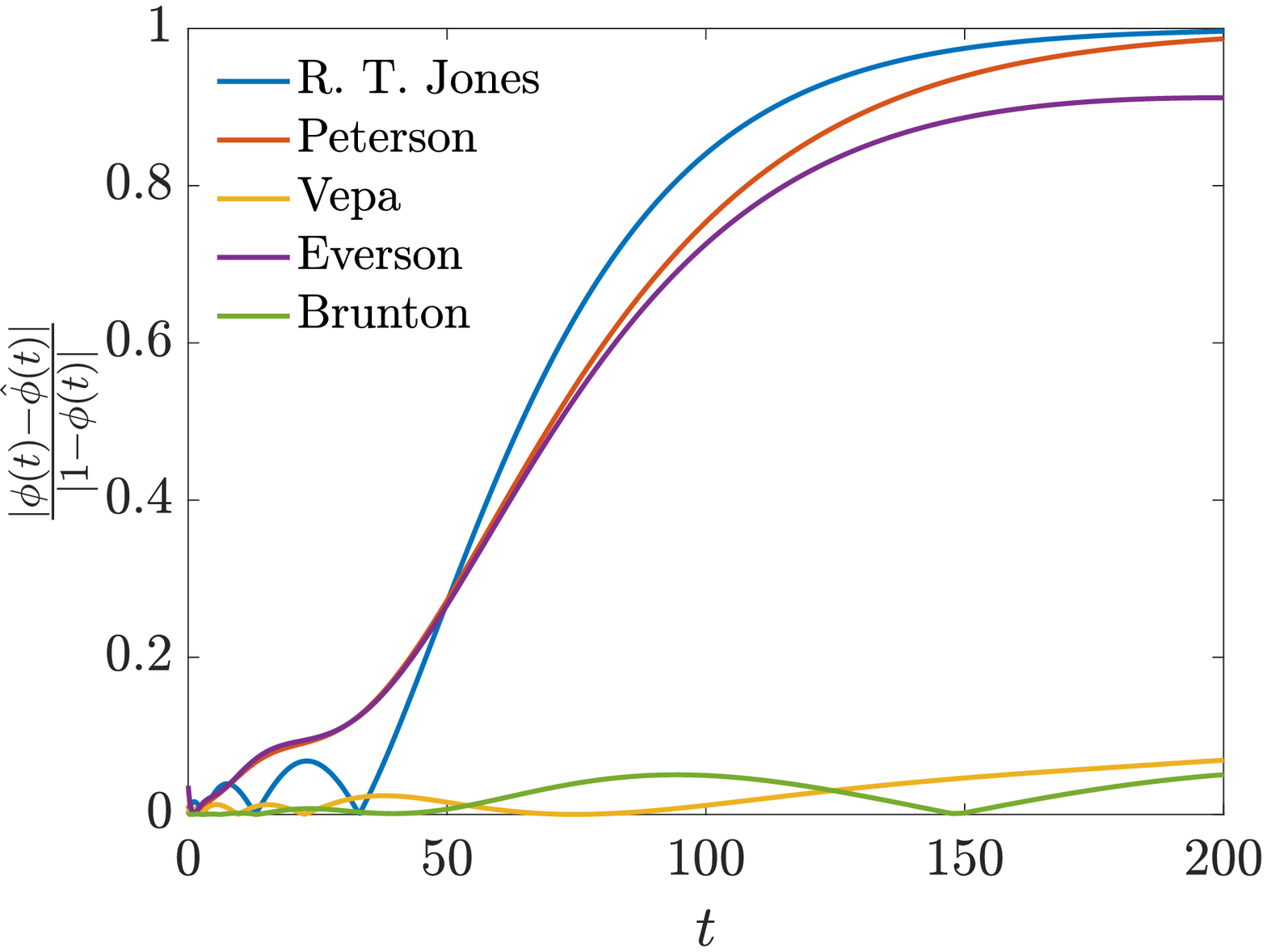}} \\
 \subfloat[]{\includegraphics[width= 0.45\textwidth]{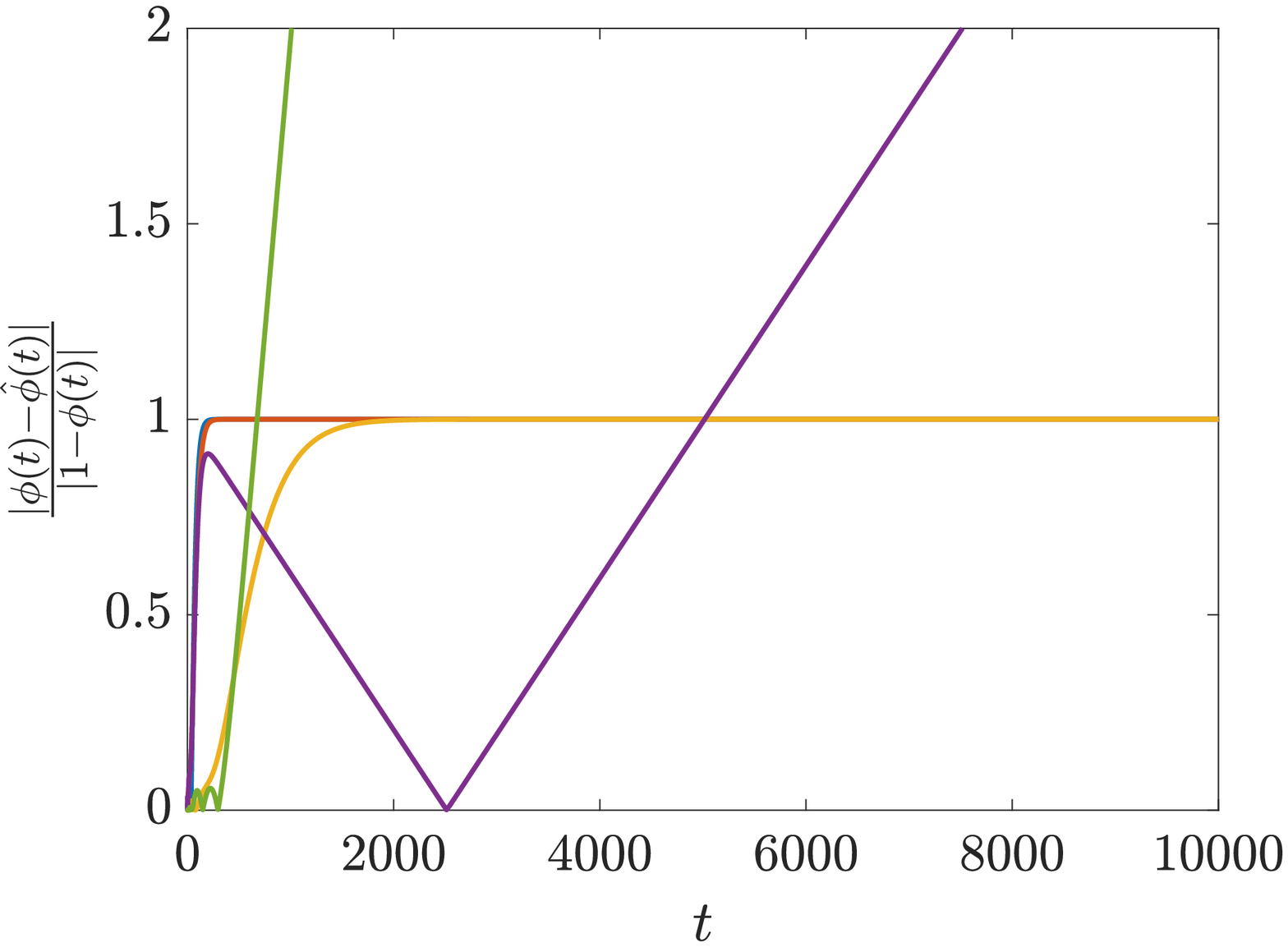}}
}
\caption{Comparison between the exact Wagner function $\phi(t)$ and various  linear approximations  over the domain (a) $0\leq t\leq 20$, and (b)  $0\leq t\leq 1000$, and (c) for the function $1-\phi(t)$ on a log scale.}
\label{fig:WagLin}
\end{figure}
Figure~\ref{fig:WagLin} plots a selection of these linear approximations, in comparison to the true Wagner function. 
For clarity we only plot some of the approximations listed in Table~\ref{tab:lincoeffs}, though we show those covering a range of behaviors. We note again that while these these approximations can exhibit a high level of accuracy at early times, it is not possible for them to capture the correct asymptotic behavior using a finite-dimensional linear system. 

As evidenced by the numerous different  approximations mentioned here, there are several methods to obtain an approximate linear model for a system, in either the time or frequency domain.  In Sect.~\ref{sec:BEM}, we will utilize one such method, the eigensystem realization algorithm (ERA), to obtain a linear model from startup flow data.

\subsection{Other approximations to the Wagner function}
\label{sec:otherApprox}
As well as the linear system approximations to the Wagner function considered in Sect.~\ref{sec:LinApprox}, there have also been several nonlinear approximations, in addition to that due to Garrick mentioned earlier.  
Note that for nonlinear approximations, the relationship between approximations to the Theodorsen and Wagner functions is more difficult to obtain in general. 

Numerous nonlinear approximations have been obtained by considering early- and late-time behavior of the Wagner function.
Vepa~\cite{vepa1977finite} shows 
 that a series expansion at small time is
\begin{equation}
\label{eq:vepashort}
\hat\phi(t) = 1 - \frac{2}{t + 4} + \frac{t^3}{768(1 +.875 t+1.28435t^2+1.84283t^3+4.09134t^4)},
\end{equation}
which can be viewed an extension of Garrick's approximation, obtained from the inverse Laplace transform of an asymptotic approximation to the Theodorsen function at large $s$. In practice this approximation is very similar to Garrick's, with the additional term having a maximum value of $1.45\times 10^{-4}$.

von K\`{a}rm\`{a}n and Sears~\cite{karman1938airfoil} also give approximations purportedly valid for small times, proposing
\begin{equation}
\label{eq:karmanshort}
\hat\phi(t) = \frac{1}{2} + \frac{t}{8} - \frac{t^2}{32} + 0.00554 t^3
\end{equation}
for $0 \leq t \leq 2$, and
\begin{equation}
\label{eq:karmanshort2}
\hat\phi(t) = 1 - \frac{\exp(-t/2) - (1+ 0.185 t)\exp(-0.185 t)}{4}
\end{equation}
for $0\leq t \leq 10$. Note that Eq.~\eqref{eq:karmanshort2} can also be viewed as the step response of a linear system with a repeated eigenvalue at $\lambda = -0.185$. 
Sears~\cite{sears1940operational} uses power series expansions of Bessel functions to find for small times 
\begin{equation}
\label{eq:searsshort}
\hat\phi(t) = \frac{1}{2}+\frac{t}{8}-\frac{t^2}{32} + \frac{7t^3}{768}+\cdots,
\end{equation}
and for large times
\begin{equation}
\label{eq:searslarge}
\hat\phi(t) = 1-\frac{1}{t}-\frac{2\log(2t)}{t^2}+\frac{2}{t^2} -\frac{6(\log(2t))^2}{t^3}+ \frac{16\log(2t)}{t^3}-\frac{7/2+\gamma-\pi^2}{t^3} + \cdots,
\end{equation}
where here $\gamma \approx 0.5772$ is the Euler-Mascheroni constant.
Note that Eq.~\eqref{eq:searsshort} matches Eq.~\eqref{eq:karmanshort} up to the quadratic term, and we find that Eq.~\eqref{eq:karmanshort} seems to give a better approximation.  
Note also that Eqns.~\eqref{eq:karmanshort} and \eqref{eq:searsshort} diverge for large times. 
 
\begin{figure}
 \centering {
 \subfloat[]{\includegraphics[width= 0.45\textwidth]{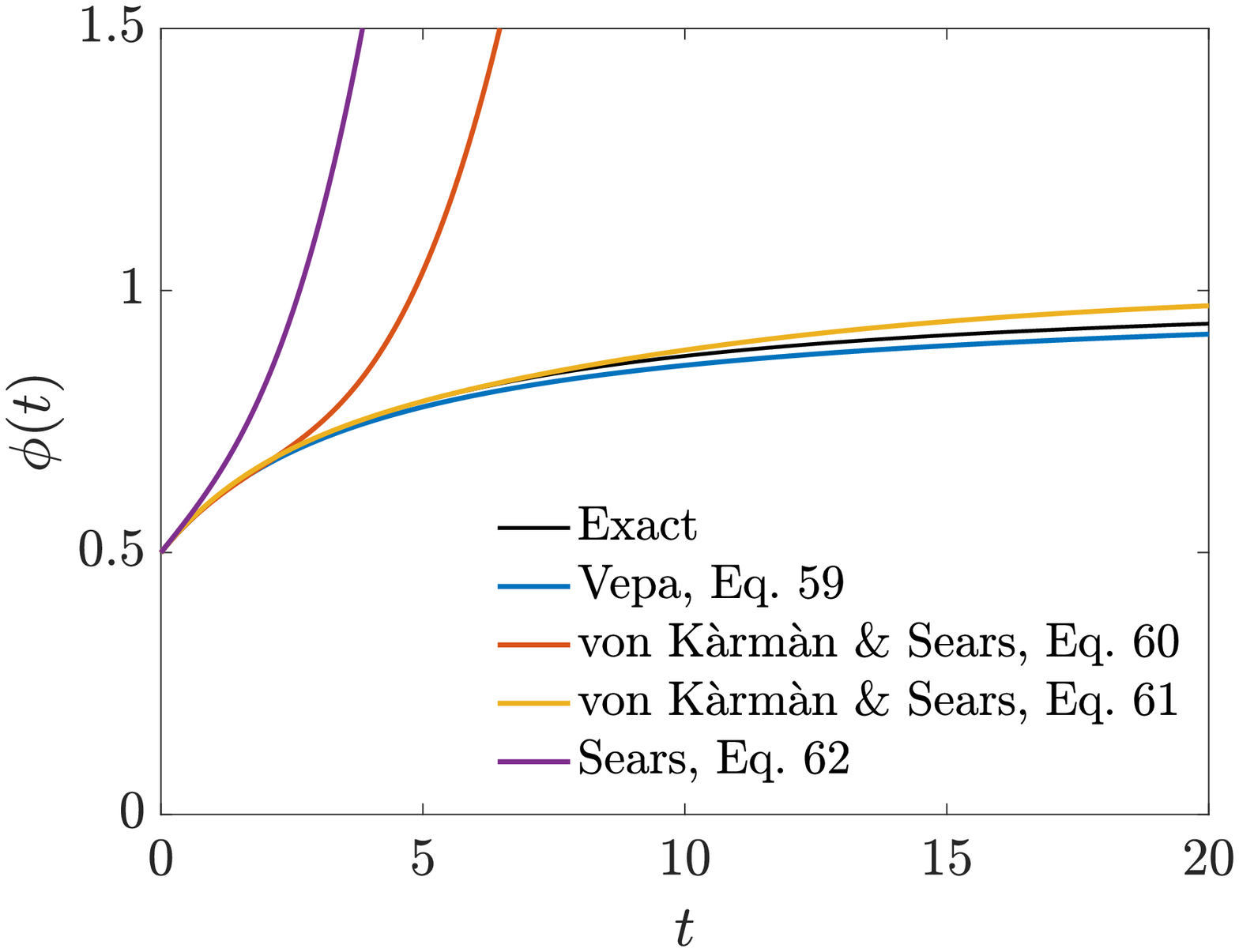}} \ \ 
 \subfloat[]{\includegraphics[width= 0.45\textwidth]{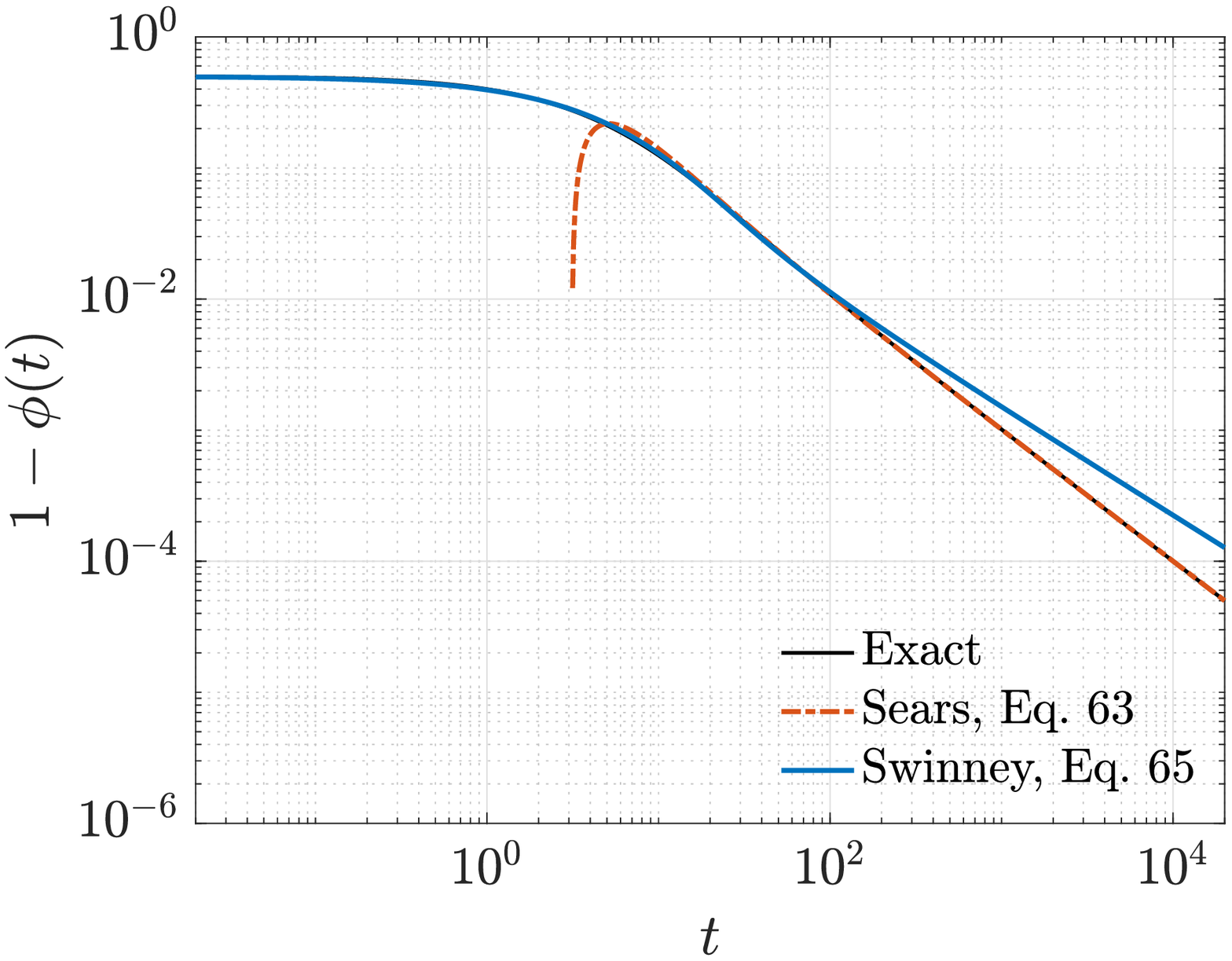}}
}
\caption{Comparison of various (a) small- and (b) large-time approximations to the Wagner function.}
\label{fig:ShortLong}
\end{figure}

Aside from short- and long-time expansions,  other nonlinear approximations can be obtained from nonlinear approximations to the Theodorsen function in the Laplace domain. 
Swinney~\cite{swinney1989fractional} and Bagley et al.~\cite{bagley1991fractional} apply fractional calculus approximations to the Theodorsen function of the form
\begin{equation}
C(s) = \frac{1 + c s^\beta}{1+2c s^\beta},
\end{equation}
where they use the parameters c = 2.19 and $\beta = 0.82$ or $5/6$, with the former value being more accurate, but the latter rational expression being used for mathematical convenience. This gives a corresponding Wagner function approximation expressible in terms of a Mittag-Leffler function~\cite{mittag1901representation} $E_\beta$:
\begin{equation}
\label{eq:MLwag}
\hat \phi(t) = 1- \frac{1}{2} E_\beta\left(\frac{t^\beta}{2c}\right),
\end{equation}
where 
\begin{equation}
E_\beta(t) = \sum_{k = 0}^\infty \frac{z^k}{{\bm{\Gamma}}(\beta k + 1)},
\end{equation}
and where here $\bm{\Gamma}$ is the gamma function.
Figure~\ref{fig:ShortLong} shows the performance of the relevant approximations at small and large times. It is observed that the small-time approximations lose accuracy within the first 20 time units.  The asymptotic expansion in Eq.~\eqref{eq:searslarge} accurately captures the behavior of the Wagner function for $ t > 50$, but this accuracy does not extend to earlier times. The approximation proposed in~\cite{swinney1989fractional,bagley1991fractional} (Eq.~\eqref{eq:MLwag}) is not as accurate at large times, but is relatively accurate at short times, and still performs better at large times than the linear models considered in Sect.~\ref{sec:LinApprox}.

This section has attempted to survey myriad of approximations that have been proposed for the Wagner function.  Overall, while these approximations capture many of the features of the Wagner function, there does not appear to exist an approximation that is accurate across all times, particularly if we are interested in capturing the correct asymptotic behavior at large times.

\section{Modeling the Wagner function with sparse system identification
}
\label{sec:sindy}

This section will introduce an alternative method for obtaining an approximation to the Wagner function, and more generally to model transient phenomena related to impulsive changes in flow over airfoils. The method will apply a variant of the sparse identification of nonlinear dynamics (SINDy) algorithm, which is described in Sect.~\ref{sec:method}. This method is subsequently applied to the Wagner function, as well as to a more general startup flow, in Sects.~\ref{sec:WagSindy} and \ref{sec:BEM}, respectively. 

\subsection{Modeling methodology}
\label{sec:method}

As demonstrated in section \ref{sec:LinApprox}, there are  fundamental limitations in the accuracy of approximating the Wagner function using the response of a linear system.  Here, we examine the use of nonlinear system identification methods to identify approximations to the Wagner function. In particular, we will identify a nonlinear system that gives an initial condition response corresponding to the Wagner function. 

 Suppose we have the response of the lift for an airfoil (in a given lift-producing state) to an impulsively-started flow, given by $L(t)$.  We work with the normalized, steady-state-subtracted lift, given by
\begin{equation}
\label{eq:liftsubvar}
\tilde{L}(t) = \frac{L(t)-L_0}{L_0},
\end{equation}
where $L_0 = \lim_{t\to\infty} L(t)$. 
It was observed in section \ref{sec:intro} that  Garrick's approximation to the Wagner function (Eq.~\eqref{eq:garrick}) was a solution to a nonlinear ordinary differential equation.  
More generally, using the variable defined in Eq.~\eqref{eq:liftsubvar}, a function of the  general form
\begin{equation}
\label{eq:Hyp}
\tilde L(t) = \frac{b}{t+c},
\end{equation}
is a solution of the nonlinear differential equation
\begin{equation}
\dot{ \tilde L}(t) =  -\frac{b}{(t+c)^2} =- \frac{ \tilde L^2}{b}.
\end{equation}
This highlights that the asymptotic behavior of the Wagner function can be represented as the response of a scalar differential equation with a quadratic nonlinearity.  
Using this observation as motivation, we seek an approximation to the Wagner function from a slightly more general class of nonlinear, scalar differential equations.

 We will assume that $\tilde L(t)$ can be approximated by the solution of a nonlinear, scalar, ordinary differential equation.  We will consider using both first and second order differential equations, with general form
\begin{align}
\label{eq:form1}
\dot{\tilde L}(t) &= \sum_{j = 0}^r c_{j} \tilde L^j(t), \\
\label{eq:form2}
\ddot{\tilde L}(t) &= \sum_{j,k\in\bf J} c_{jk} \tilde L^j(t) \dot{\tilde L}^k(t),
\end{align}
respectively, where  $\{j,k\} \in {\bf J}$ are non-negative integers, and $\dot\tilde$ denotes a time derivative. 
Note that we can also express Eq.~\eqref{eq:form2} as two first order differential equations by introducing $\dot{\tilde L}$ as an additional variable. 
Loiseau et al.~\cite{loiseau2018sparse} showed that it is possible to model nonlinear fluid flow dynamics, such as the viscous flow past a cylinder, with the lift force and its derivatives as the state vector.  
After selecting a set of candidate monomials $\tilde L^j$ or $\tilde L^j \dot{\tilde L}^k$ by specifying $r$ or ${\bf J}$ respectively, we find the coefficients $c_j$ or $c_{jk}$ using a regularized, sparsity-promoting least-squares algorithm~\cite{brunton2016sindy}. Letting $\bm c$ be a vector of the coefficients $c_j$ or $c_{jk}$, one way of formulating this is to find
\begin{equation}
\label{eq:sindymin}
{\bm c} = \argmin_{\bm c}\left[ \int_{t = 0}^\infty e(t)^2 {\rm d}t + \beta_2^2 \|\bm{c}\|^2_2 + \beta_0 \|\bm{c}\|_0\right],
\end{equation}
for specified regularization parameters $\beta_0$, $\beta_2 \geq 0$, where $e(t)$ denotes the residual error of the models given by Eqns.~\eqref{eq:form1}--\eqref{eq:form2}, respectively
\begin{align}
\label{eq:error1}
e(t) &= \left|\dot{\tilde L}(t) - \sum_{j = 0}^r c_{j} \tilde L^j(t)\right| \\
\label{eq:error2}
e(t) &= \left|\ddot{\tilde L}(t) - \sum_{\bf J} c_{jk} \tilde L^j(t) \dot{\tilde L}^k(t)\right|.
\end{align}
The $\| \bm{c} \|_0$ term in Eq.~\eqref{eq:sindymin} promotes sparsity of the coefficient vector $\bm{c}$, while the inclusion of the $\|\bm{c}\|_2$  term provides a regularization that ensures that the nonzero entries of $\bm{c}$ do not grow unreasonably large, which can be a consequence of overfitting the training data. 
In practice, the nonconvexity of the $\ell_0$ pseudo-norm makes this optimization problem intractable to solve directly. 
Instead, we apply an iterative thresholding algorithm that finds a regularized least-squares fit, sets to zero the coefficients below a certain threshold $\eta$, and then repeats the least squares fit. 
This can be described as follows
 \begin{enumerate}
 \item Identify a coefficient vector ${\bm c}$ from solving Eq.~\eqref{eq:sindymin} without the $\ell_1$ term, i.e.
 \begin{equation}
 \label{eq:sindymin2}
{\bm c} = \argmin_{\bm c}\left[\int_{t = 0}^\infty e(t)^2 {\rm d}t + \beta_2^2 \|\bm{c}\|_2 \right].
\end{equation}      
 \item Remove coefficients $c_{jk}$ that are smaller than the threshold $\eta$, and recompute Eq.~\eqref{eq:sindymin2} with these terms excluded.
 \item Repeat step 2 until a converged solution is obtained.
 \end{enumerate}
 More details of the SINDy procedure are given in~\cite{brunton2016sindy}, while discussion of the additional $\ell_2$ ridge regularization is given in~\cite{rudy2019data}.
 
\subsection{Approximating the Wagner function using SINDy}
\label{sec:WagSindy}

This section details the results from applying the methods described in section \ref{sec:method} to the Wagner function. 
We obtain accurate data for the Wagner function using the methods described in section \ref{sec:compute}, and identify models on data computed for $t \in [0,2000]$, with a uniform timestep $dt = 0.02$. 
 
We start by identifying models of the form given by Eq.~\eqref{eq:form1}, and consider a series of models, where the highest order of nonlinearity is taken to be $r\in\{2,3,\cdots, 8\}$. Table~\ref{tab:coef1} shows the identified coefficients, identified with the regularization parameters $\eta = 0.1$ and $\beta_2 = 10^{-5}$. 
 We observe that the constant and linear terms disappear from the model, and that the coefficients of the higher order terms appear to be relatively stable as the order of the nonlinearity increases. 
 The constant term disappearing is consistent with $\tilde L = 0$ being an equilibrium solution  to the identified differential equation. The fact that the coefficient of the linear term is zero is consistent with the known asymptotic behavior of the Wagner function. For example, for the case where $r=2$, we obtain a solution of the form given in Eq.~\eqref{eq:Hyp}, without a linear term, but with a nonzero linear term the differential equation 
\begin{equation}
\dot{\tilde L}(t) = c_1 \tilde L(t) + c_2 \tilde L^2(t)
\end{equation}
 with the initial condition $\tilde L (0) = 0.5$ has the general solution 
 \begin{equation}
\tilde L(t) = \frac{c_1\exp(c_1 t)}{2 c_1+c_2-c_2\exp(c_1 t)},
\end{equation}
which does not match the known asymptotic behavior. 
This highlights the importance of promoting sparsity to set certain terms to zero, as even a very small nonzero $c_1$ would thus give the incorrect asymptotic behavior. 
 The specific values of the coefficients given in Table~\ref{tab:coef1} are dependent on the choice of regularization parameters, though the overall behavior is generally robust to various choices. 
 Without enforcing sparsity (i.e.~letting $\eta = 0$, and with the same value of $\beta_2$) the constant coefficients are of order $10^{-8}$--$10^{-5}$ for the various models considered in Table 2, and the linear terms  range from $10^{-5}$--$0.03$. Thus, as long as $\eta$ is between 0.03 and the smallest nonzero term identified (approximately $0.4$), the identified coefficients will be identical.  The choice of $\beta_2$ does affect the identified coefficients (particularly the larger ones), though the results obtained are very similar for $0 \leq \beta \leq 10^{-3}$. 
 
\begin{figure}
 \centering {
 \subfloat[]{\includegraphics[width= 0.45\textwidth]{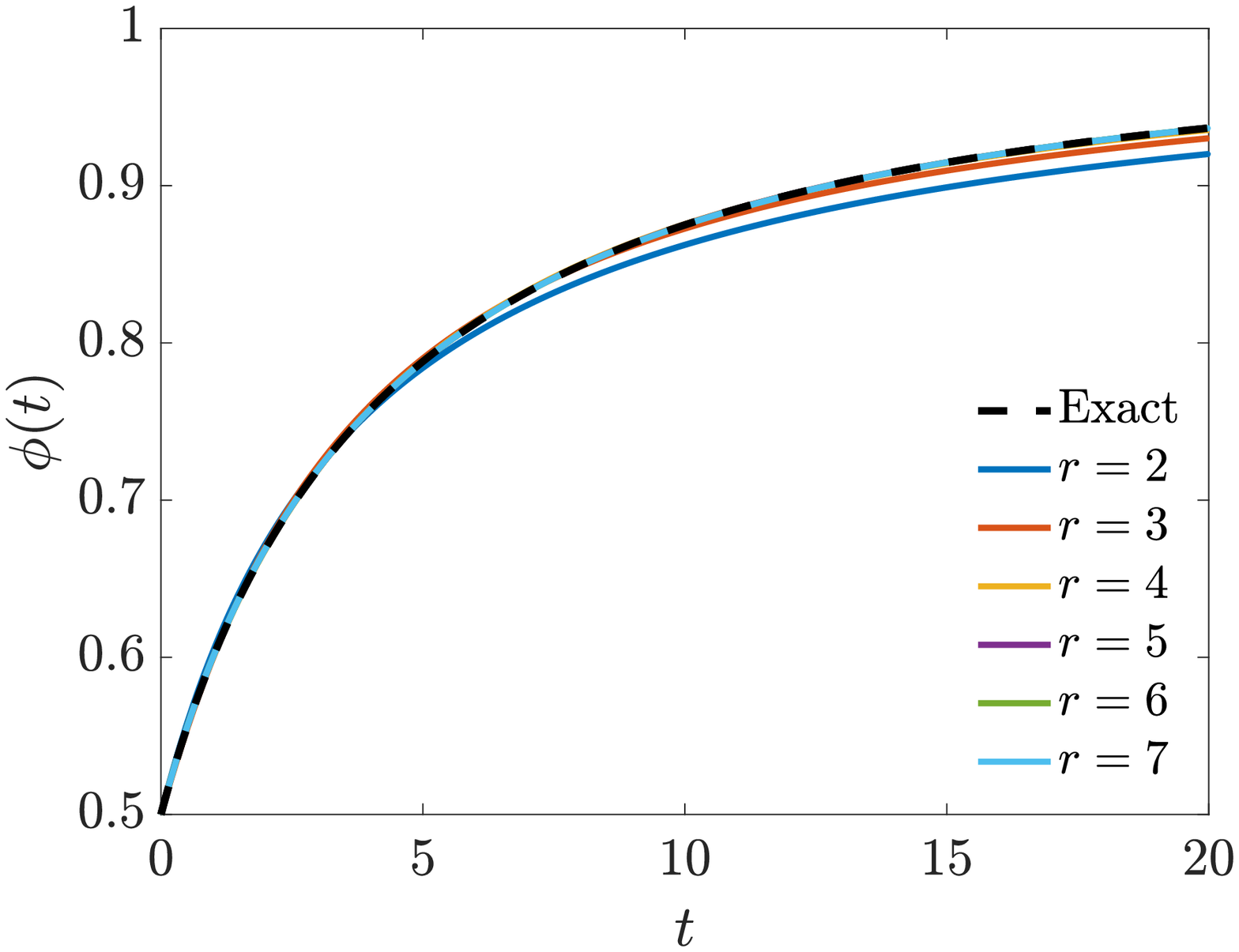}}
 \subfloat[]{\includegraphics[width= 0.45\textwidth]{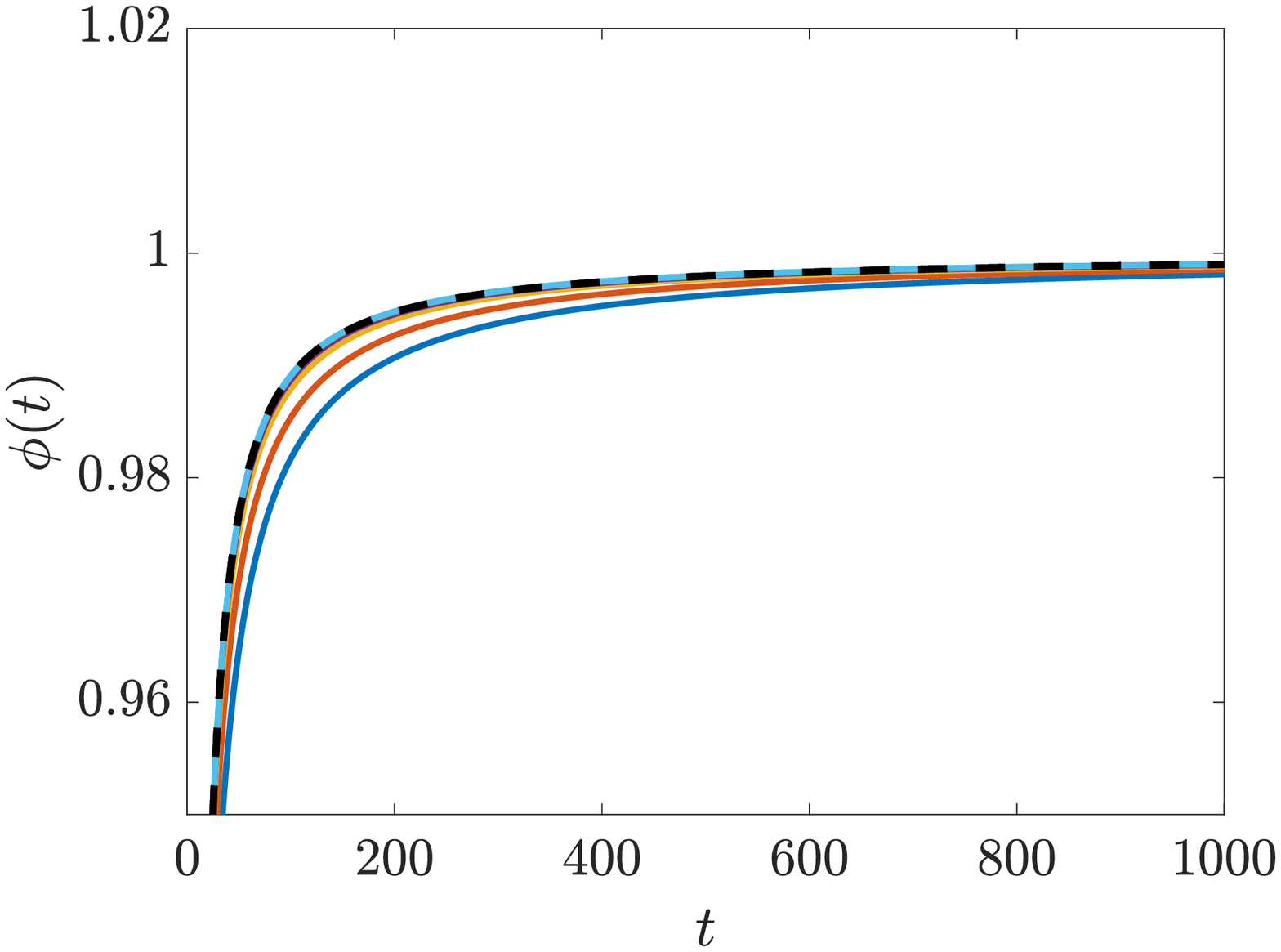}}\\
 \subfloat[]{\includegraphics[width= 0.45\textwidth]{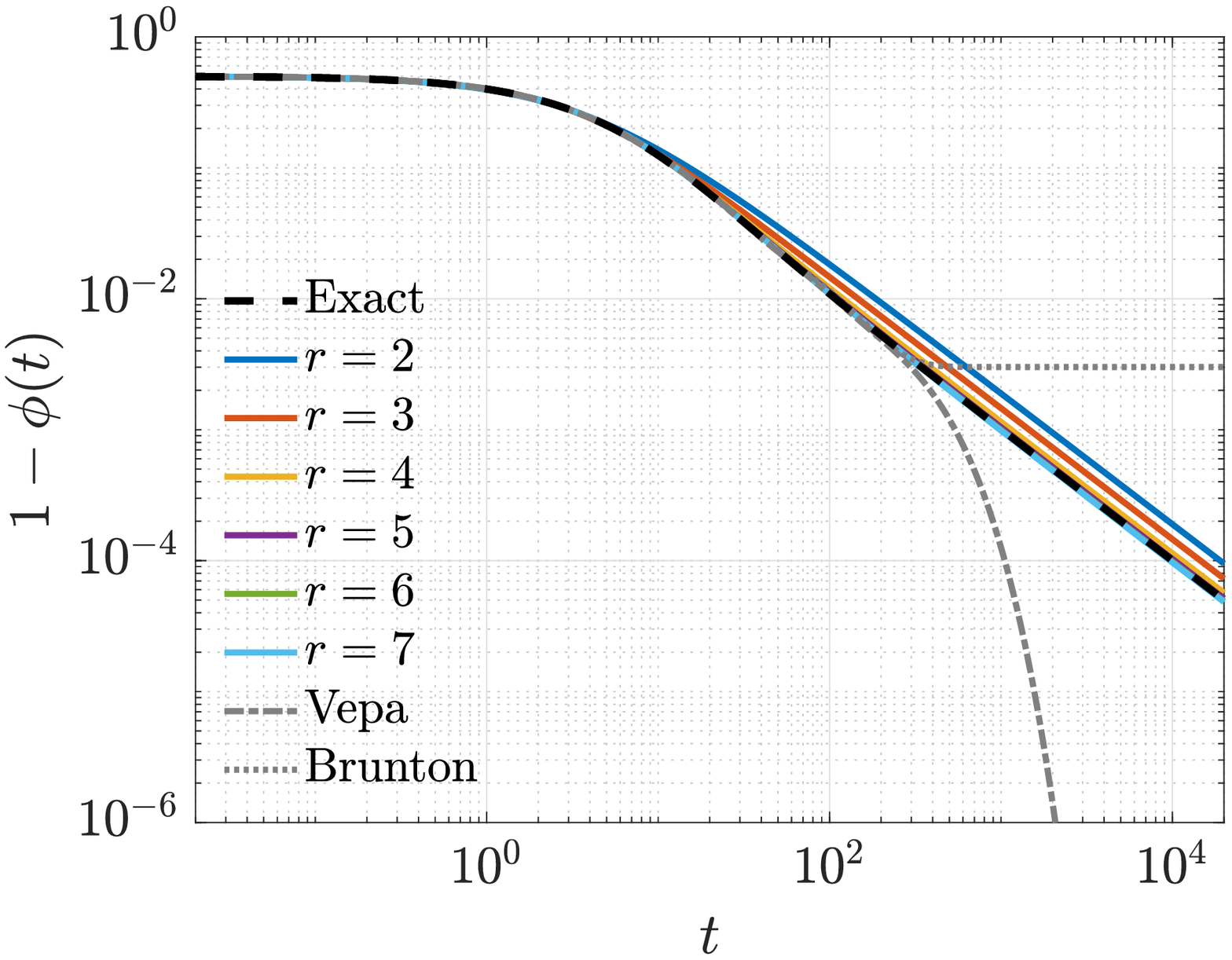}}
  \subfloat[]{\includegraphics[width= 0.45\textwidth]{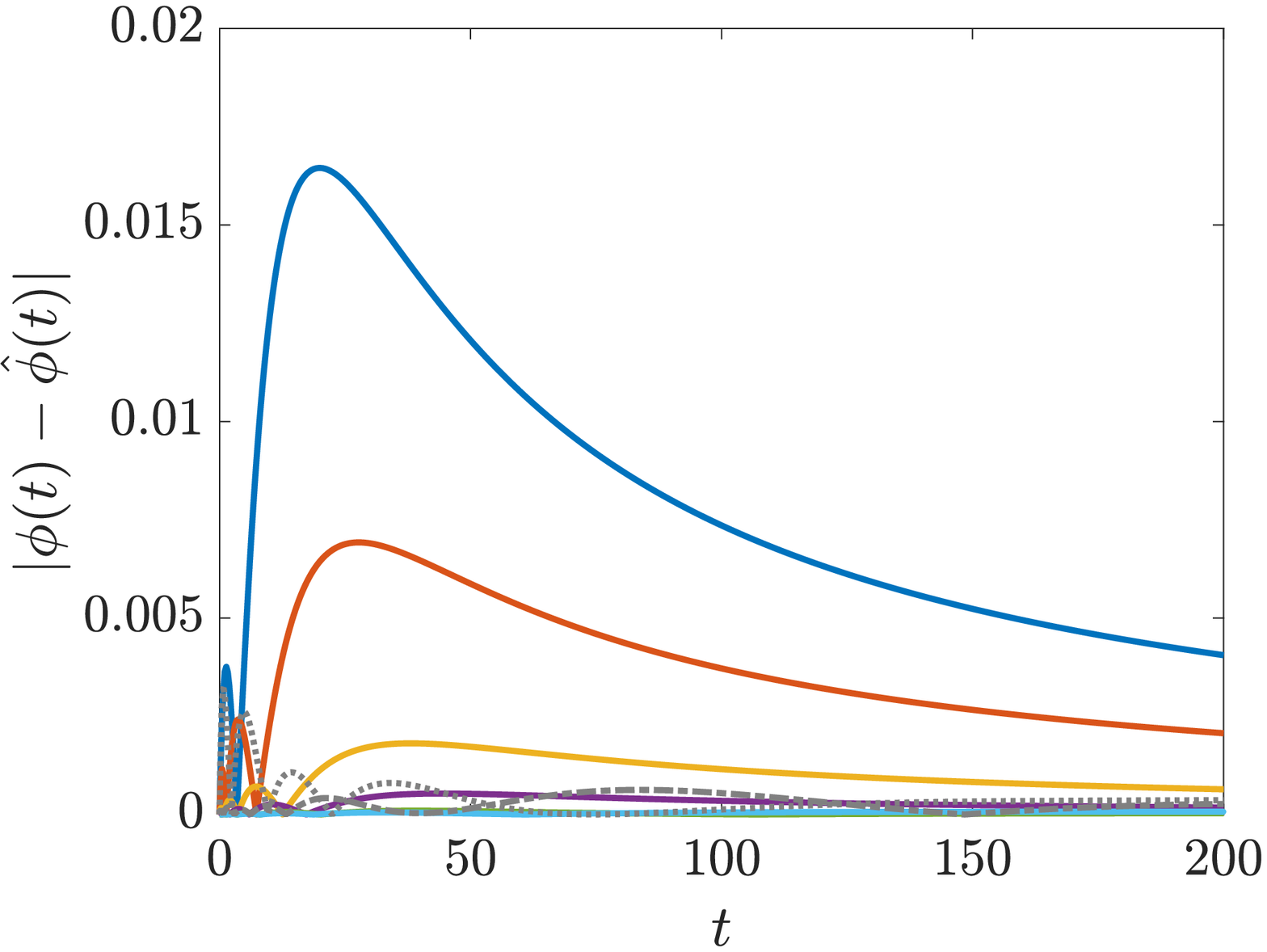}}\\
   \subfloat[]{\includegraphics[width= 0.45\textwidth]{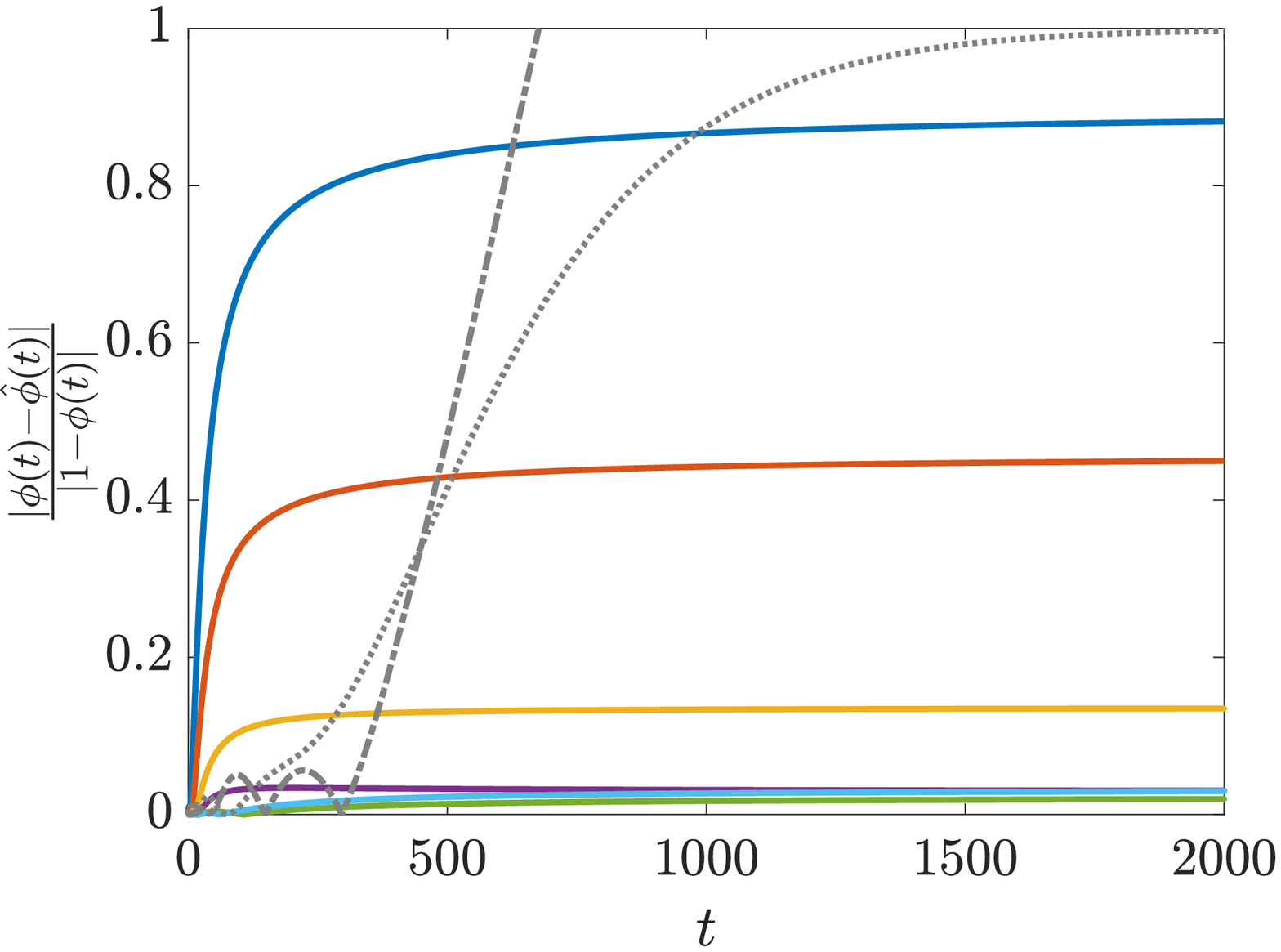}}
}
\caption{Comparison between the exact Wagner function $\phi(t)$ and approximations taking the form of Eq.~\eqref{eq:form1} for various values of $r$, identified using the SINDy method, over the domain (a) $0\leq t\leq 20$, and (b)  $0\leq t\leq 1000$, and (c) for the function $1-\phi(t)$ on a log scale. Subplots (d) and (e) show the absolute and relative (to asymptote) error of this approximation, respectively. Subplots (c)-(e) also show the predictions/error of the Vepa and Brunton models, as originally shown in Fig.~\ref{fig:WagLin}.}
\label{fig:WagSindy1}
\end{figure}

Figure~\ref{fig:WagSindy1} shows the accuracy of these models in reconstructing the Wagner function across a range of time intervals. We observe that all models appear to have the correct asymptotic behavior, with the accuracy improving as $r$ increases from 2 to 6.  In particular, the absolute and relative error shown in Figs.~\ref{fig:WagSindy1}(d)--(e) is typically lower than those observed for those shown in Figs.~\ref{fig:WagError1} and \ref{fig:WagLin}, particularly for larger $r$ values. For direct comparison, the two best performing linear models are also plotted again here. The error increases again slightly for $r = 7$ in comparison to $r=6$. While not shown, performance of the $r=8$ model is similar to the $r=7$ case. 
It can be verified that for large times, since $\tilde L(t) \to 0$, the quadratic term $c_2\tilde L^2$ becomes the dominant contributor to the sum on the right hand side of Eq.~\eqref{eq:form1}, which is responsible for providing the correct asymptotic behavior.
 
  \begin{table}[hbt!]
  \caption{\label{tab:coef1} Identified coefficients $c_{j}$ of the terms  $\tilde L^j(t)$ for Wagner function models taking the form of  Eq.~\eqref{eq:form1}, for $2\leq r \leq 8$.  Models are identified with the regularization parameters $\eta = 0.1$, $\beta_2 = 10^{-5}$.}
\centering
\begin{tabular}
{  l c c c c c c c c c   }
\hline
$r$ & $c_{0}$ &$c_{1}$ & $c_{2}$ & $c_{3}$  & $c_{4}$  & $c_{5}$  & $c_{6}$ & $c_7$ & $c_8$ 
\\
\hline
2 & 0  & 0 & 0.5265 &  - & - &  -  & - & - & - \\ 
3 & 0  & 0 & 0.6858 & 0.4161 & - &  -  & - & - & - \\ 
4 & 0  & 0 & 0.8803 & 1.6676 & 1.8349 &  -  & - & - & - \\ 
5 & 0  & 0 & 0.9722 & 2.7234 & 5.4262 &  3.7528  & - & - & - \\ 
6 & 0  & 0 & 1.0236 & 3.6396 & 10.7535 & 16.2454 & 10.2251 & - & - \\ 
7 & 0  & 0 & 1.0347 & 3.9252 & 13.2502 & 26.0199 & 27.8458 & 11.9184 & - \\ 
8 & 0  & 0 & 1.0356 & 3.9257 & 12.9819 & 23.2178 & 16.5324 & -8.388 & -13.5316 \\ 
\hline
\end{tabular}
\end{table}

\begin{figure}
 \centering {
 \subfloat[]{\includegraphics[width= 0.45\textwidth]{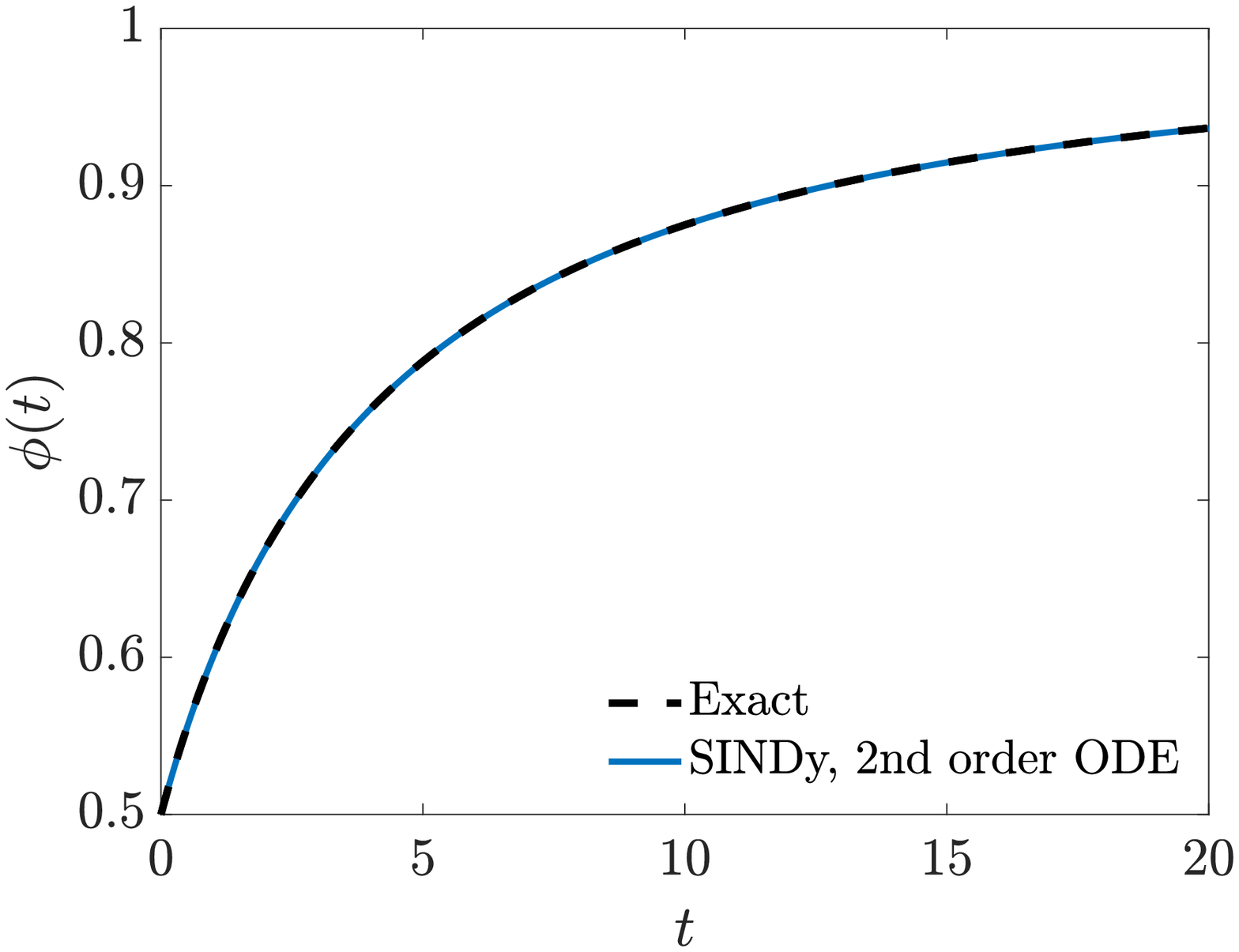}}
 \subfloat[]{\includegraphics[width= 0.45\textwidth]{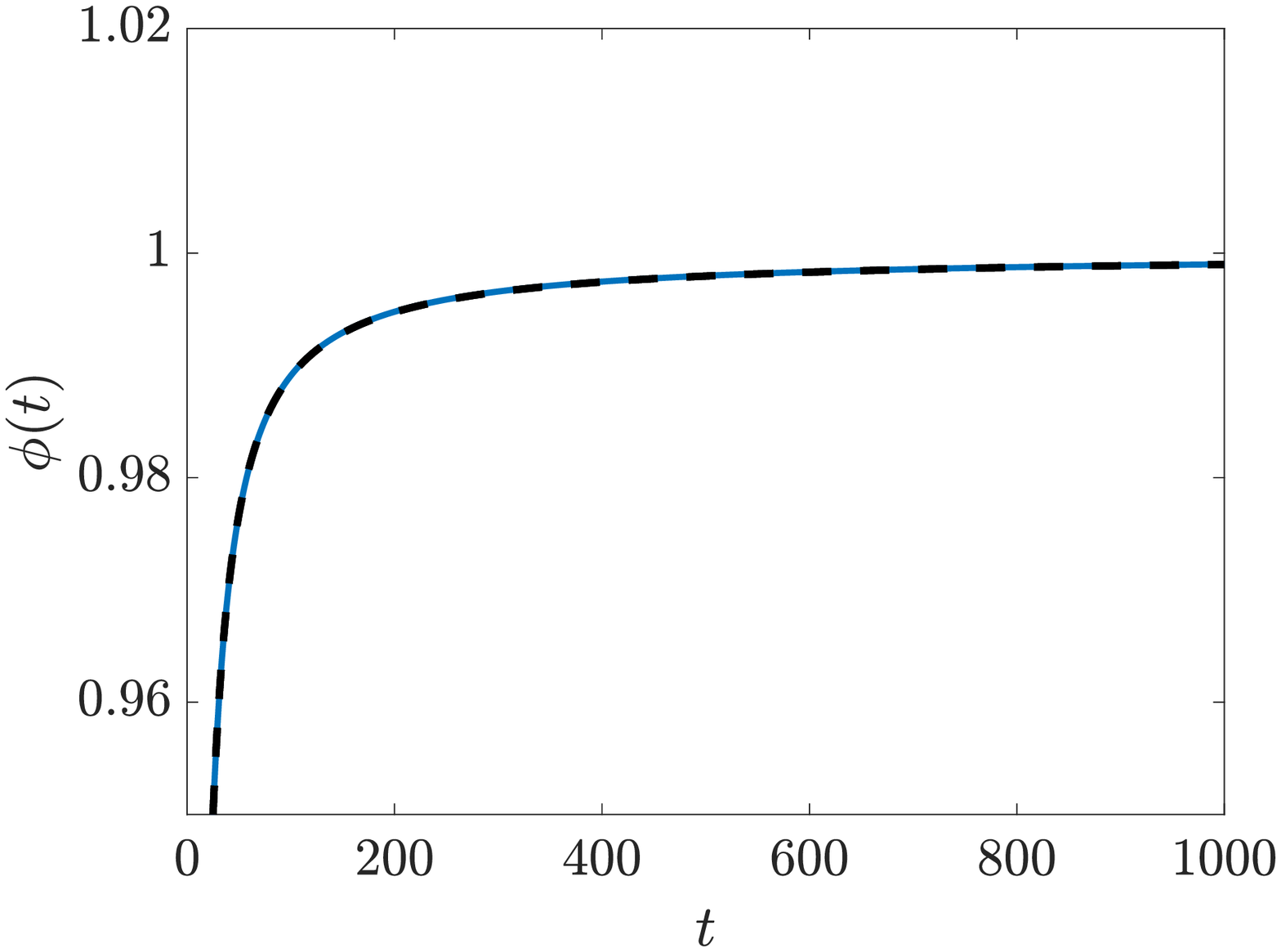}}\\
 \subfloat[]{\includegraphics[width= 0.45\textwidth]{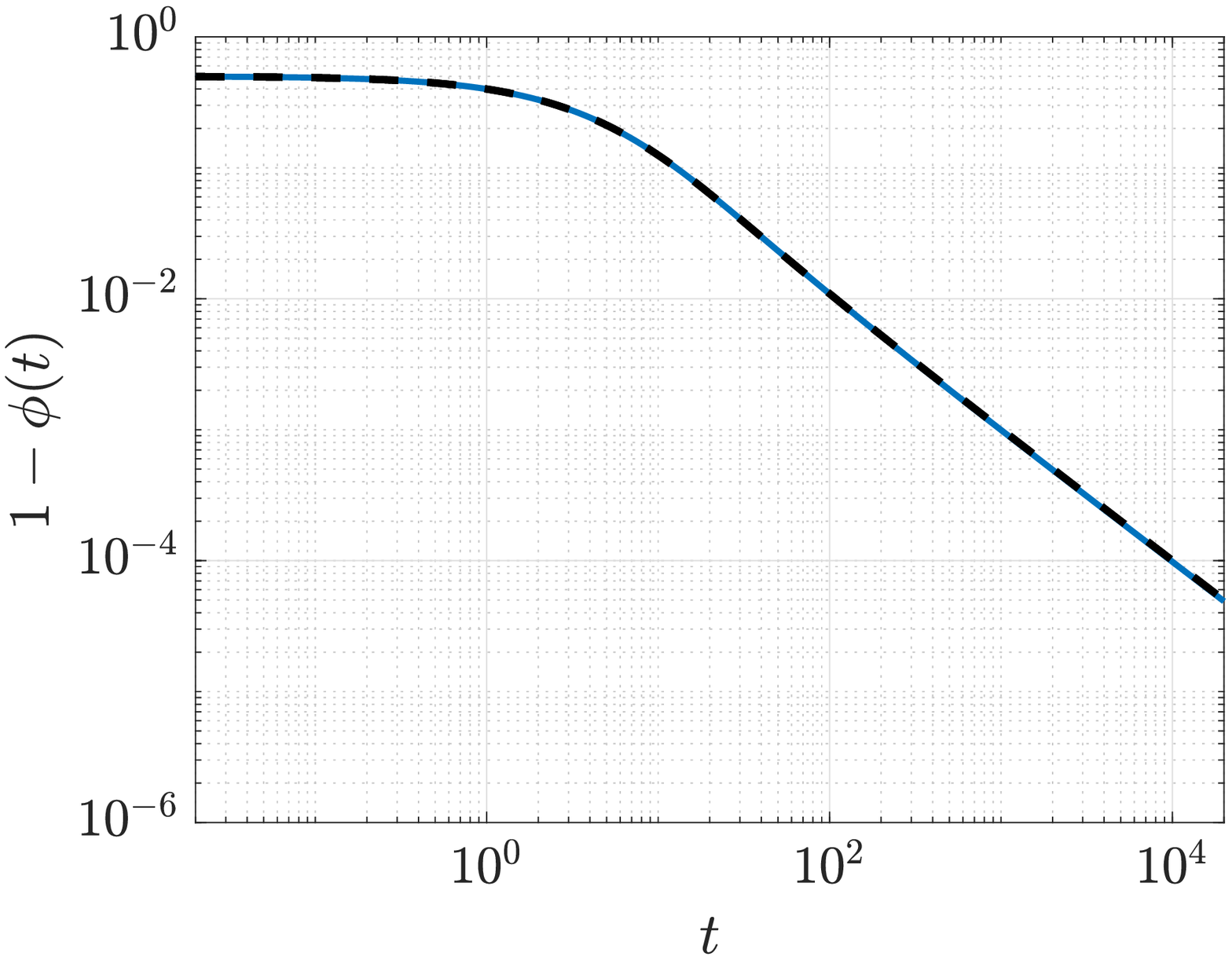}}
  \subfloat[]{\includegraphics[width= 0.45\textwidth]{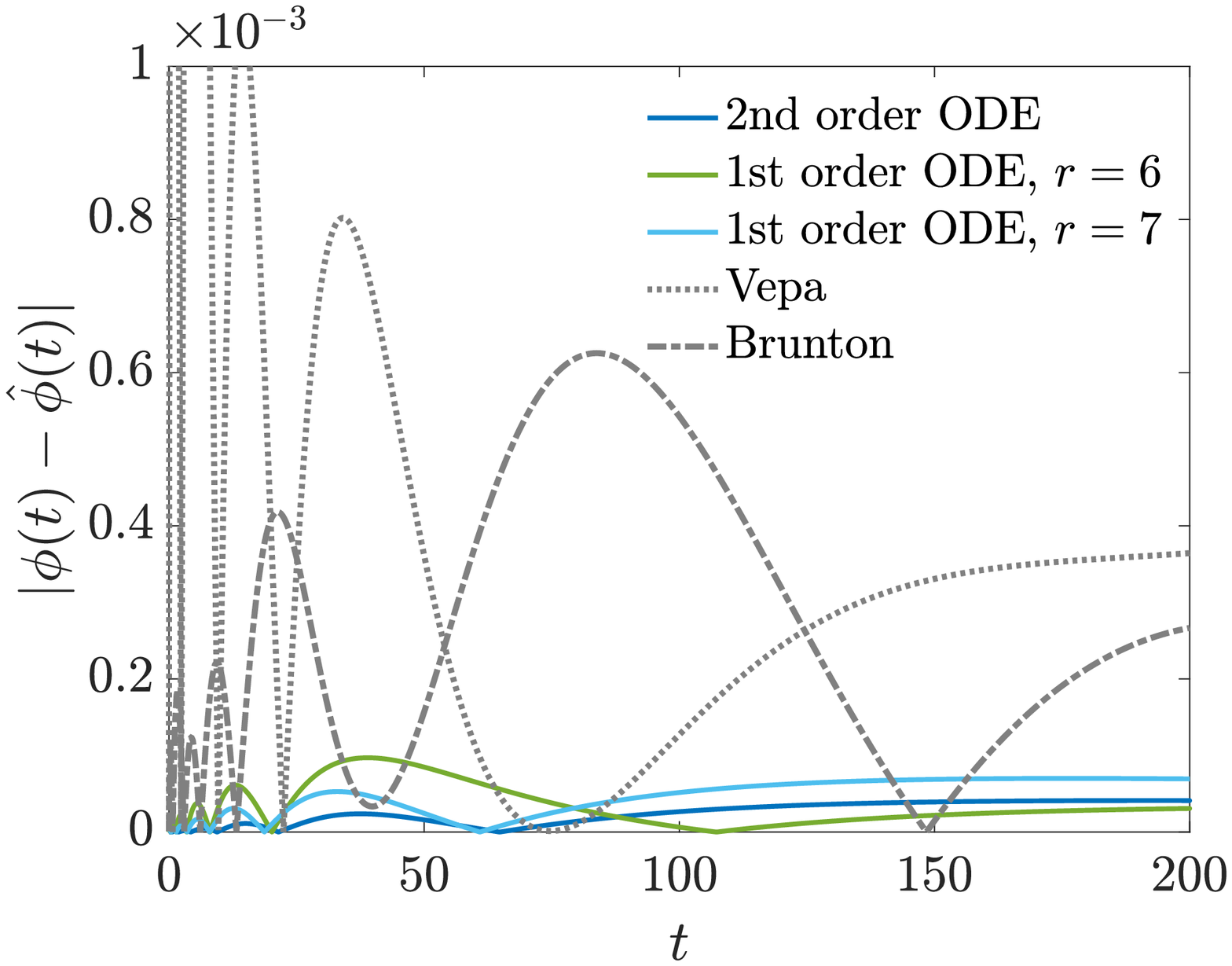}}\\
   \subfloat[]{\includegraphics[width= 0.45\textwidth]{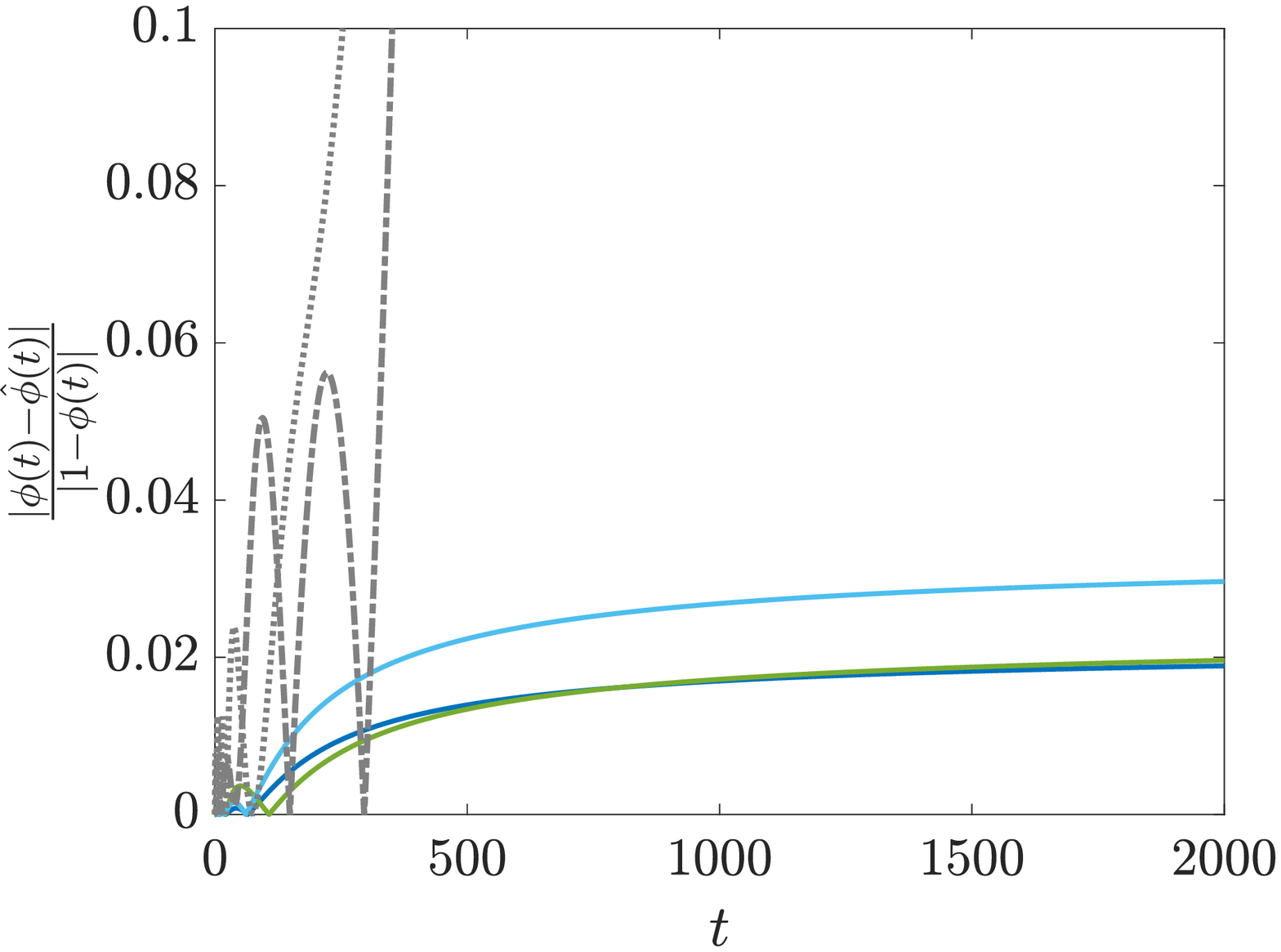}}
}
\caption{
As for Fig.~\ref{fig:WagSindy1}, but for a model identified using a second order differential equation of the general form given in Eq.~\eqref{eq:form2}. For comparison, subplots (d) and (e) also show the error of the first order models with $r=6$, and $7$,  as well as the two most accurate linear models from Fig.~\ref{fig:WagLin}.}
\label{fig:WagSindy2}
\end{figure}

We now assess the accuracy of a model that takes the general form given in Eq.~\eqref{eq:form2}, where $j$, $k\in\{0,1,2,3\}$ with $j+k \leq 3$ (that is, we include up to cubic terms). The regularization parameters are again set to be $\beta_2 = 10^{-5}$ and $\eta = 0.1$, and here, unlike for the first order differential equations, we find that a nonzero $\beta_2$ is required to identify a stable model. 
 The identified coefficients for the resulting model are given in Table~\ref{tab:coef2}. Similar to the models considered in Table~\ref{tab:coef1}, the constant term, as well as the term linearly proportional to the $\tilde L$, are set to zero due to the sparsity-promoting step. 
Figure~\ref{fig:WagSindy2} shows that this approximation again very closely matches the true Wagner function across all times.  In Figs.~\ref{fig:WagSindy2}(d)--(e), we also show a comparison of the absolute and relative error to the $r=6$ and 7 cases plotted in Figs.~\ref{fig:WagSindy1}(d)--(e), as well as again showing the Brunton and Vepa linear models. We find that the second order model has a smaller maximum absolute error over the interval $t\in[0,200]$, with the relative error over a longer time horizon similar to the first order model with $r = 6$. We note again that these errors are several orders of magnitude smaller than any of the other approximations considered previously.  Moreover, the number of coefficients required to specify the models (8 nonzero coefficients for the second order model, 5 coefficients for the first order model with $r=6$) is comparable to the number of parameters required to specify many of the other approximations. 

  \begin{table}[hbt!]
  \caption{\label{tab:coef2} Coefficients $c_{jk}$ of the terms  $\tilde L^j(t) \dot{\tilde L}^k(t)$ in Eq.~\eqref{eq:form2}. The model is identified with the regularization parameters $\eta = 0.1$, $\beta_2 = 10^{-5}$.}
\centering
\begin{tabular}
{  l c c c c c c c c c   }
\hline
$c_{00}$  & $c_{10}$ &$c_{01}$ & $c_{20}$ & $c_{11}$  & $c_{02}$  & $c_{30}$  & $c_{21}$  & $c_{12}$  & $c_{03}$  \\
\hline
$0$ & 0  & -0.3773 & 0.3857 &  3.7246 & 5.4840 &  -0.4893  & 0.2268 & 3.1434 &-4.2629 \\
\hline
\end{tabular}
\end{table}

\begin{figure}
 \centering {
  \subfloat[]{\includegraphics[width= 0.45\textwidth]{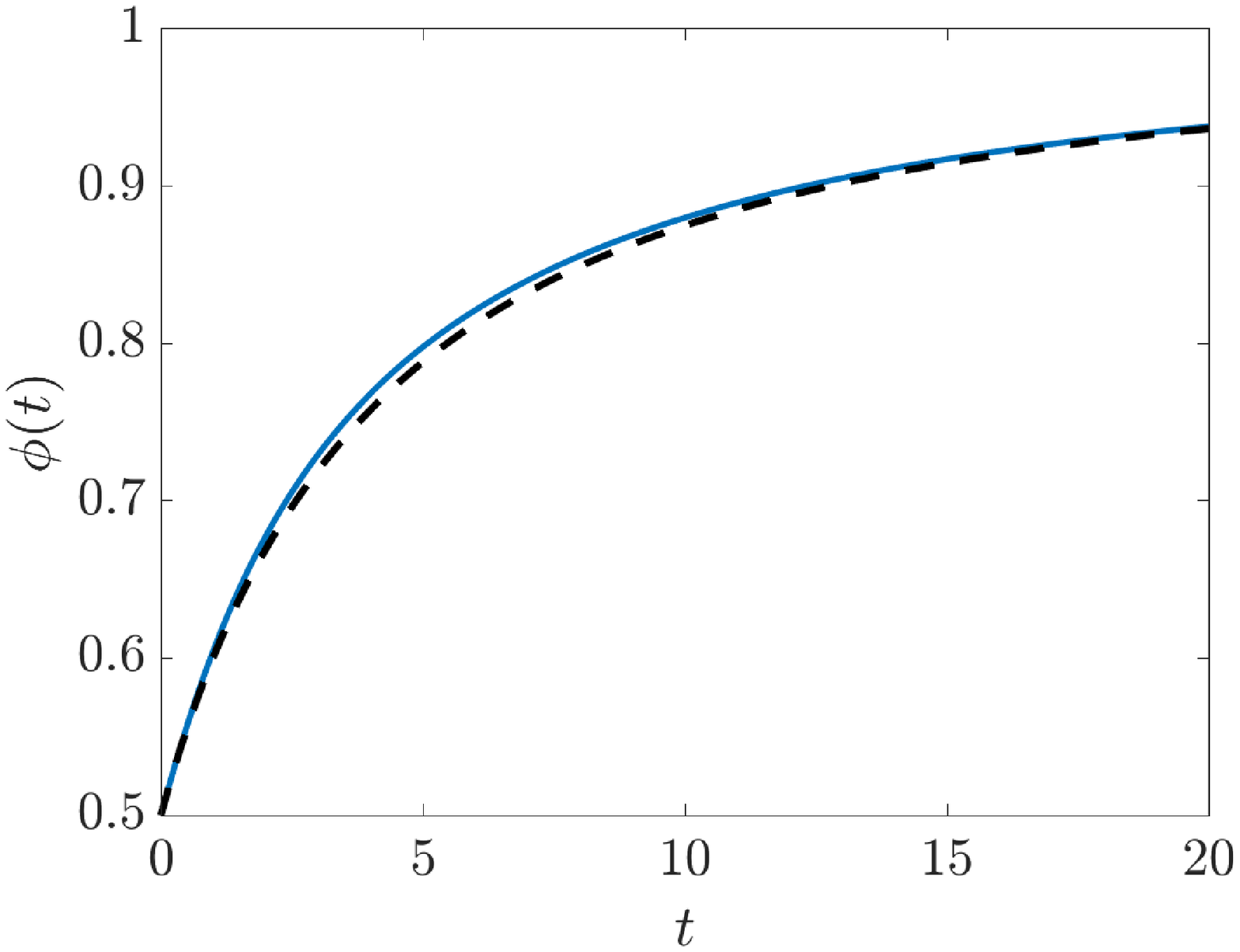}}
 \subfloat[]{\includegraphics[width= 0.45\textwidth]{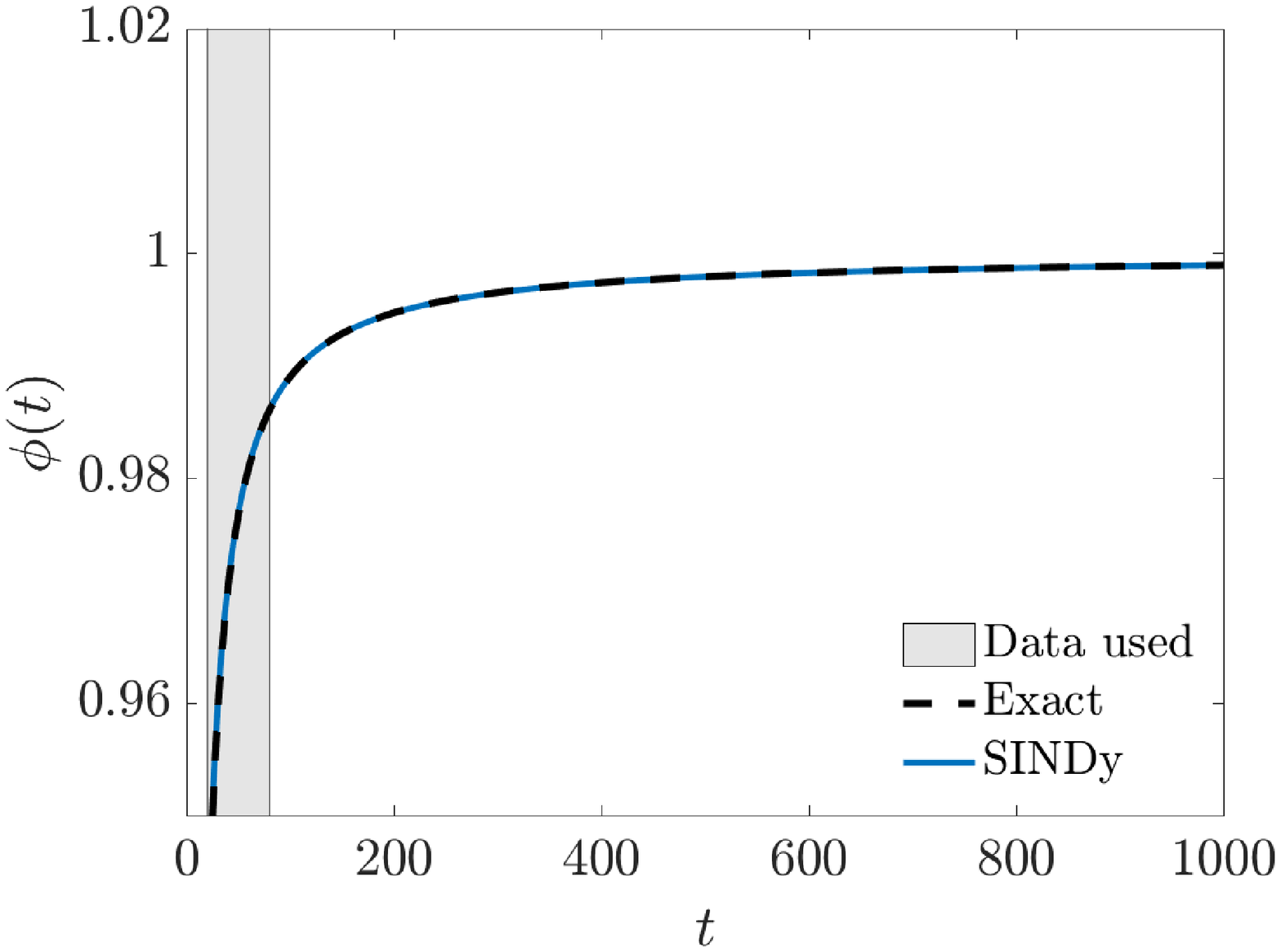}} \\
 \subfloat[]{\includegraphics[width= 0.45\textwidth]{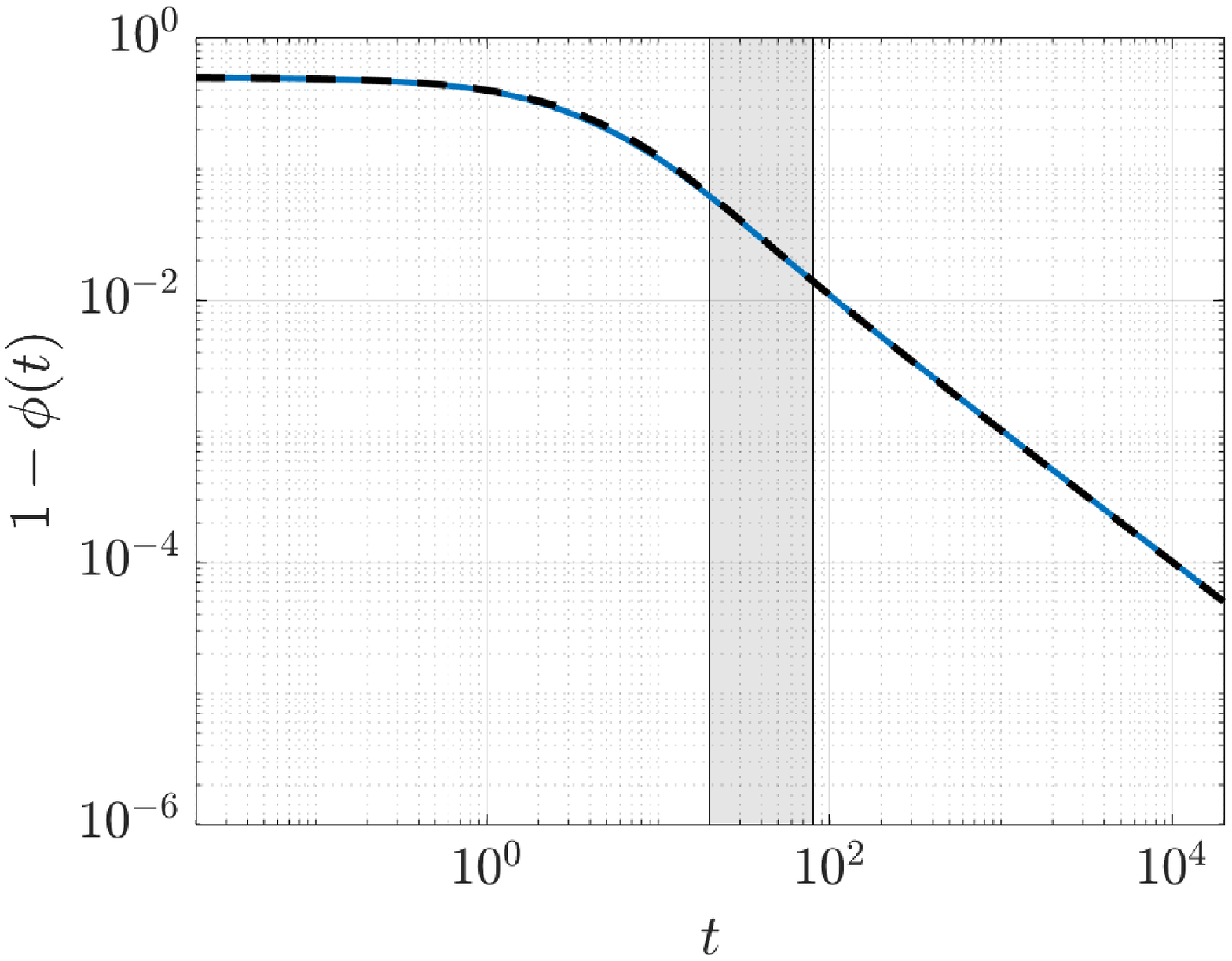}} 
 \subfloat[]{\includegraphics[width= 0.45\textwidth]{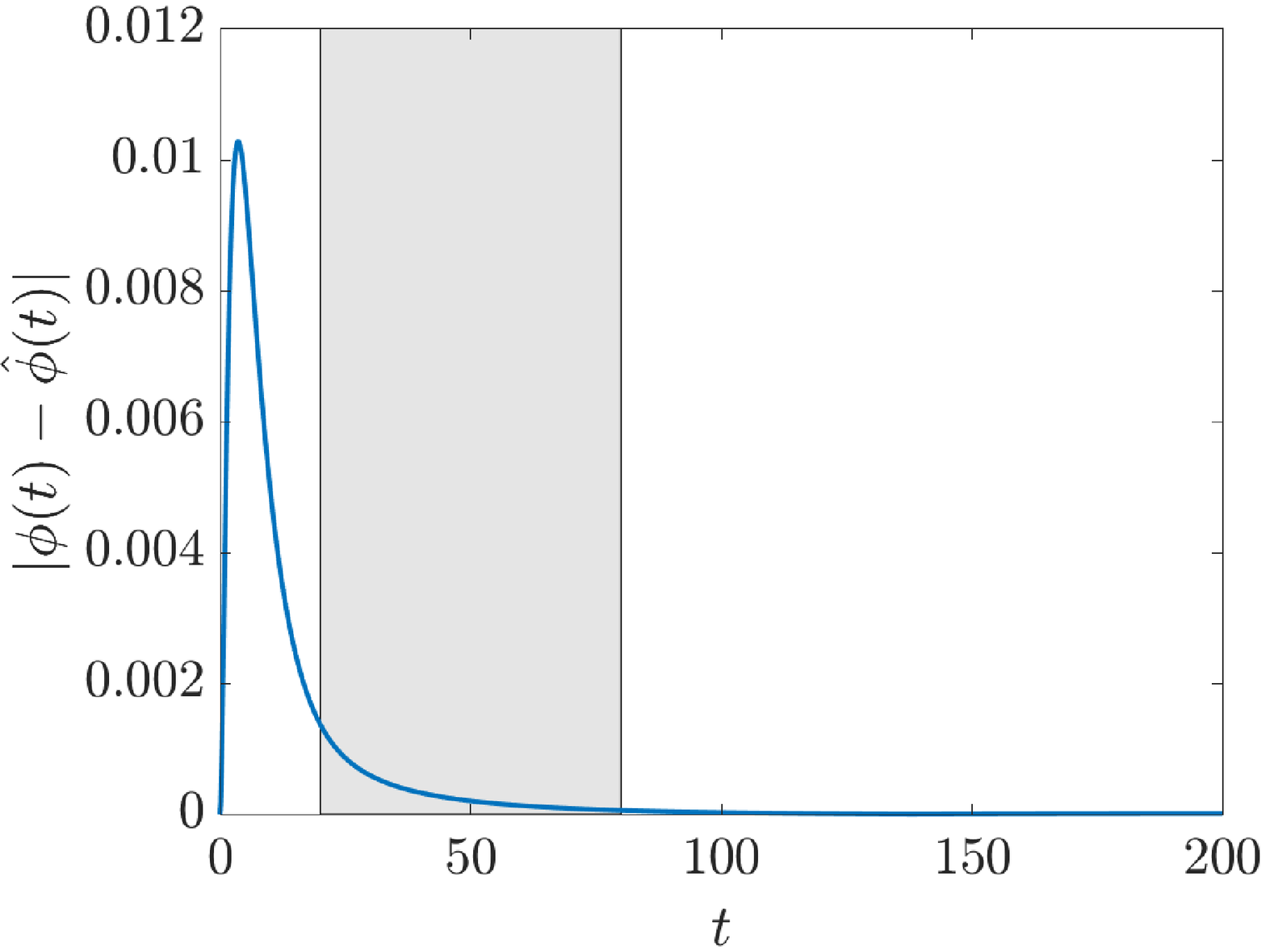}}\\
 \subfloat[]{\includegraphics[width= 0.45\textwidth]{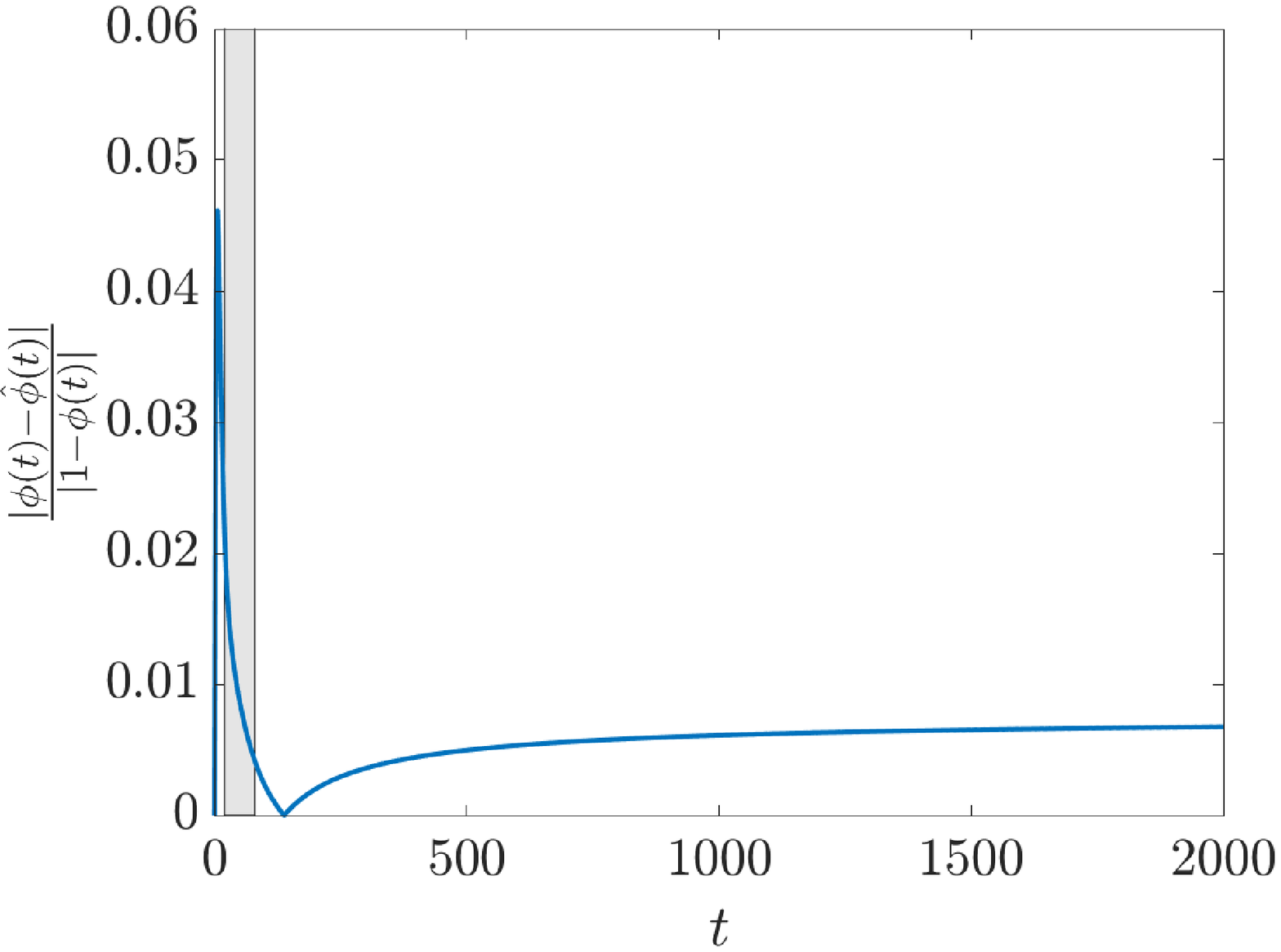}}
}
\caption{As for Fig.~\ref{fig:WagSindy2}, but where the  SINDy model for the Wagner function using a second order differential equation is identified over the interval $t \in [20, 80]$, which is indicated by the grey region of each subplot.}
\label{fig:WagErrorSindyLimited}
\end{figure}

To further demonstrate the utility and robustness of this method, we now consider identifying a model using a limited set of data.  In particular, we apply the same identification method to identify the parameters for a second order differential equation (Eq.~\eqref{eq:form2} on Wagner function data collected over the interval $t \in [20, 80]$, using the same values for the regularization parameters $\beta_2$ and $\eta$ as before. Figure~\ref{fig:WagErrorSindyLimited} shows that the resulting model again accurately captures the features of the Wagner function across all times, including times well outside the training data interval. 
The error and relative error are largest at early times, with a maximum error of approximately 0.010 at $t \approx 3.6$ and a maximum relative error of approximately 0.046 at $t \approx 6.0$, which is perhaps unsurprising given that the model has no training data for $t<20$.   
While this error is substantially larger than the approximation identified from data at all times, it is still more accurate overall than the approximations described in section \ref{sec:approx}. 
While not shown here, we found that a larger window of data was typically required to identify an accurate first order model of the form given in Eq.~\eqref{eq:form1}. 
Note that the ability to identify an accurate model from a restricted set of data is of practical importance for more realistic aerodynamic systems, where the data available may be limited.
   
It is worth noting that while the Wagner function has been approximated from the dynamics of a set of nonlinear systems, any additional physical interpretation of these models should be treated with caution.  
In particular, note that the prediction of more general system responses to changes in conditions (for example, continuously changing freestream velocity and/or angle of attack) should still proceed using linear superposition of the indicial response (the Wagner function), rather than using the nonlinear differential equation directly.  

      \begin{figure}
 \centering {
  \subfloat[]{\includegraphics[width= 0.5\textwidth]{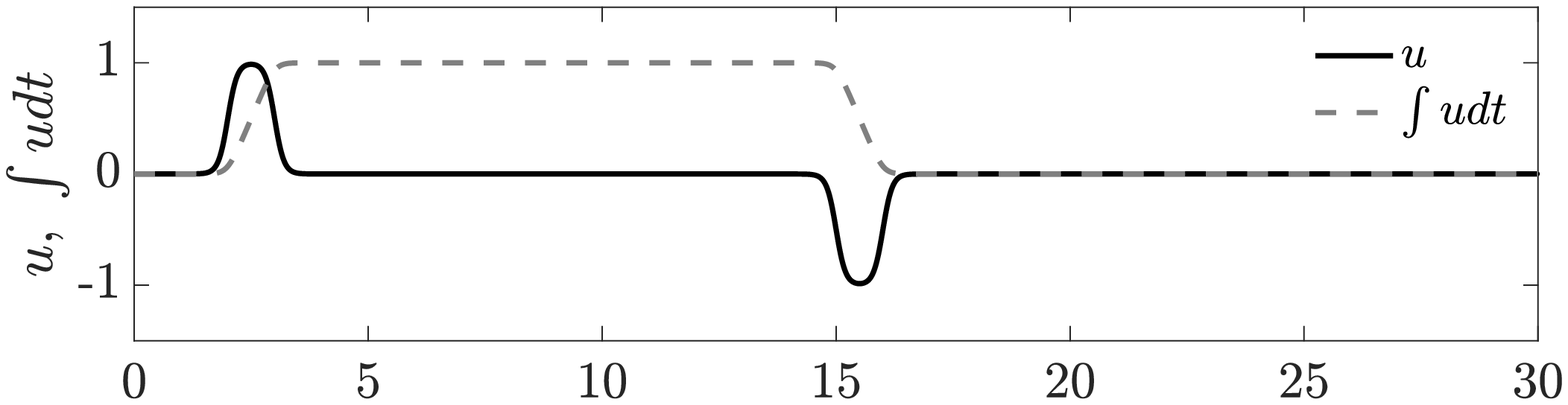}} \\
 \subfloat[]{\includegraphics[width= 0.5\textwidth]{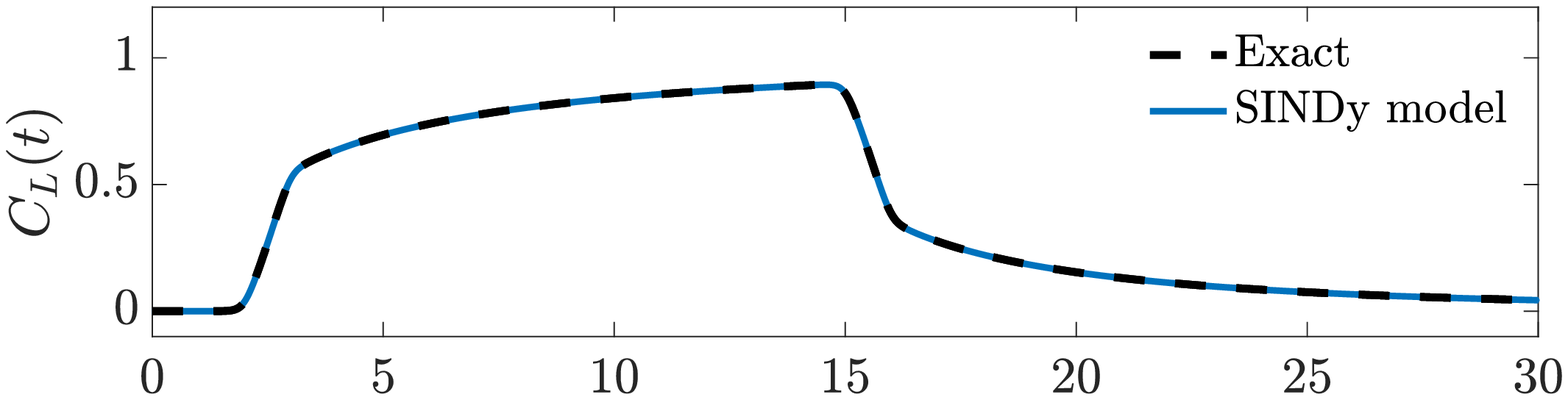}} \\
 \subfloat[]{\includegraphics[width= 0.5\textwidth]{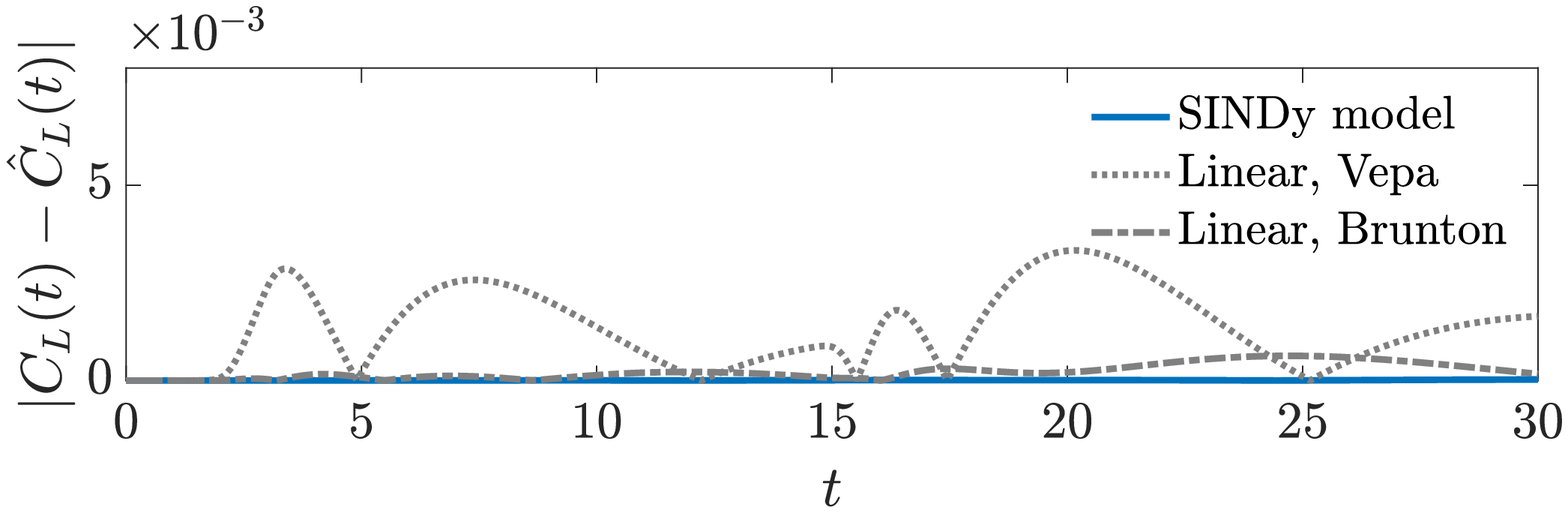}} 
}
\caption{Performance of the (second-order ODE) SINDy-based model in predicting the response to a more general input signal, shown in (a). (b) compares the response using the true and approximated Wagner function, while (c) compares the accuracy of a second-order SINDy-based model to the Vepa and Wagner linear models.}
\label{fig:man}
\end{figure}

To show the performance of the SINDy-based  models more generally, we show in Fig.~\ref{fig:man} the results from using this model to predict the response to a more general time-varying input, by applying the convolution of the input signal with the approximated Wagner function. The maneuver shown in Fig.~\ref{fig:man}(a) could, for example, represent a pitch-up, hold, pitch-down maneuver, with $u = \dot\alpha$ and $\int u dt = \alpha$. For clarity, we only show the circulatory component of the lift associated with the Wagner function (i.e.~we exclude  added mass contribution). It can be observed that the SINDy-based model (here using the second-order ODE with coefficients as identified in Table \ref{tab:coef2}) accurately predicts the response, and is more accurate than the two best-performing linear models from Fig.~\ref{fig:WagLin} (though all models give low error throughout the maneuver, and are largely indistinguishable if all plotted on Fig.~\ref{fig:WagLin} (a)).  While much of this work has focused on achieving correct asymptotic behavior of Wagner function, this example demonstrates that the SINDy-based models can also give very accurate predictions across all times, for more general input signals. 

 \subsection{Modeling startup flow over a finite-thickness airfoil with a nonplanar wake}
\label{sec:BEM}
\begin{figure}
 \centering {
  \subfloat[]{\includegraphics[width= 0.33\textwidth]{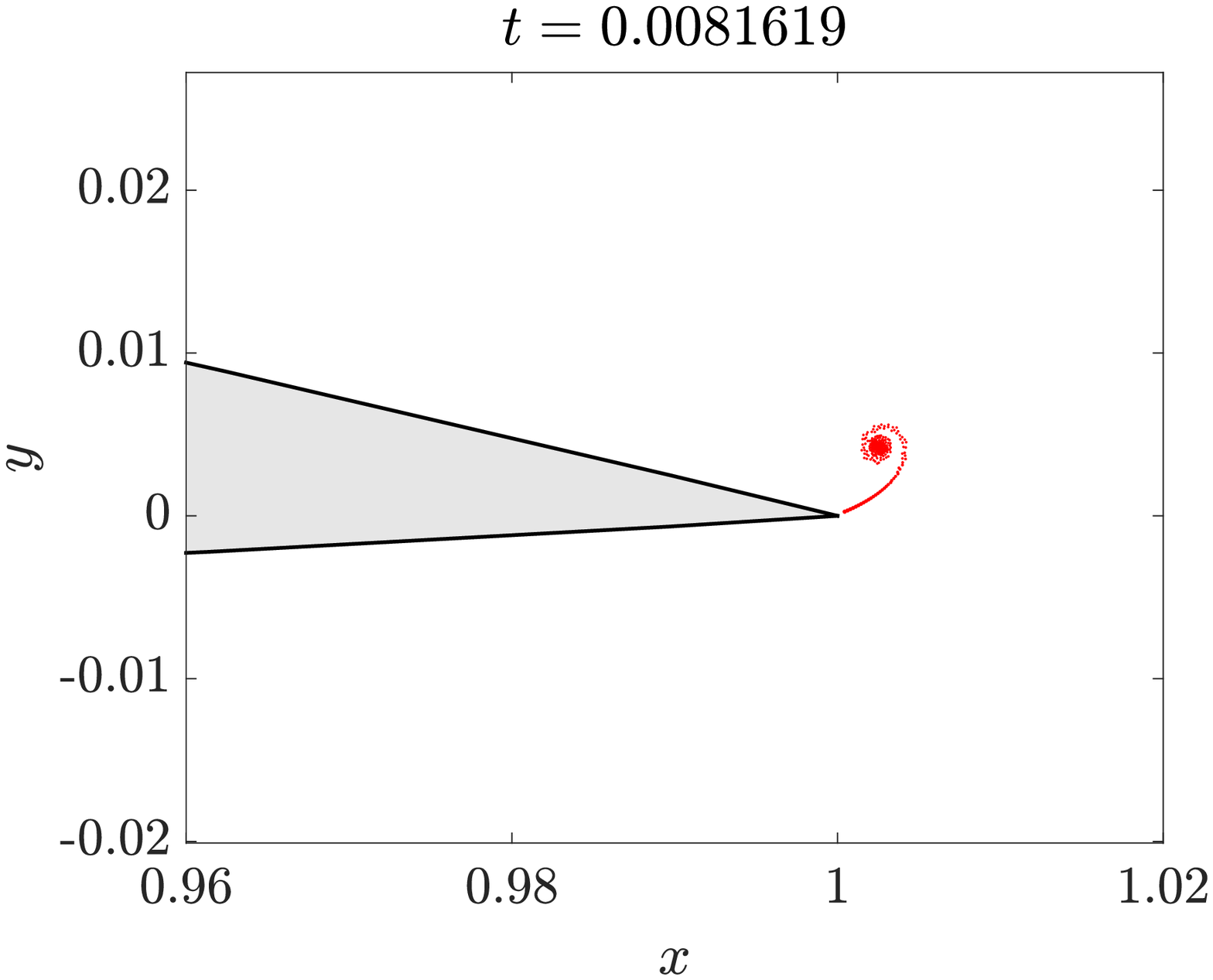}}
 \subfloat[]{\includegraphics[width= 0.33\textwidth]{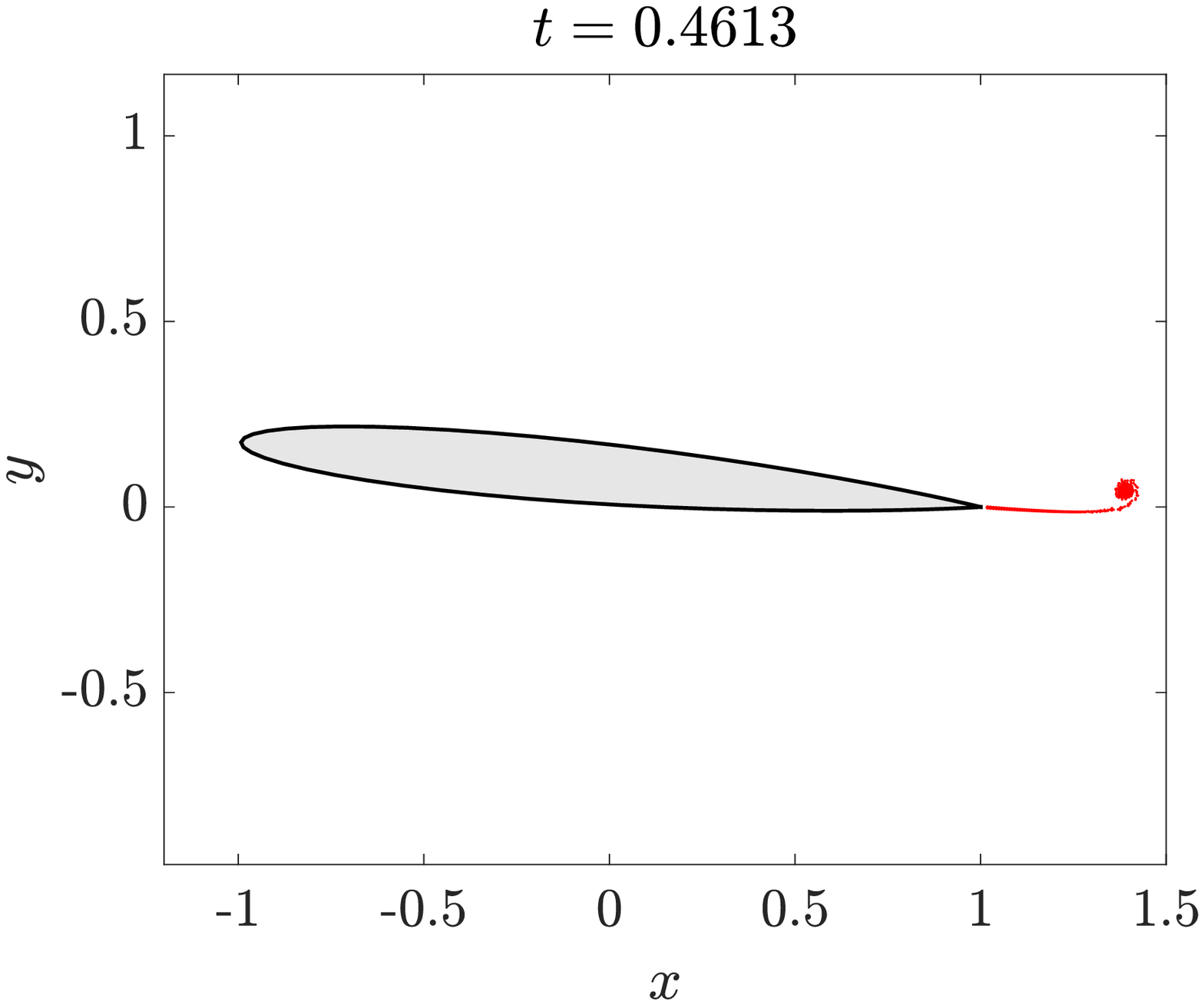}}
 \subfloat[]{\includegraphics[width= 0.32\textwidth]{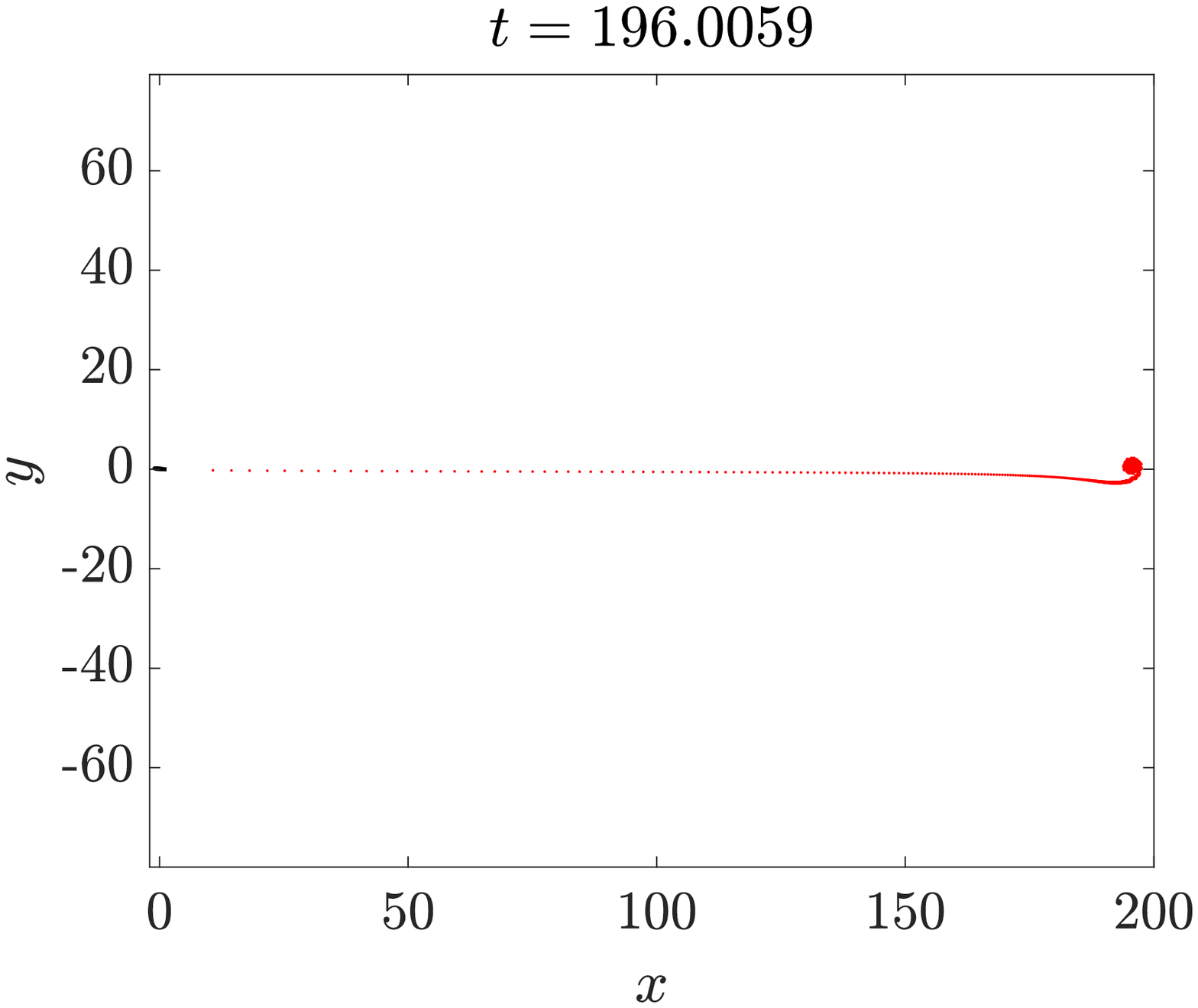}} 
 }
\caption{Visualization of the evolution of the wake vorticity for startup flow over a 8.4\% thickness von Mises airfoil at angle of attack of $5^\circ$. Each red circle represents the vorticity shed from the trailing edge at a given timestep.}
\label{fig:BEMvort}
\end{figure}

Given the well-known limitations in the accuracy of classical unsteady aerodynamic theory, it is desired that the methods developed for approximating the Wagner function in section \ref{sec:method} could also be applied for modeling more general aerodynamic systems and data. 
In this section, we  apply this method to model impulsively started flow over airfoil, but without a number of the assumptions used in the derivation of the Wagner function. 
 We consider flow over a von Mises airfoil of $8.4\%$ thickness at angle of attack of $5^\circ$ using an inviscid unsteady boundary element code. The code employs the methods developed and described in~\cite{hess1967calculation} and~\cite{basu1978unsteady}. This methodology assumes no flow separation, with a trailing-edge Kutta condition that determines the direction of the trailing-edge velocity by requiring zero pressure difference immediately downstream of the trailing edge.  The method also allows for the interaction between vortex elements shed into the wake, leading to a nonplanar wake. 
 This nonplanar wake results in a substantial departure from the wake dynamics assumed in the Wagner function for startup flow,  particularly at early times.  
 These effects were first studied in early computational work~\cite{giesing1968nonlinear}, though the very early time behavior ($t < 0.1$) was not accurately computed until  the work of~\cite{graham1983lift}, owing to the very small timesteps required to resolve the early-time dynamics.  In Fig.~\ref{fig:BEMvort}, we show the wake vorticity at several instances of time, highlighting that the wake is highly nonplanar at early times, owing to the rollup and vertical deflection of the starting vortex, but becomes increasingly planar as the starting vortex convects downstream.  The computation is performed using 800 logarithmically-spaced timesteps between $t = 10^{-5}$ and $10^3$, with the flow impulsively started at $t= 0$. For consistency, we again nondimensionalize time by the semichord and freestream velocity.
  
\begin{figure}
 \centering {
  \subfloat[]{\includegraphics[width= 0.45\textwidth]{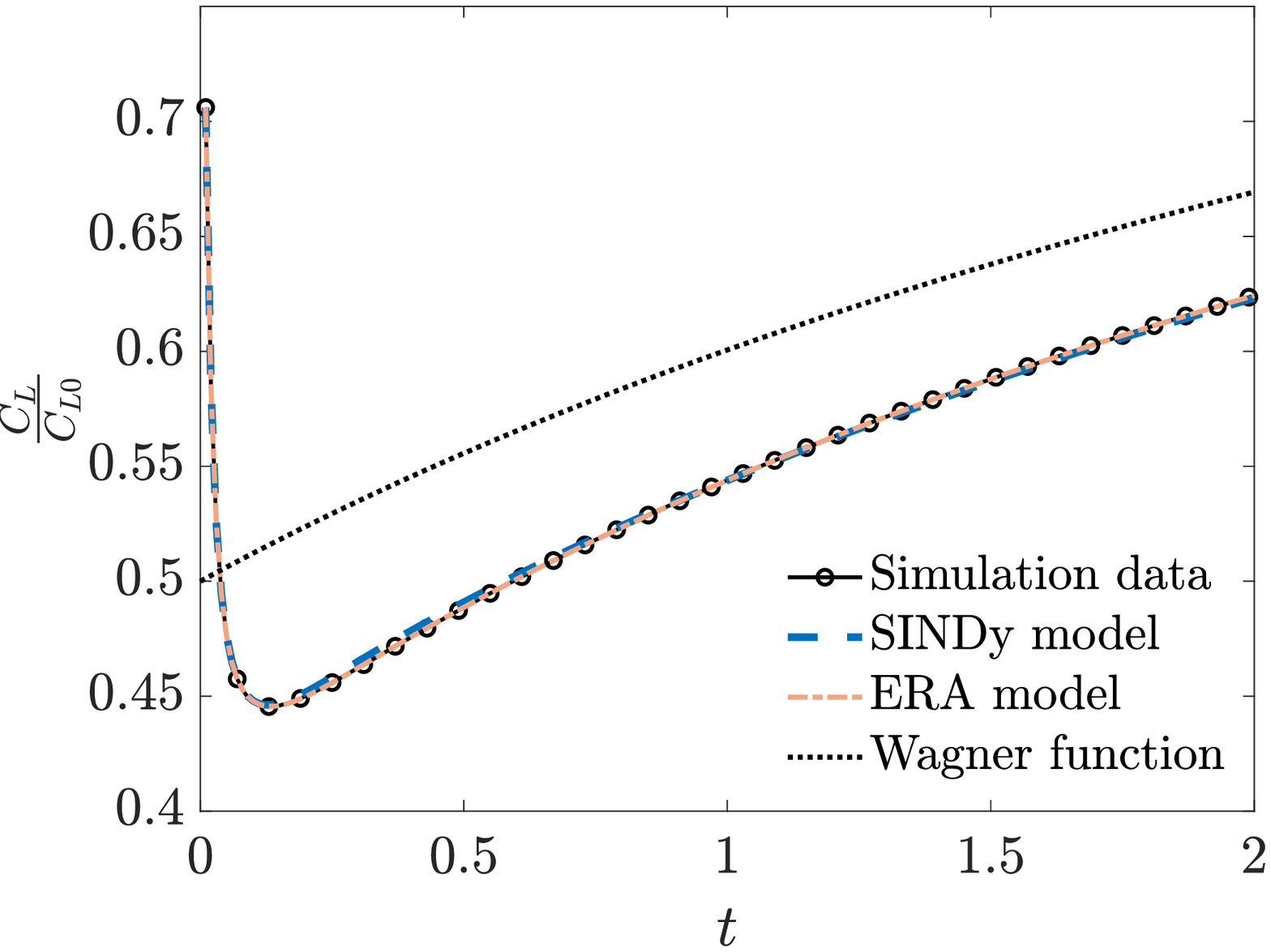}}
 \subfloat[]{\includegraphics[width= 0.45\textwidth]{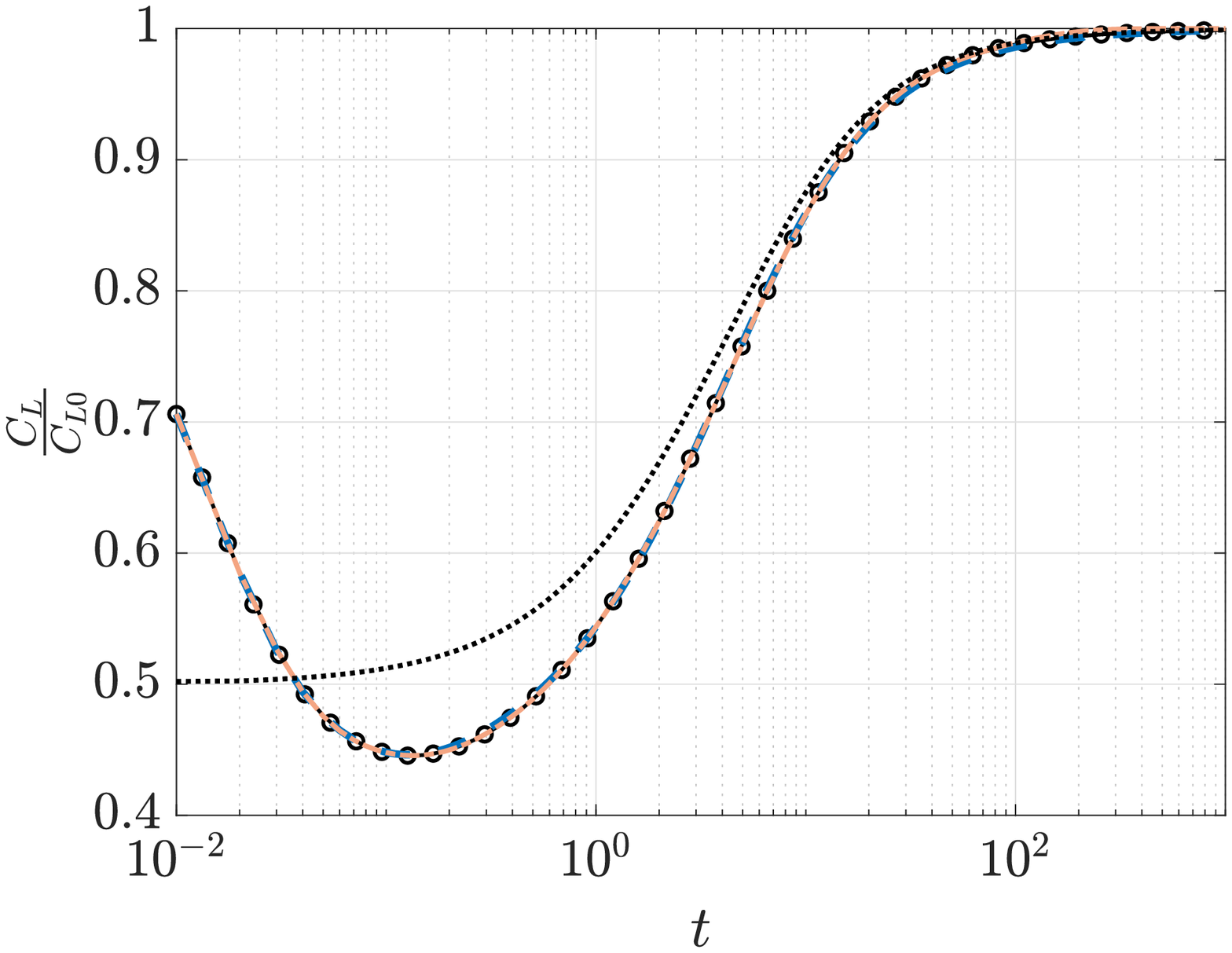}} \\
 \subfloat[]{\includegraphics[width= 0.45\textwidth]{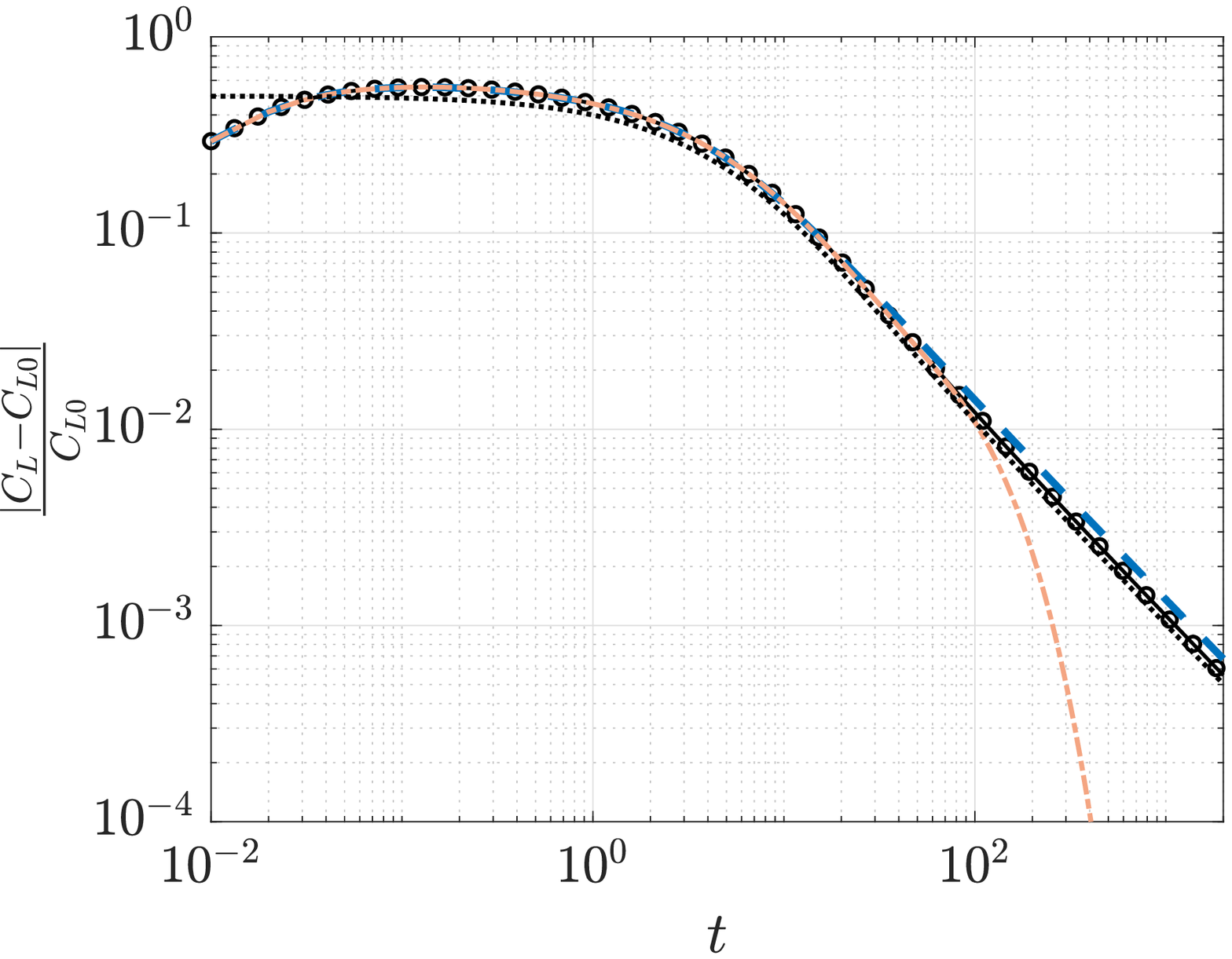}} 
 \subfloat[]{\includegraphics[width= 0.45\textwidth]{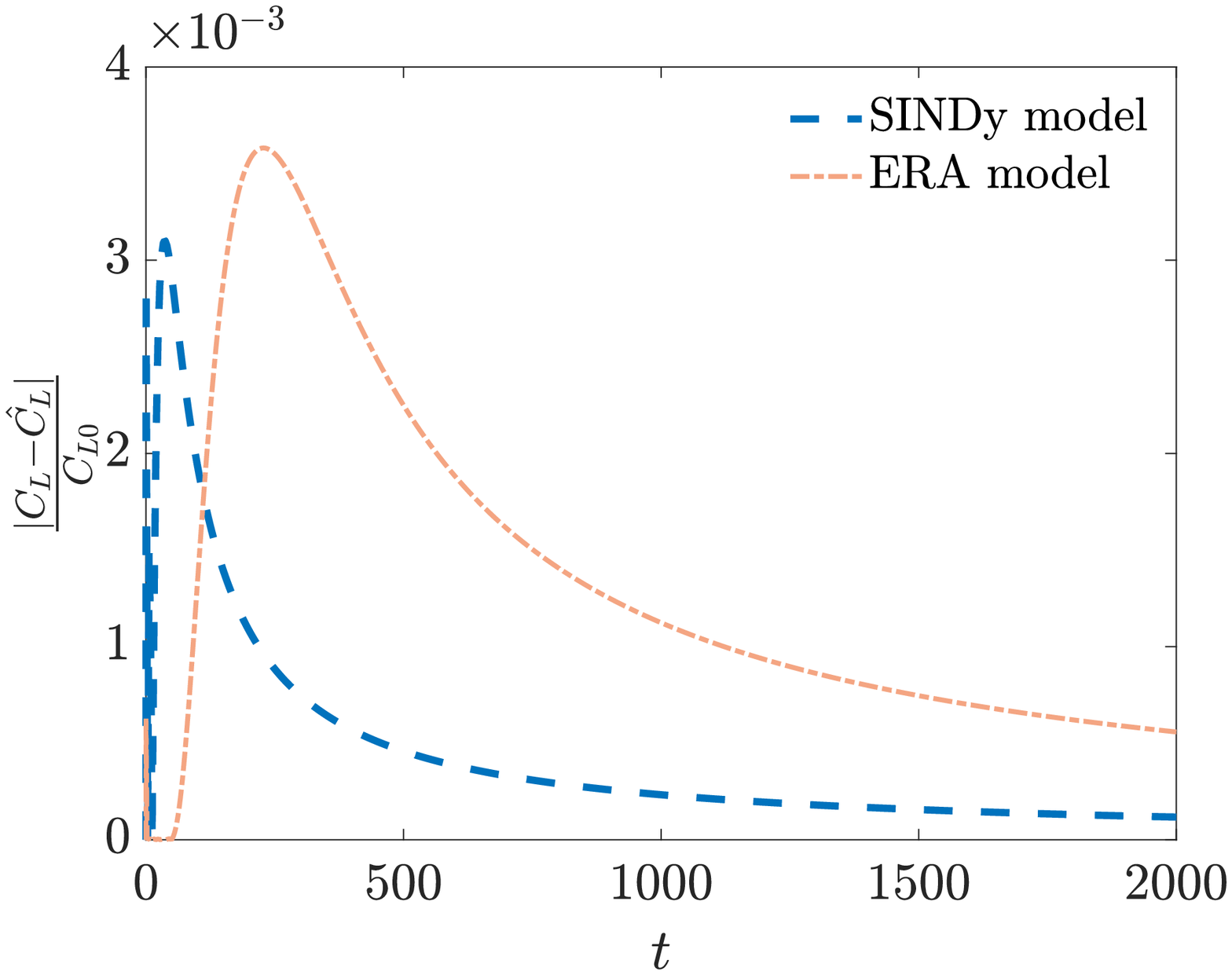}} \\
 \subfloat[]{\includegraphics[width= 0.45\textwidth]{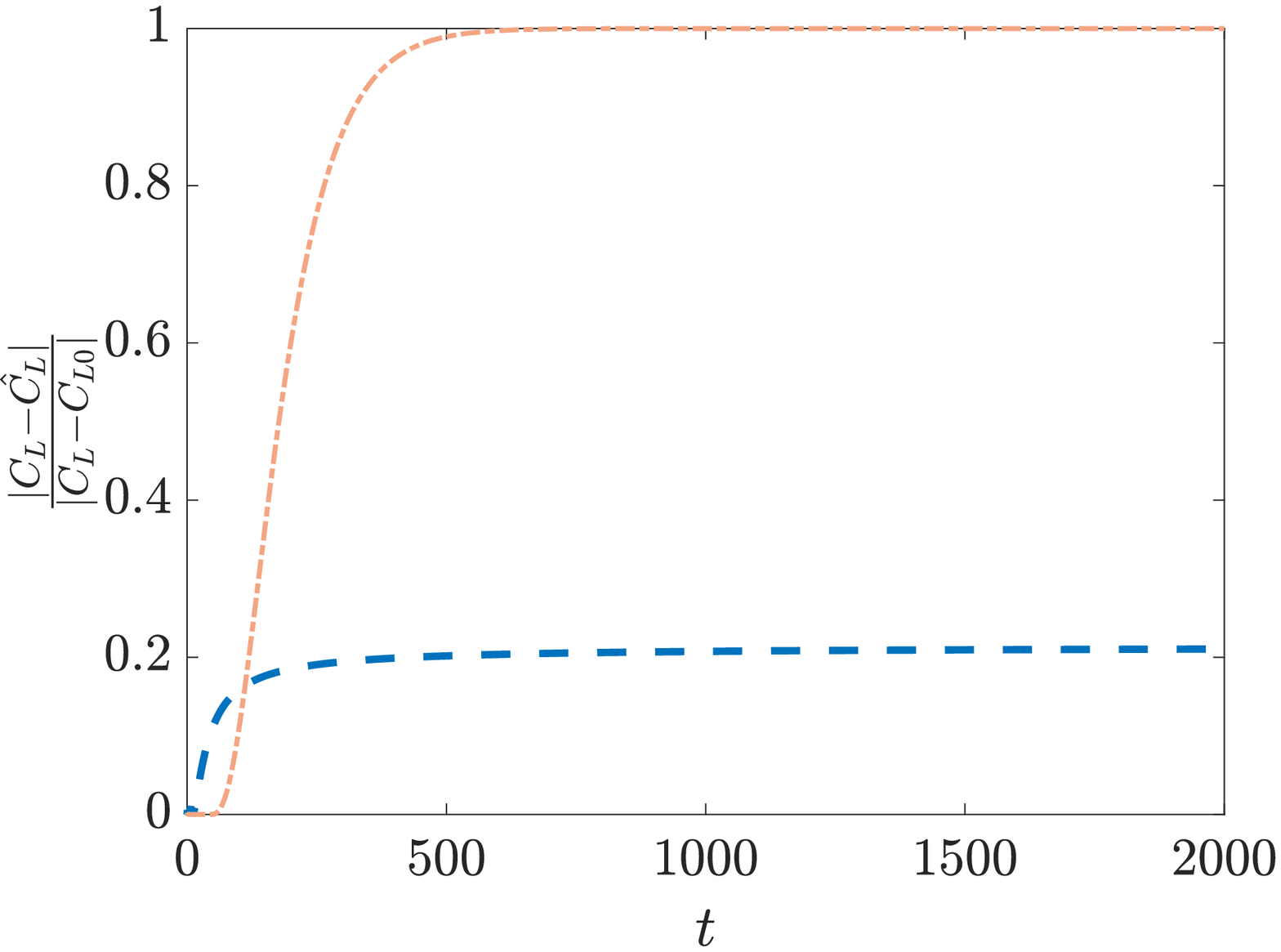}}
}
\caption{Lift response on an impulsively-started airfoil at  an angle of attack of $5^\circ$ computed using an inviscid boundary element method, compared with the Wagner function  and models identified using SINDy-based methodology and ERA. Subplots (a)--(c) plot the lift response on different domains and axis scalings, while (d) and (e) show the absolute and relative (to the asymptote) error for the two models.}
\label{fig:BEMlift}
\end{figure}

Figure~\ref{fig:BEMlift} plots the unsteady lift on the airfoil from this startup flow. Note that we do not include the lift at $t=0$ directly due to the impulsively-started freestream (first described analytically for finite-thickness airfoils in~\cite{chow1982initial}).  We observe that the lift response varies substantially from the Wagner function at early times, though has similar asymptotic behavior, as is consistent with the flow physics observed in Fig.~\ref{fig:BEMvort}. Also shown in Fig.~\ref{fig:BEMlift} is the result of modeling the lift response using the SINDy-based method described in section \ref{sec:method}, using a second order differential equation, (Eq.~\eqref{eq:form2}, again with up to cubic nonlinearities).  This model is identified on data collected for $t \leq 50$ with a fixed timestep $dt = 0.01$ and regularization parameters $\eta = 2$ and $\beta_2 = 0.01$, though very little difference is observed if data at later times is also used. For comparison, we also identify a seventh order linear dynamical system model from the same data using the eigensystem realization algorithm (ERA)~\cite{kung1978era,ERA:1985}, as described in the Appendix. For further details concerning the application of ERA to unsteady aerodynamic systems, see~\cite{brunton2013jfm,brunton2014state}.  
Subplots \ref{fig:BEMlift}(a)--(c) show this comparison over different domains and scales, to emphasize the early- and late-time behavior of the system and the two models. The absolute and relative approximation errors are plotted in subplots \ref{fig:BEMlift}(d)--(e). 
 As is the case with the Wagner function approximations considered in section \ref{sec:LinApprox}, this linear model does not capture the non-exponential asymptotic behavior, though it is very accurate for early times.  We emphasize again that this is a fundamental limitation of linear dynamical systems, independent of the specific system identification method or amount of data used. 
  
\section{Conclusions}

This work has proposed a novel method for approximating both the Wagner function, and the general response of the transient lift experienced by an airfoil that is subject to impulsively-changing conditions.  
We have shown that models for both cases can be accurately modeled by using a SINDy-based method to identify nonlinear scalar differential equations from data, which, when given an appropriate initial condition, result in a trajectory that accurately captures the transient lift. We find that the Wagner function can be approximated in this matter using both first and second order differential equations. For first order differential equations, we find that the approximation improves as the highest order nonlinearity increases up to sixth powers of the lift. The second order differential equations with third order nonlinearities give similar accuracy, and also perform better with limited data, and when applied to more general lift responses.
 
 These nonlinear differential equation models can capture the correct long-time behavior of these systems, which feature hyperbolic asymptotes in time.  This can be advantageous compared to models based on linear dynamical systems, which are not capable of modeling this asymptotic behavior.  
 This highlights the fact that, while classical unsteady aerodynamic theory is linear in the sense that the response to arbitrary inputs can be computed entirely from an indicial response via an appropriate convolution integral with the input signal, the indicial response itself does not come from a finite-dimensional linear system. 
 While many applications of nonlinear techniques for modeling aerodynamic systems are motivated primarily by the need for accuracy over a range of operating conditions (e.g.~Volterra series models~\cite{lucia2005volterra,Prazenica:2007, Balajewicz:2012}, state-space models incorporating nonlinear functions~\cite{goman:94ss,williams2015gk,luchtenburg2015gk}, linear parameter varying~\cite{hemati2016aiaa} and switched linear models~\cite{dawson2015data}), here we have highlighted that even in  the most simple unsteady aerodynamic systems without high angles of attack or large amplitude motions, linear models can be insufficient. 
 That being said, many of the previously-developed approximations to the Wagner function are likely to be sufficiently accurate for many typical applications, such as when predicting the short-time transient aerodynamic response to unsteady airfoil maneuvers by convolving the Wagner function with a specified input signal. However, even in such cases, having a comprehensive understanding the limitations in accuracy of such approximations can be helpful for understanding this potential source of error.

While this paper has proposed a method for approximating the Wagner function that is accurate at all times, it is likely that using a direct numerical computation of the Wagner function is still the safest approach, to ensure that no errors are introduced.  
However, if one wishes to obtain a convenient low dimensional representation for transient responses for aerodynamic systems that differ from the Wagner function, then the sparse identification framework detailed in section \ref{sec:sindy} provides a method to achieve this, avoiding the pitfalls of linear modes described in section \ref{sec:LinApprox}.  
Further work will extend these approaches for a wider range of aerodynamic configurations, that include, for example, viscous effects, and a wider range of angles of attack. We further will seek  to identify parametrized models that are accurate over a range parameters such as the angle of attack, Reynolds number, and airfoil shape.

\section*{Appendix}\label{sec:appendix}
This appendix briefly describes the application of the eigensystem realization algorithm to the data considered in section~\ref{sec:BEM}. For simplicity, we describe the identification of a single-input, single-output linear system in discrete time, of the form
\begin{equation}
  \label{eq:dlti}
  \begin{aligned}
   \bx_{j+1} &= \mA_d \bx_j + \mB_d u_j \\
   y_j &= \mC_d \bx_j+ \mD_d u_j.
  \end{aligned}
  \end{equation}
We start with a time-series of scalar-valued data $\{y_0,y_1,\cdots,y_m\}$ arising from a sudden change in conditions (i.e.~an impulse response), evenly spaced in time with a timestep $\Delta t$. For the data considered in Section \ref{sec:BEM}, $y_j = \frac{C_L(t_0+j\Delta t)}{C_{L0}}$. We next form the Hankel matrices
   \begin{align*} 
   \bm{H}_1 &= \begin{bmatrix} y_0   & y_1   & y_2 & \cdots & y_{m_c }  \\
 					 y_1 & y_{2} & y_{3} & \cdots & y_{m_c + 1} \\
					 \vdots & \vdots & \vdots    & \ddots & \vdots \\
					 y_{m_o} &   y_{m_o+1} &  y_{m_o+2} & \cdots & y_{m_o +m_c}  \end{bmatrix}, \\
  \bm{H}_2 &= \begin{bmatrix} y_1   & y_{2}    & y_{3} & \cdots & y_{m_c+1}  \\
 					 y_{2}  & y_{3} & y_{4} & \cdots & y_{m_c +2} \\
					 \vdots & \vdots & \vdots    & \ddots & \vdots \\
					 y_{m_o +1} &   y_{m_o+2} &  y_{m_o+3} & \cdots &y_{m_o +m_c+1}  \end{bmatrix}.
\end{align*}
For the data in section \ref{sec:BEM}, there are a total of $m = 5000$ timesteps of data available for model identification. We choose $m_o = m_c = 2499$ (meaning $\bm{H}_1$ and $\bm{H}_2$ are square) to utilize all of this data. ERA proceeds by taking the truncated singular value decomposition (SVD)
\begin{equation*}
    \bm{H}_1 \approx \mU_r \mSigma_r\mV_r^T, 
\end{equation*}
where the leading $r$ singular values and vectors are retained. From this, the matrices in Eq.~\eqref{eq:dlti} are found from
\begin{align*}
     \mA_r &= \mSigma_r^{-1/2} \mU_r^T \mH_2 \mV_r\mSigma_r^{1/2}, \\
 \mB_r &= \text{first column of }\mSigma_r^{1/2}  \mV_r^T, \\
 \mC_r &= \text{first row of }\mU_r \mSigma_r^{1/2}, \\
 \mD_r &= y_0.
\end{align*}
For the results plotted in Fig.~\ref{fig:BEMlift}, the ERA model prediction is given by the impulse response of this linear system.

\section*{Acknowledgements}
The authors thank M.~Fairchild for sharing his unsteady boundary element method code, J.-C.~Loiseau for valuable discussions about the implementation of nonlinear system identification algorithms, and K.~Asztalos for enlightening discussions on the derivation of classical relationships in unsteady aerodynamics. 
SLB acknowledges support from the Army Research Office ({ARO W}911{NF}-19-1-0045) and the Air Force Office of Scientific Research (AFOSR {FA}9550-18-1-0200).

\bibliographystyle{abbrvnat}
\bibliography{Master}

\end{document}